\documentclass[a4paper,english,11pt]{article}
\usepackage{jheppub}

\usepackage{epsfig,multicol}
\usepackage{amssymb,amsmath,bm, mathrsfs, mathtools}
 \usepackage{amsfonts}
\usepackage{amstext}
\usepackage{epsf}
\usepackage{graphicx}
\usepackage{longtable}
\usepackage{afterpage}
\usepackage{placeins}
\usepackage{color}
\usepackage{bbm}
\usepackage{afterpage}
\usepackage{amsfonts}
\newcommand{\overbar}[1]{\mkern 1.5mu\overline{\mkern-1.5mu#1\mkern-1.5mu}\mkern 1.5mu}

\usepackage{slashed}
\usepackage{textcomp}
\usepackage{placeins}

\usepackage{color}

\usepackage{tikz}
\usetikzlibrary{arrows,decorations.markings,decorations.pathmorphing}

\usepackage{subfig}
\usepackage{latexsym}
\usepackage{array}

\DeclareMathOperator{\IM}{Im}
\newcommand{\be}{\begin{equation}}
\newcommand{\ee}{\end{equation}}
\newcommand{\bea}{\begin{eqnarray}}
\newcommand{\eea}{\end{eqnarray}}
\newcommand{\nn}{\nonumber}
\newcommand{\bi}{\begin{itemize}}
\newcommand{\ei}{\end{itemize}}

\newcommand{\Lag}{\mathcal{L}}
\newcommand{\Heff}{{\cal H}_\text{NP}}
\newcommand{\Hfull}{{\cal H}_\text{full}}
\DeclareMathOperator{\diag}{diag}
\newcommand{\eps}{\epsilon}
\newcommand{\ord}{{\cal O}}
\newcommand{\Br}{{\rm Br}}

\title{Flavor from the Electroweak Scale}

\author[a,d]{Martin Bauer,}
\author[a,b,c]{Marcela Carena,}
\author[a]{Katrin Gemmler}

\affiliation[a]{Fermilab, P.O. Box 500, Batavia, IL 60510, USA}
\affiliation[b]{Enrico Fermi Institute, University of Chicago, Chicago, IL 
60637, USA}
\affiliation[c]{Kavli Institute for Cosmological Physics,University of Chicago, 
Chicago, IL 60637, USA}
\affiliation[d]{Institut f\"ur Theoretische Physik, Universit\"at Heidelberg, Germany }

\emailAdd{m.bauer@thphys.uni-heidelberg.de}
\emailAdd{carena@fnal.gov}
\emailAdd{katrin@fnal.gov}

\abstract{
We discuss the possibility that flavor hierarchies arise 
from the electroweak scale in a two Higgs doublet model, in which the two Higgs doublets jointly act as the flavon.
Quark masses and mixing angles are explained by effective Yukawa couplings, generated by higher dimensional operators involving quarks and Higgs doublets. Modified Higgs couplings yield important effects on the production cross sections and decay rates of the light Standard Model like Higgs. In addition, flavor changing neutral currents arise at tree-level and lead to strong constraints from meson-antimeson mixing.
Remarkably, flavor constraints turn out to prefer a region in parameter space that is in excellent agreement with the one preferred by recent Higgs precision measurements at the Large Hadron Collider (LHC). Direct searches for extra scalars at the LHC lead to further constraints. Precise predictions for the production and decay modes of the additional Higgs bosons are derived, and we present benchmark scenarios for searches at the LHC Run II. Flavor breaking at the electroweak scale as well as strong coupling effects demand a UV completion at the scale of a few TeV, possibly within the reach of the LHC.
}


\preprint{\\ FERMILAB-PUB-15-251-T\\ EFI-15-19}

\begin{document}

\maketitle

\section{Introduction \label{sec:intro}}

The origin of the observed hierarchies in fermion masses and mixings remains one of the most intricate puzzles of the Standard Model (SM) of particle physics. The sizes of the Yukawa couplings range over at least six orders 
of magnitude, and the magnitude of the CKM matrix elements varies between $1$ and $10^{-3}$. 
Various extensions of the SM have been proposed in order to explain these hierarchies.
In a seminal paper, Froggatt and Nielsen introduced an abelian flavor symmetry by which only the top Yukawa 
coupling is allowed as a renormalizable operator \cite{Froggatt:1978nt}. The remaining Yukawa couplings are 
generated as higher order effective operators, schematically given by
\begin{equation}\label{eq:FN1}
\mathcal{O}=y~\left(\frac{S}{\Lambda}\right)^{n}~ \bar Q~ 
H ~ q_{R} \,,
\end{equation}
where lighter fermion masses require additional insertions of 
the Froggatt-Nielsen scalar, or flavon $S$. At a given energy scale, the flavon acquires a vacuum expectation value $\langle S\rangle =f$ 
and breaks the flavor symmetry. The fundamental Yukawa couplings $y$ are anarchic and hierarchies in the 
effective Yukawas are generated by the exponents $n$ of the ratio $f/\Lambda<1$, where $\Lambda$ is the scale at which new physics sets in.
 While the Froggatt-Nielsen paradigm does neither specify the flavor breaking scale $f$ nor the new physics scale $\Lambda$, the later implementation of this mechanism by Babu and Nandi \cite{Babu:1999me} and Giudice and Lebedev \cite{Giudice:2008uua} relate the flavor breaking scale to the electroweak scale. In particular, they propose $S/\Lambda\rightarrow H^\dagger H/\Lambda^2$ in \eqref{eq:FN1}. This interesting idea however has the shortcoming that the bilinear $H^\dagger H$ is a singlet under all symmetries, in particular it cannot carry a flavor charge. As a consequence, the number of flavon insertions needed in order to generate the observed fermion mass hierarchies is ad hoc and not related to a flavor symmetry. As briefly mentioned in \cite{Giudice:2008uua}, such a connection between the electroweak and the flavor breaking scale can however be motivated in a supersymmetric model featuring two Higgs doublets. Phenomenological constraints from the SM Higgs mass and signal strengths measurements 
exclude both the original Babu-Nandi-Giudice-Lebedev model as well as a possible (minimal) supersymmetric extension. 

\begin{figure}[h]\centering
\includegraphics[scale=.4]{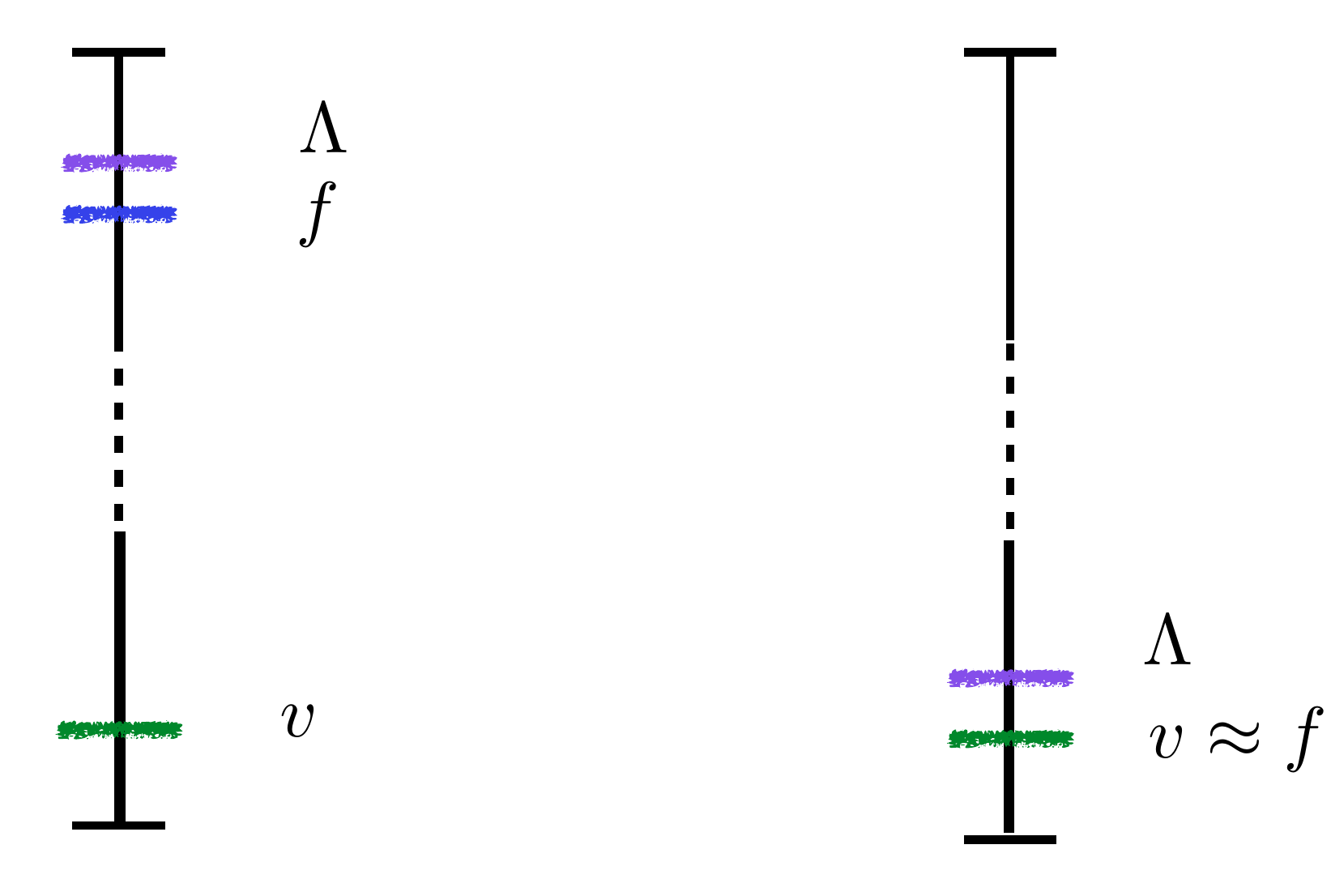}
\caption{Mass scales in a generic Froggatt-Nielsen model (left) compared to the model proposed here (right).}
\label{fig:scales}
\end{figure}

In this article, we propose a two Higgs doublet model, in which the two scalars $H_u$ and $H_d$ act jointly as the 
flavon field, such that $S/\Lambda\rightarrow H_u H_d/\Lambda^2$. As a consequence, the flavor breaking scale is set by the electroweak scale, $v\approx f$, and the new physics scale is in the ballpark of a few TeV, as sketched in Figure \ref{fig:scales}.

In the present study we concentrate on the quark sector and include the tau Yukawa couplings, reserving a full treatment of the lepton sector for future work. We discuss Higgs phenomenology, as well as its connection to flavor physics and show the potential for distinctive discovery signals that point towards an explanation of flavor at the electroweak scale.

In our model, the Higgs dependent effective Yukawa couplings induce 
tree-level flavor changing neutral currents (FCNCs) mediated by the Higgs bosons. These
FCNCs, although naively very large, turn out to be under control for a sizable 
region of the parameter space.  To this end we perform a careful study of FCNC effects in $K-\bar K$, $B_{d,s}- \bar B_{d,s}
$ mixing and estimate effects in the inclusive $B_s \rightarrow X_s  \gamma$ decay as well as in the flavor-violating top decay $t \rightarrow h c$.  
Flavor diagonal couplings of the SM-like Higgs to quarks, as well as couplings between the Higgs and 
electroweak gauge bosons, are modified with respect to the SM. While the former are unique to our model, the latter are equivalent to the Higgs couplings to gauge 
bosons in generic two Higgs doublet models \cite{Gunion:1989we,Craig:2013hca}. 
This leads to deviations in both the Higgs production cross section and decay rates and we compute these 
effects for all relevant channels to compare them with current bounds from both the ATLAS and CMS experiments. 
We perform a global fit to all SM Higgs LHC data and we can accommodate the experimental data at a $2\sigma$ level for a sizable range of model parameters. It is most  
remarkable, that the parameter space preferred by flavor observables has a significant overlap with the region preferred by the SM-like Higgs global fit.

A characteristic feature of this two Higgs doublet flavor model is, that both the constraints from Higgs signal strength measurements and 
flavor physics point to a parameter region far from the alignment/decoupling limit, such that the additional Higgs bosons
cannot be arbitrarily heavy. Furthermore, electroweak precision observables favor a large mass splitting between charged and neutral scalars, while the neutral scalar masses are preferred to be almost degenerate. As a result, direct collider searches for the additional Higgs bosons are very powerful in probing this model. We analyze the LHC results from direct searches for the CP-even and CP-odd Higgs scalars as well as for the charged Higgs boson in various production and decay modes and identify the most promising channels for a discovery.  
Although the bosonic Higgs couplings parametrically correspond to the ones in a generic two Higgs doublet model, the parameter space singled out by flavor constraints and Higgs precision measurements leads to distinctive predictions for future searches at the LHC.

Altogether, the two Higgs doublet flavor model presented in this work provides an explanation for quark masses and mixing angles from physics at the electroweak scale, while providing new opportunities for Higgs phenomenology at the LHC. The model can be tested by high precision measurements of meson-antimeson mixing and implies a UV completion at a scale that can be probed at the LHC.

This paper is organized as follows: In Section \ref{sec:setup} we introduce our model, discuss the relevant 
parameters in the Yukawa sector and constraints from quark masses and mixing angles. We subsequently 
compute the Higgs couplings to quarks in Section \ref{sec:Higgscouplings}.  In Section 
\ref{seq:Higgsprod&decay}, \ref{sec:flavor} and \ref{sec:EWPT} we investigate constraints from Higgs, flavor and electroweak 
precision observables and map out the parameter space in agreement with these constraints. Section 
\ref{sec:colliderboundsforscalars} contains a detailed analysis of present and future collider searches for the extra scalars. 
We comment on a possible UV completion Section \ref{seq:tevcompletion}. In Section \ref{sec:bench} we present benchmarks for our model, before we   
summarize our main results in Section \ref{seq:conclusions}. 

\section{Flavor from the Electroweak Scale \label{sec:setup}}
We consider a two Higgs doublet model in which fermion masses are generated by a Froggatt-Nielsen 
mechanism. We assume that the combination of the two scalar doublets $H_u H_d$ carries a non-zero flavor 
charge such that the flavon is replaced by
\begin{align}
 \frac{S}{\Lambda}\rightarrow \frac{H_u H_d}{\Lambda^2} \equiv \frac{H_u^T (i \sigma_2) H_d}{\Lambda^2}   \,.
\end{align}
We assign opposite hypercharges to the two Higgs doublets and parametrize them as 
\begin{equation}\label{eq:Higgses}
H_u=\frac{1}{\sqrt{2}}\begin{pmatrix}
v_u+\mathrm{Re}\, H_u^0+ i\, \mathrm{Im}\, H_u^0\\
\sqrt{2}\, H_u^-
\end{pmatrix}\,,\quad H_d=\frac{1}{\sqrt{2}}\begin{pmatrix}
\sqrt{2}\, H_d^+\\
v_d+\mathrm{Re}\, H_d^0+ i\, \mathrm{Im}\, H_d^0
\end{pmatrix}\,.
\end{equation}
In this setup the electroweak scale sets the flavor breaking scale by
\begin{align}
\frac{f}{\Lambda}\rightarrow \frac{\langle H_u H_d \rangle}{\Lambda^2}  = \frac{v_u v_d}{2\Lambda^2}\,,
\end{align}
where
\begin{align}
v^2=v_u^2+v_d^2\,,\qquad\frac{v_u}{v_d}=\tan\beta \,,
\end{align}
with $v=246$ GeV and $0\leq \beta \leq \pi/2$, such that $v_u$ and $v_d$ are positive. We define the expansion parameter
\begin{equation}
\varepsilon = \frac{v_u v_d}{2 \Lambda^2}= \frac{\tan\beta}{1+\tan\beta^2}\,\frac{v^2}{2\Lambda^2}\,.
\label{eq:defeps}
\end{equation} 
We choose $\varepsilon=m_b/m_t\approx 1/60$, such that the Yukawa coupling for the bottom quarks 
corresponds to an effective operator with one insertion of the Higgs doublets ($n=1$ in terms of equation 
\eqref{eq:FN1}). Therefore for $\tan\beta=1$, the new physics scale is approximately $\Lambda \approx 4 \,v \approx 1$ TeV.  If the fundamental Yukawa couplings in the UV completion are slightly larger than 1, this bound becomes weaker, and values of $\tan \beta >1$ are possible with a UV scale of the order of a TeV. Therefore, an ultraviolet completion at the TeV scale and  $\tan\beta$ of $\mathcal{O}(1)$ are
predictions of this model. We further discuss such a UV completion in Section \ref{seq:tevcompletion}.\\ 

We consider the quarks and scalars in our model to be charged under a global $U(1)_F$ symmetry. Therefore in 
the flavor eigenbasis the Yukawa sector of the SM is replaced by the effective Lagrangian (to leading order in 
powers of $\varepsilon$)
\begin{align}\label{eq:yuksector}
 \hspace{-.2cm} \Lag_\mathrm{Yuk}=\,& y_{ij}^u \left(\frac{H_u H_d}{\Lambda^2} \right)^{a_{i}-a_{u_j}-a_{H_u}}\!\bar {Q}_i H_u 
{u_R}_j + y_{ij}^d \, \left(\frac{H_u H_d}{\Lambda^2} \right)^{a_{i}-a_{d_j}-a_{H_d}}  \!\bar {Q}_i H_d {d_R}_j+h.c.\,,
\end{align} 
in which $a_{u_j}=a_u ,a_c, a_t$, and $a_{d_j}=a_d, a_s, a_b$ denote the flavor charges of the three generations of up- and 
down-type quark singlets, $a_{i}=a_1 ,a_2, a_3$ the flavor charges of the three generations of quark doublets and 
$a_{H_u}$, $a_{H_d}$ the flavor charges of the Higgs doublets. The leading order Yukawa couplings in equation \eqref{eq:yuksector} reduce to the Yukawa sector of a two Higgs doublet model of type II in the limit of vanishing flavor charge $a_i, a_{u_j}, a_{d_j}\rightarrow 0$. Couplings of $H_u (H_d)$ to the down- (up-) type quarks are suppressed by additional powers of $\varepsilon$.\footnote{We also explored choices of flavor charges in which both up- and down-type quarks couple to one of the Higgs doublets at leading order (based on a two Higgs doublet model of type I),  which will be discussed in a separate publication \cite{inprep}.}
The fundamental Yukawa couplings $y_{ij}^u$ and $y_{ij}^d$ are considered to be anarchic and of $\ord(1)$.
In writing 
equation \eqref{eq:yuksector} we normalized the sum of the Higgs charges to $a_{H_u}+a_{H_d}=1$. 

The 
effective Yukawa couplings are then given by 
\begin{align}\label{eq:effuse}
 \left(Y_u\right)_{ij} = y_{ij}^u \, {\varepsilon}^{a_{i}-a_{u_j}-a_{H_u}} \,,\qquad
  \left(Y_d\right)_{ij}=  y_{ij}^d \, \varepsilon^{a_i-a_{d_j}-a_{H_d}}\,.
\end{align}
In \eqref{eq:yuksector} and \eqref{eq:effuse}, repeated indices between $y_{ij}$ and $\varepsilon^{a_{i}-a_{u_j}-a_{H_u}}$ are \emph{not} summed over, \emph{i.e.},  for example  $\left(Y_u\right)_{12} = y_{12}^u \,
\varepsilon^{a_1-a_c-a_{H_u}}$. 
Thus the hierarchy of the effective Yukawa couplings is determined by the structure of the exponents of $
\varepsilon$. 
The rotation to the mass eigenbasis is performed via
\begin{align}
 Y_{u,d}=U_{u,d}\, 
\lambda_{u,d}\, W_{u,d}^\dagger \,,
\label{eq:quarkmassrot}
\end{align}
with diagonal matrices given by
\begin{align}
 \lambda_u=\frac{\sqrt{2}}{v_u}\,\diag(m_u, m_c, m_t )\,,\quad
\lambda_d=\frac{\sqrt{2}}{v_d}\,\diag (m_d, m_s, m_b )\,,
\end{align}
and unitary rotation matrices  $U_{u,d}, W_{u,d}$. 
 
In the following we fix the flavor charges of the quarks and Higgs bosons by imposing constraints from quark 
masses and the CKM matrix.
If the charges of the three generations of quark doublets and singlets are ordered such that
\begin{equation}
a_1 \geq a_2 \geq a_3\,,\quad a_t \geq a_c \geq a_u\,,\quad  a_b \geq a_s \geq a_d\,,
\end{equation}
one can derive the $\mathcal{O}(
\varepsilon)$ dependence for the quark masses and rotation matrices \cite{Froggatt:1978nt},
\begin{align}
\label{eq:ordmag}
m_{u_j}&\propto\frac{v_u}{\sqrt{2}}\, \varepsilon^{a_j-a_{u_j}-a_{H_u}}\,,\qquad m_{d_j}\propto\frac{v_d}{\sqrt{2}}\, 
\varepsilon^{a_j-a_{d_j}-a_{H_d}}\\[2pt]
 (U_q)_{ij}\propto&\,\varepsilon^{|a_{i}-a_{j}|}\,,\qquad  
(W_u)_{ij}\propto\varepsilon^{|a_{u_i}-a_{u_j}|}\,, \qquad 
(W_d)_{ij}\propto\varepsilon^{|a_{d_i}-a_{d_j}|}\,,\notag
\end{align}
for $i,j=1,2,3$. In the numerical analysis we will use the full unitary rotation matrices and include a scanning of anarchic Yukawa couplings with arbitrary phases and absolute 
values $| y_{ij}^{u,d
}| \in [0.5,1.5]$. 
Six of the 11 flavor charges are fixed by the quark masses. We choose
\begin{align}\label{eq:qmass}
m_t\approx \frac{v_u}{\sqrt{2}}\,,\quad 
 \frac{m_b}{m_t}\approx \frac{m_c}{m_t}\approx \varepsilon^1\,, \quad  \frac{m_{s}}{m_t} \approx \varepsilon^2\,,
\quad\frac{m_{d}}{m_t}\approx \frac{m_{u}}{m_t}\,\approx \varepsilon^3\,.
\end{align}
Additional conditions follow from the CKM matrix,
\begin{align}
V_\mathrm{CKM}=U_u^\dagger \, U_d\,,
\end{align}
by fixing
\begin{equation}\label{eq:ckmcond}
(V_\mathrm{CKM})_{12}\approx \varepsilon^0\,,\qquad  (V_\mathrm{CKM})_{13}\approx 
(V_{\mathrm{CKM}} )_{23}\approx \varepsilon^1\,.
\end{equation}
These conditions end up fixing only two parameters. 
Including the normalization of the Higgs charges $a_{H_u}+a_{H_d}=1$ and our choice of $a_{H_u}=1$, we 
have 10 conditions on 11 parameters\footnote{Different choices for the normalization condition or the Higgs charges, \textit{e.g.} $a_{H_d}=1, a_{H_u}=0$, do not change the physics of this model but will only imply different assignments for the quark flavor charges.}. The remaining choice allows for an overall shift of quark flavor charges. 
Physical quantities 
however only depend on invariant differences. Thus the 
remaining choice
does not have any phenomenological consequences and we set
\begin{align}\label{eq:FCharges}
\begin{matrix}
a_{H_u}\,=\,1\,,\\[2pt]
 a_{H_d}\,=\,0\,,
\end{matrix}\qquad
\begin{matrix}
a_1\,=\,2\,,\\[2pt]
a_2\,=\,2\,,\\[2pt]
a_3\,=\,1\,,
\end{matrix}\qquad
\begin{matrix}
a_u\,=-2\,,\\[2pt]
a_c\,=\phantom{-}0\,,\\[2pt]
a_t\,=\phantom{-}0\,,
\end{matrix}
\qquad
\begin{matrix}
a_d\,=-1,\\[2pt]
a_s\,=\phantom{-}0\,,\\[2pt]
a_b\,=\phantom{-}0\,.
\end{matrix}
\end{align}
If the last condition \eqref{eq:ckmcond} is replaced by
\begin{equation}
(V_{\mathrm{CKM}})_{12}\approx(V_
{\mathrm{CKM}})_{13}\approx (V_{\mathrm{CKM}})_{23} \approx \varepsilon^0\,,
\end{equation}
only the structure of the quark masses is explained by the flavor charges, while the hierarchical form of the CKM matrix is determined by the fundamental Yukawas $y_{ij}^u$, $y_{ij}^d$. In this case, a suitable choice of flavor charges read
\begin{align}\label{eq:FCharges2}
\begin{matrix}
a_{H_u}\,=\,1\,,\\[2pt]
 a_{H_d}\,=\,0\,,
\end{matrix}\qquad
\begin{matrix}
a_1\,=\,2\,,\\[2pt]
a_2\,=\,2\,,\\[2pt]
a_3\,=\,2\,,
\end{matrix}\qquad
\begin{matrix}
a_u\,=-2\,,\\[2pt]
a_c\,=\phantom{-}0\,,\\[2pt]
a_t\,=\phantom{-}1\,,
\end{matrix}
\qquad
\begin{matrix}
a_d\,=-1,\\[2pt]
a_s\,=\phantom{-}0\,,\\[2pt]
a_b\,=\phantom{-}1\,.
\end{matrix}
\end{align}
This choice of charges is motivated by considerably weaker constraints from flavor observables due to the aligned charges for the left-handed quark fields. \\

A detailed implementation of lepton masses and mixing angles is beyond the scope of this work. We will however define the couplings of the tau leptons to the scalars in our model, since they are important for the Higgs phenomenology. We set  
\begin{align}
 \mathcal{O}_\tau^b= y_\tau \, \frac{H_u H_d}{\Lambda^2}\,\bar \tau_L\, H_d \tau_R \, ,
\end{align}
such that $m_\tau/m_t\approx \varepsilon$.

\section{Higgs Couplings \label{sec:Higgscouplings}}

The Yukawa interactions give rise to modifications to flavor diagonal Higgs couplings as well as potentially 
dangerous flavor changing neutral currents. 
In the flavor eigenbasis the interaction between quarks and the real neutral components of the Higgs doublet 
scalars follows from \eqref{eq:yuksector} and we obtain
\begin{align}
   \Lag_0=& \left(Y_u\right)_{ij}
\left[(1+a_i-a_{u_j}-a_{H_u^0})\,\mathrm{Re} \,H_u^0+
(a_i-a_{u_j}-a_{H_u^0})\,\tan\beta\, \mathrm{Re}\, H_d^0
\right] \bar 
u_{L_i}u_{R_j}\\
& \,+\left(Y_d\right)_{ij} 
\left[(1+a_i-a_{d_j}-a_{H_d^0})\,\mathrm{Re}\,H_d^0 +
(a_i-a_{d_j}-a_{H_d^0})\,\cot\beta\, \mathrm{Re}\,H_u^0
\right] \bar d_{L_i}d_{R_j}\,+h.c..\notag
\end{align}
We rotate to the quark mass eigenbasis, according to equation \eqref{eq:quarkmassrot} and introduce the Higgs 
mass eigenstates as defined in Appendix \ref{app:Higgspotential}. The rotation of the scalars gives rise to the 
following couplings between the scalar mass eigenstates and quarks 
\begin{align}
  \Lag_0=&\, \left(b_u\right) \bar 
u_{L_i}\,h\,u_{R_j} + \left(b_d\right) \bar 
d_{L_i}\,h\,d_{R_j} 
+ \left(B_u\right) \bar 
u_{L_i}\,H\,u_{R_j} + \left(B_d\right) \bar 
d_{L_i}\,H\,d_{R_j}\,+h.c.,
\end{align}
in which
\begin{align}
\left(b_u\right)_{ij}&= \notag
\left(Y_u\right)_{ij}\,\left[(1+a_i-a_{u_j}-a_{H_u})\cos\alpha-(a_i-a_{u_j}-a_{
H_u})\sin\alpha\tan\beta\right]\,,\\
\left(b_d\right)_{ij}&= 
\left(Y_d\right)_{ij}\,\left[-(1+a_i-a_{d_j}-a_{H_d})\sin\alpha+(a_i-a_{d_j}-a_{
H_d})\cos\alpha\cot\beta\right]\,,\notag\\
\left(B_u\right)_{ij}&= 
\left(Y_u\right)_{ij}\,\left[(1+a_i-a_{u_j}-a_{H_u})\sin\alpha+(a_i-a_{u_j}-a_{
H_u})\cos\alpha\tan\beta\right]\,,\notag\\
\left(B_d\right)_{ij}&= 
\left(Y_d\right)_{ij}\,\left[(1+a_i-a_{d_j}-a_{H_d})\cos\alpha+(a_i-a_{d_j}-a_{
H_d})\sin\alpha\cot\beta\right]\,.
\end{align}
After rotating to the quark mass eigenbasis, 
\begin{align}
g_{hu_iu_j}&= (U_u^\dagger )_{ik}\, \left(b_u\right)_{kl}\, (W_u )_{lj}\,,\qquad g_{Hu_iu_j}= (U_u^\dagger
)_{ik}\, \left(B_u\right)_{kl}\, (W_u )_{lj}\,,\notag\\
g_{hd_id_j}&= (U_d^\dagger )_{ik}\, \left(b_d\right)_{kl}\, (W_d)_{lj}\,,\qquad g_{Hd_id_j}= (U_d^\dagger
)_{ik}\, \left(B_d\right)_{kl}\, (W_d )_{lj}\,,
\end{align}
we find for the couplings of the light neutral scalar $h$,
\begin{align}
 g_{h {u_i}{u_j}}&= \left(\frac{m_u}{v}\right)_{ij} \delta_{ij} \left[\frac{c_\alpha}{s_\beta}
- a_{H_u}\, f(\alpha,\beta)\right]
+ f(\alpha,\beta)\,\left[ \mathcal{Q}^u_{ij}\left(\frac{m_u}{v}\right)_{jj}-\left(\frac{m_u}{v}
\right)_{ii}\mathcal{U}_{ij}\right]\,,\notag\\[2pt]
  g_{h {d_i}{d_j}}&=\left(\frac{m_d}{v}\right)_{ij}  \delta_{ij} \left[-\frac{s_\alpha}{c_\beta}-a_{H_d}\, f(\alpha,\beta)
\right]
+ f(\alpha,\beta) \left[ \mathcal{Q}^d_{ij}\left(\frac{m_d}{v}\right)_{jj}-\left(\frac{m_d}{v}\right)_{ii}\mathcal{D}_{ij}
\right]\,,\label{eq:hcoupling}
\end{align}
and for the heavy neutral scalar  $H$,
\begin{align}
g_{H {u_i}{u_j}}&= \left(\frac{m_u}{v}\right)_{ij} \delta_{ij} \left[\frac{s_\alpha}{s_\beta} 
-a_{H_u}\,F(\alpha,\beta)\right]
+F(\alpha,\beta)\left[ \mathcal{Q}^u_{ij}\left(\frac{m_u}{v }\right)_{jj}-\left(\frac{m_u}{v}
\right)_{ii}\mathcal{U}_{ij}\right]\,,\notag\\[2pt]
  g_{H {d_i}{d_j}}&=\left(\frac{m_d}{v}\right)_{ij}  \delta_{ij} \left[\frac{ c_\alpha}{c_\beta}-a_{H_d}\, F(\alpha,\beta) 
\right]
+ F(\alpha,\beta) \left[\mathcal{Q}^d_{ij}\left(\frac{m_d}{v}\right)_{jj}-\left(\frac{m_d}{v}\right)_{ii}\mathcal{D}_{ij}
\right]\,,\label{eq:Hcoupling}
\end{align}
in which $m_u=\mathrm{diag} (m_u, m_c, m_t)$, $m_d=\mathrm{diag} (m_d, m_s, m_b)$ and we define $s_\varphi=\sin \varphi, c_\varphi=\cos \varphi$ and $t_\varphi=\tan \varphi$, for any angle $\varphi$.
In both \eqref{eq:hcoupling} and \eqref{eq:Hcoupling}, repeated indices are not summed over and we suppress 
the chirality index of the fermions $q_i \equiv q_{L_i}$, $q_j \equiv q_{R_j}$.
We make use of the following trigonometric functions 
\begin{align}\label{eq:trig}
f(\alpha,\beta)=\frac{c_\alpha}{s_\beta}-\frac{s_\alpha}{c_\beta}=c_{\beta-\alpha}\left(\frac{1}{t_\beta}-t_\beta
\right)  +2 s_{\beta-\alpha}\,,\notag\\
F(\alpha,\beta)=\frac{c_\alpha}{c_\beta}+\frac{s_\alpha}{s_\beta}=2 c_{\beta-\alpha}  + s_{\beta-\alpha}\left(t_
\beta-\frac{1}{t_\beta}\right)\,,
\end{align}
which are universal for up- and down-type quarks. We also define the matrices
\begin{align}
 \mathcal{Q}^u_{ij}&=\sum_{\ell=1}^3\,(U_u)_{\ell i}^\ast\,(U_u)_{\ell j}  \,a_\ell\,,\qquad \,\,\quad\mathcal{Q}
^d_{ij}=\sum_{\ell=1}^3\,(U_d)_{\ell i}^\ast\,(U_d)_{\ell j}\,a_\ell\,,\notag\\
  \mathcal{U}_{ij}&=\sum_{k=1}^3 \,(W_u)_{k i}^\ast\, (W_u)_{k j} \,a_{u_k}\,,\qquad 
   \mathcal{D}_{ij}=\sum_{k=1}^3 \,(W_d)_{k i}^\ast\,(W_d)_{k j}  \,a_{d_k}\,.
   \label{eq:quarkmix}
\end{align}
The structure of these matrices is fixed by the flavor charges, as given at the end of Section \ref{sec:setup}. We find for the flavor charges in 
\eqref{eq:FCharges},
\begin{align}\label{eq:struc}
 \mathcal{Q}^u\sim  \mathcal{Q}^d \sim
 \begin{pmatrix}
 2                            &\,\,\varepsilon^2 &   \varepsilon\\
\,\,\varepsilon^2      &2			 &\varepsilon\\
\,\,\varepsilon	      &\,\, \varepsilon    &1
\end{pmatrix}\,, \qquad 
 \mathcal{U} \sim
 \begin{pmatrix}
 -2			                   &\,\,\varepsilon^2	        &\,\varepsilon^2\\
		\,\,  \varepsilon^2\,      &\,\,\varepsilon^2    &\,\,\varepsilon^4\\
		  \,\,\varepsilon^2 & \,\,\varepsilon^4   &\,\,\varepsilon^4
\end{pmatrix}\,, \qquad  \mathcal{D} \sim\begin{pmatrix}
-1			&\,\,\varepsilon&\,\,\varepsilon\\
\,\,\varepsilon	&\,\,\varepsilon^2&\,\,\varepsilon^2\\
\,\,\varepsilon	& \,\,\varepsilon^2 &\,\,\varepsilon^2
\end{pmatrix}\,.
\end{align}
For completeness, we also give the expressions for these matrices in the case of the flavor charges
\eqref{eq:FCharges2},
\begin{align}\label{eq:struc2}
 \mathcal{Q}^u\sim  \mathcal{Q}^d \sim\begin{pmatrix}\,\,2&\,\,0&\,\,0\\
\,\,0&\,\,2&\,\,0\\
\,\,0& \,\,0&\,\,2
\end{pmatrix}\,, \qquad \,\,
 \mathcal{U} \sim\begin{pmatrix}
 -2&\varepsilon^2&\,\,\varepsilon^3\\
\varepsilon^2&\,\,\varepsilon^2&\varepsilon\\
\,\,\varepsilon^3& \varepsilon &1
\end{pmatrix}\,,\,\, \qquad  \mathcal{D} \sim\begin{pmatrix}-1&\varepsilon&\,\,\varepsilon^2\\
\varepsilon&\,\,\varepsilon^2&\varepsilon\\
\,\,\varepsilon^2& \varepsilon &1
\end{pmatrix}\,.
\end{align}
Note that all flavor off-diagonal Higgs couplings  are proportional to these matrices. In the limit of degenerate 
flavor charges $a_i$, $a_{u_i}$ or $a_{d_i}$,  these matrices become diagonal and do not induce any flavor violating couplings. For the flavor charges \eqref{eq:FCharges2}, therefore only $\mathcal{U}$ and $\mathcal{D}$ generate FCNCs. 

 In addition, all flavor violating 
couplings of the scalars in \eqref{eq:hcoupling} and  \eqref{eq:Hcoupling}  are proportional to the trigonometric 
functions in  \eqref{eq:trig}. In the limit $f(\alpha,\beta)=0$, all flavor off-diagonal couplings of the light Higgs 
vanish and the diagonal couplings  are independent of both $c_{\beta-\alpha}$ and $t_\beta$, and approach 
their SM values (up to a sign). It should be noted that this sign difference corresponds to the wrong-sign Yukawa coupling in a generic two Higgs doublet model \cite{Ferreira:2014naa,Dumont:2014wha}. 
We will come back to these observations when we discuss flavor observables in 
Section \ref{sec:flavor}. The limit $c_{\beta-\alpha}=0$, associated with decoupling \cite{Gunion:2002zf, Asner:2013psa, Haber:2013mia}
or alignment \cite{Asner:2013psa, Haber:2013mia, Craig:2013hca, Carena:2013ooa} is not the SM, but corresponds to 
the model proposed by Babu, Nandi \cite{Babu:1999me}, and Giudice and Lebedev \cite{Giudice:2008uua}. \\

The  pseudoscalar mass eigenstate $A$ is obtained through the rotation \eqref{eq:Amasses} and its couplings 
to quark mass eigenstates can be derived from \eqref{eq:Hcoupling}, by replacing
\begin{align}\label{eq:Acoups}
  g_{A {q_i} {q_j}}&= i\,g_{H {q_i}{q_j}}\Big\vert _{
                                                 c_\alpha \rightarrow 
s_\beta,\, s_\alpha \rightarrow c_\beta}\,.
\end{align}

Finally, the charged Higgs couplings can also be obtained from \eqref{eq:yuksector}  and are independent of the 
flavor charges. After rotation to quark and Higgs mass eigenstates, see \eqref{eq:Hpmmasses}, we obtain
\begin{align}
\Lag_\mathrm{\pm} = \frac{\sqrt{2}}{v}\frac{1}{t_\beta}\left( m_{u}\right)_{kj}  \,\left(V_\mathrm{CKM}^\dagger
\right)_{ik} \,\bar d_{L_i}\, H^-\, u_{R_j}\,+ \frac{\sqrt{2}}{v}t_\beta\, \left(m_{d}\right)_{kj} \left(V_\mathrm{CKM}
\right)_{ik}\,\bar u_{L_i}\, H^+\, d_{R_j} +h.c.\,.\label{eq:Hpmcoupling}
\end{align}
The couplings of the charged Higgs to quarks are therefore equivalent to the ones in the two Higgs doublet 
model of type II, see for example \cite{Branco:2011iw}.\\

\section{Higgs Production and Decay \label{seq:Higgsprod&decay}}

A light SM-like Higgs has been discovered at the LHC in various decay channels.  While observations are 
mainly in the ballpark of SM expectations,  there is still room for new physics. The modified flavor diagonal fermion
couplings of the light Higgs $h$ introduced in the previous section as well as modified gauge boson couplings lead to 
deviations in both production cross section and decay rates. In the following we compute these deviations and 
compare the results with the proton-proton collision data at $\sqrt{s} = 7$ and $8$ TeV  obtained from the ATLAS \cite{ATLAS:combinedHiggsdata} and CMS \cite{CMS:Moriond} experiments.\\

For a given Higgs boson production channel and decay rate into specific final states $X$, normalized to the SM 
values, we define the signal strength parameter 
 \begin{equation}
 \mu_X = \frac{\sigma_{\text{prod}}}{\sigma_{\text{prod}}^{\text{SM}}}\, \frac{\Gamma_{h\to X}}{\Gamma_{h\to X}
^{\mathrm{SM}}}\,
     \frac{\Gamma_{h,\,\rm tot}^{\mathrm{SM}}}{\Gamma_{h}}\,.
 \label{eq:higgsstrength2}
\end{equation}
New physics can enter each of these three quantities: the production cross section $\sigma_\mathrm{prod}$, the 
partial decay rate $\Gamma_{h\rightarrow X}$ and the total width $\Gamma_{h ,\mathrm{tot}}$.
We quantify the changes in flavor diagonal couplings of the Higgs to fermions 
$f=t,b,\tau$ and to vector bosons $V=W^\pm,Z$ with respect to the SM by 
\begin{align}\label{eq:norm}
 g_{h {f}{f}}&= \kappa_f  \, g_{h {f}{f}}^{\text{SM}}=  \kappa_f \frac{m_f}{v} \, ,\notag\\
 g_{h {V}{V}}&= \kappa_V  \, g_{h {V}{V}}^{\text{SM}}=  \kappa_V \frac{2 m_V^2}{v} \,,
\end{align}
such that $\kappa_f=\kappa_V=1$ in the SM limit. 

\begin{figure}[t!]
\centering
\begin{tabular}{cc}
\includegraphics[width=.46\textwidth]{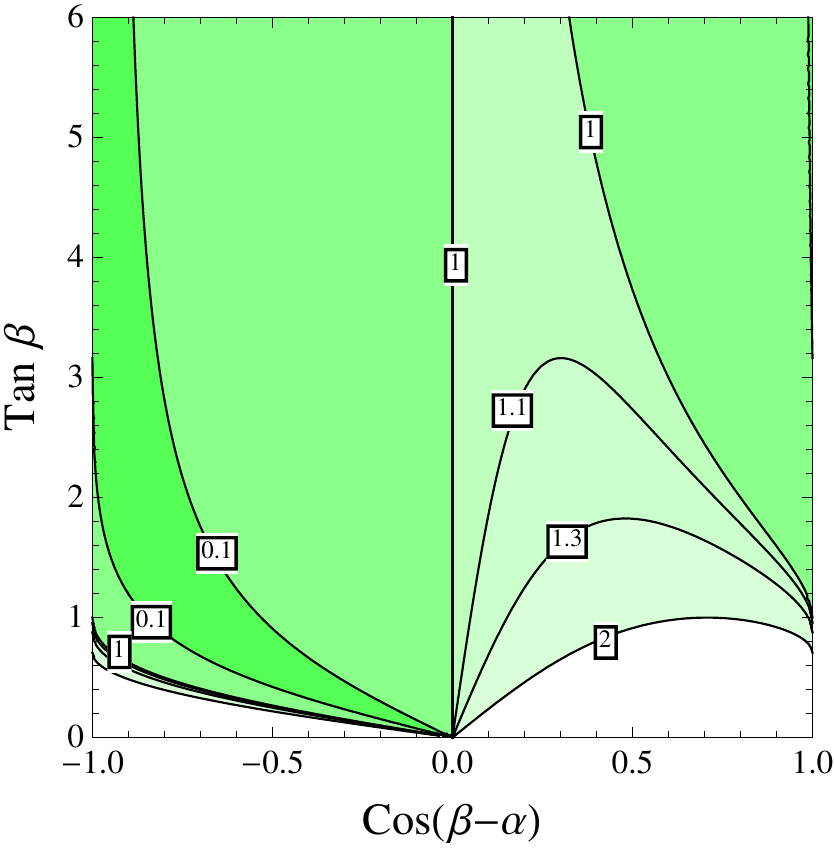}&
\includegraphics[width=.46\textwidth]{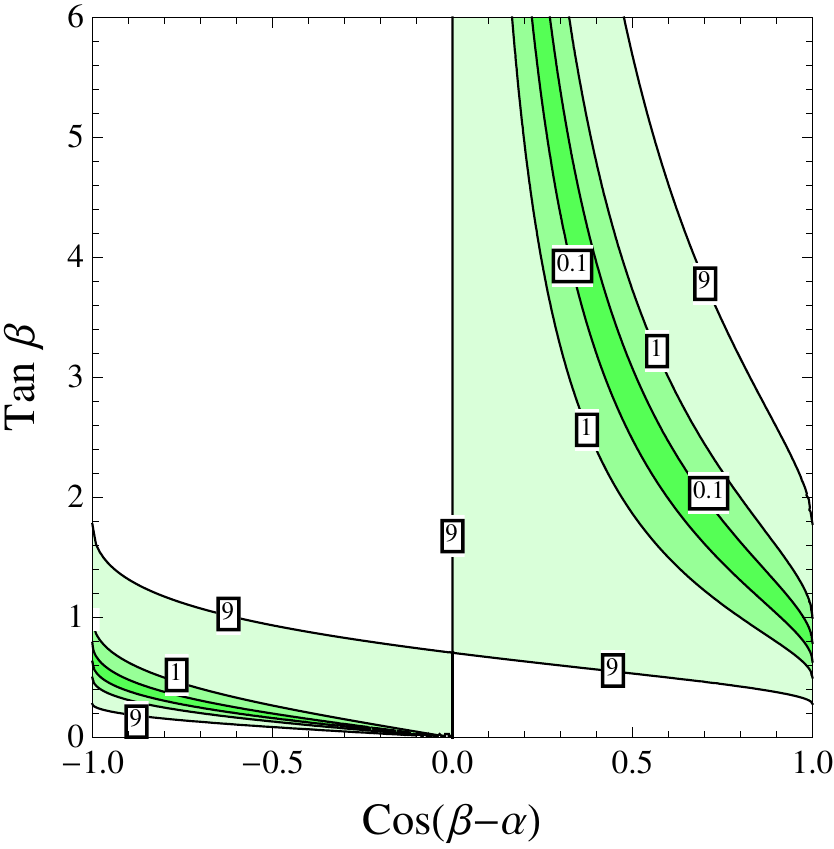}
\end{tabular}
\caption{Contours of $\kappa_t^2$ (left) and $\kappa_b^2$ (right) in the $\cos(\beta-\alpha)-\tan{\beta}$ plane.  
$\kappa_t^2=\kappa_b^2=1$ corresponds to the SM limit, for  $\kappa_b$ up to a sign in the right upper (lower left) corner for $\cos(\beta-\alpha)>0$ ($\cos(\beta-\alpha)<0)$. The decoupling/alignment limit corresponds to the Babu-Nandi-Giudice-Lebedev model.
\label{fig:kappabt}}
\end{figure}
It follows from equation \eqref{eq:hcoupling}, that the coupling of the light Higgs to the top quark is rescaled by 
\begin{equation}\label{eq:kapt}
\kappa_t= \frac{c_\alpha}{s_\beta } =\frac{c_{\beta-\alpha}}{t_\beta} + s_{\beta-\alpha}\,.
\end{equation}
 As a result, these couplings 
 are modified 
in the same way as in two Higgs doublet models of type II, see for example \cite{Branco:2011iw,Altmannshofer:2012ar,Craig:2013hca}.  However, couplings to the other flavors significantly differ from 
the couplings in generic two Higgs doublet models because of the Higgs dependent effective Yukawas, such 
that the Higgs-bottom coupling is rescaled by
 \begin{align}\label{eq:kapb}
\kappa_b=-2\,\frac{s_\alpha}{c_\beta }+\frac{c_\alpha}{s_\beta}= 3s_{\beta-\alpha}+ c_{\beta-\alpha}\left(\frac{1}{t_\beta}-2t_\beta\right)\,.
\end{align}
Note, that
for $f(\alpha,\beta)=0$, any dependence on $c_{\beta-\alpha}$ and $t_\beta$ cancels in $\eqref{eq:kapt}$ and $
\eqref{eq:kapb}$ and we find that
$\kappa_t=1$ and $\kappa_b=-1$ and therefore the light Higgs has couplings to fermions of SM strength. We illustrate the 
parameter dependence of the square of these couplings in Figure \ref{fig:kappabt}. In the right panel of Figure \ref{fig:kappabt} the value of $\kappa_b^2$ goes through zero signalizing $\kappa_b$ changes sign and becomes negative in the upper right (lower left) corner for $\cos(\beta-\alpha)>0$ ($\cos(\beta-\alpha)<0$).
The structure of these couplings has 
significant impact on the Higgs boson production cross sections and decay rates.
Further, the coupling of the light Higgs boson to charm quarks is rescaled by
\begin{equation}
 \kappa_c =  3 s_{\beta-\alpha}+ c_{\beta-\alpha} \left(\frac{2}{t_\beta}-t_\beta \right)\,.
\end{equation}
In general, fermion mixing effects generate corrections to the couplings, since the flavor charges of the quarks 
are not universal. These effects are encoded in the matrices $\mathcal{Q}^{u,d}$, $\mathcal{U}$ and $
\mathcal{D}$ given in equation \eqref{eq:quarkmix}.  For flavor-diagonal Higgs couplings to fermions we neglect 
corrections of $\ord(\varepsilon)$.  
For couplings  of the light Higgs boson to tau leptons we assume that a mechanism similar to our findings in the quark sector is responsible for generating masses, such that

\begin{equation}
 \kappa_\tau = \kappa_b\,.
\end{equation}
For the couplings of the light Higgs to vector bosons we obtain
\begin{equation}
 \kappa_V = s_{\beta - \alpha}\,,
\end{equation}
which is the same as in generic two Higgs doublet models.\\

 \begin{table}[t!]
\centering{
\begin{tabular}{|l|l|l|l|}
\hline
Decay Mode & Production Channels & Production Channels & Experiment \\
 & $\sigma_{gg\rightarrow h}$,  $\sigma_{t\bar t\rightarrow h}$ & $\sigma_{VBF}$, $\sigma_{VH}$ &  \\[2pt]
 \hline
 &&&\\[-12pt]
$h \to WW^\ast$ & $\mu_W= 1.02^{+0.29}_{-0.26}$ \cite{ATLAS:2014aga} & $\mu_W =1.27^{+0.53}_{-0.45}$ \cite{ATLAS:2014aga} & ATLAS \\[2pt]
& $\mu_W \simeq  0.75 \pm 0.35 $ \cite{Chatrchyan:2013iaa} & $\mu_W \simeq  0.7 \pm 0.85$ \cite{Chatrchyan:2013iaa} & CMS\\[2pt]
\hline
 &&&\\[-12pt]
$h \to ZZ^\ast$ & $\mu_Z =1.7^{+0.5}_{-0.4}$ \cite{Aad:2014eva} & $\mu_Z =0.3^{+1.6}_{-0.9}$ \cite{Aad:2014eva} 
& ATLAS\\
 & $\mu_Z =0.8^{+0.46}_{-0.36}$ \cite{Chatrchyan:2013mxa} & $\mu_Z =1.7^{+2.2}_{-2.1}$ \cite{Chatrchyan:2013mxa} & CMS\\[2pt]
\hline
 &&&\\[-12pt]
$h \to \gamma \gamma$ & $\mu_\gamma = 1.32 \pm 0.38$ \cite{Aad:2014eha} & $\mu_\gamma = 0.8 \pm 0.7$ 
\cite{Aad:2014eha} & ATLAS\\[2pt]
 & $\mu_\gamma = 1.13^{+0.37}_{-0.31}$ \cite{Khachatryan:2014ira} & $\mu_\gamma = 1.16^{+0.63}_{-0.58}$  
\cite{Khachatryan:2014ira} & CMS \\[2pt]
\hline
 &&&\\[-12pt]
$h \to \bar bb$ &  $\mu_b = 1.5 \pm 1.1$ \cite{Aad:2015gra} & $\mu_b = 0.52 \pm 0.32 \pm 0.24$  \cite{Aad:2014xzb} &  ATLAS\\[2pt]
 & $\mu_b=0.67^{+1.35}_{-1.33}$ \cite{CMS:2014jga} & $\mu_b = 1.0 \pm 0.5$  \cite{Chatrchyan:2013zna} &  
CMS\\[2pt]
\hline
&&&\\[-12pt]
$h \to  \tau\tau$ &  $\mu_\tau = 2.0 \pm 0.8 ^{+1.2}_{-0.8} \pm 0.3$  \cite{Aad:2015vsa} &  $\mu_\tau = 1.24 ^{+0.49 \,\,+0.31}_{-0.45 \,\,-0.29} \pm 0.08$  \cite{Aad:2015vsa} &  ATLAS\\[2pt]
 & $\mu_\tau \simeq 0.5^{+0.8}_{-0.7}$ \cite{Chatrchyan:2014nva} & $\mu_\tau \simeq 1.1^{+0.7}_{-0.5}$  \cite{Chatrchyan:2014nva} &  
CMS\\[2pt]
\hline
\end{tabular}}
\caption{\label{tab:Higgsconstraints} Input data for the global $\chi^2$-fit of Higgs production and decay with 
references. The data includes all updated results of the  pp collision data at $\sqrt{s} = 7$ and $8$ TeV  obtained from the ATLAS \cite{ATLAS:combinedHiggsdata} and CMS \cite{CMS:Moriond} experiments.}
\end{table}

The gluon fusion initiated Higgs production, neglecting light quark contributions in the fermion loops, is defined normalized to the SM value as  
 \begin{align}
 \frac{\sigma_{gg\rightarrow h}}{ \sigma_{gg\rightarrow h}^\text{SM}}=  \kappa_t^2 \left| 1+ \xi_b\, \frac{\kappa_b}
{\kappa_t}\right|^2\,,
\end{align}
where $\xi_b= -0.032+0.035\,i$ depends on the loop functions given in \cite{Gunion:1989we}. Therefore for values of $\kappa_b$ of $\mathcal{O}(1)$, the main Higgs production channel is to leading order indistinguishable from a type II two Higgs doublet model. 
Vector Boson Fusion (VBF) and Higgsstrahlung (VH) are both rescaled by $\kappa_V$,  while associated Higgs boson production with a top pair is modified by $\kappa_t$, 
\begin{align}
\qquad\frac{\sigma_{t\bar t\rightarrow h}}{ \sigma_{t\bar t\rightarrow h}^\text{SM}}=
\kappa_t^2\,\qquad \text{and}\qquad 
\frac{\sigma_\text{VBF} }{ \sigma_\text{VBF}^\text{SM} }=\frac{\sigma_\text{VH} }{ \sigma_\text{VH}^\text{SM} }= 
\kappa_V^2 \,.
\end{align}
Therefore the three production processes rescale with the same
factors as in generic two Higgs doublet models,  as given \emph{e.g.} 
in \cite{Branco:2011iw,Altmannshofer:2012ar,Craig:2013hca}.

The partial decay widths of the light Higgs into SM fermions $f$ and gauge bosons $V=W^\pm,Z$ can similarly 
be written as
\begin{align}
 \frac{\Gamma_{h \rightarrow{f}{f}}}{ \,\Gamma_{h\rightarrow {f}{f}}^{\text{SM}}}=\kappa_f^2 \,, \qquad \text{and} 
\qquad  
\frac{\Gamma_{h \rightarrow{V}{V}}}{ \,\Gamma_{h \rightarrow{V}{V}}^{\text{SM}}}=\kappa_V^2 \,.
\end{align}
Both top quark and $W^\pm$ boson loops enter the diphoton decay width \cite{Carena:2012xa}, 
\begin{equation}
 \frac{\Gamma_{h\rightarrow {\gamma}{\gamma}}}{ \Gamma^{\text{SM}}_{h \rightarrow{\gamma}{\gamma}}} = 
\big| 0.28 \kappa_t-1.28 \kappa_W   +\delta \big|^2\,,
\end{equation}
in which contributions from light fermions are neglected and contributions from charged scalar loops are 
encoded in $\delta$.  We find for $M_{H^\pm}\gtrsim 300$ GeV a contribution of less than $\delta \lesssim 0.04$ 
and set it to zero in the following \cite{Carena:2012xa,Gunion:2002zf}.

Expressed in terms of the rescaling factors $\kappa_t$, $\kappa_b$, $\kappa_c$, $\kappa_\tau$  and $\kappa_V$, the total Higgs boson width is 
given by  \cite{Denner:2011mq,Archer:2014jca} 
\begin{equation}\label{totrate}
   \frac{\Gamma_{h}}{\Gamma_{h}^{\mathrm{SM}}}
   \approx 0.57\,\kappa_b^2    + 0.25\,\kappa_V^2
    + 0.09\,\kappa_t^2
    + 0.06\,\kappa_\tau^2  + 0.026\,\kappa_c^2 +0.004 \,,
\end{equation}
where  $\Gamma_{h}^{\mathrm{SM}}=4.07$ MeV \cite{Agashe:2014kda} and we assume $h\to Z\gamma$ and even rarer modes to be SM-like. These contributions are collected in the constant term $0.004$.

The partial decay width into bottom quarks has a very different dependence on $\tan\beta$ and $\cos(\beta-\alpha)$ than in the generic type II two Higgs doublet model. This plays a relevant role in defining the allowed region in parameter space, since the bottom quark partial decay width dominates the total decay width, that in turn importantly affects the signal strength for all channels.
\\

\begin{figure}[t!]
\centering
\begin{tabular}{cc}
\quad ATLAS&\quad CMS\\
\includegraphics[width=.46\textwidth]{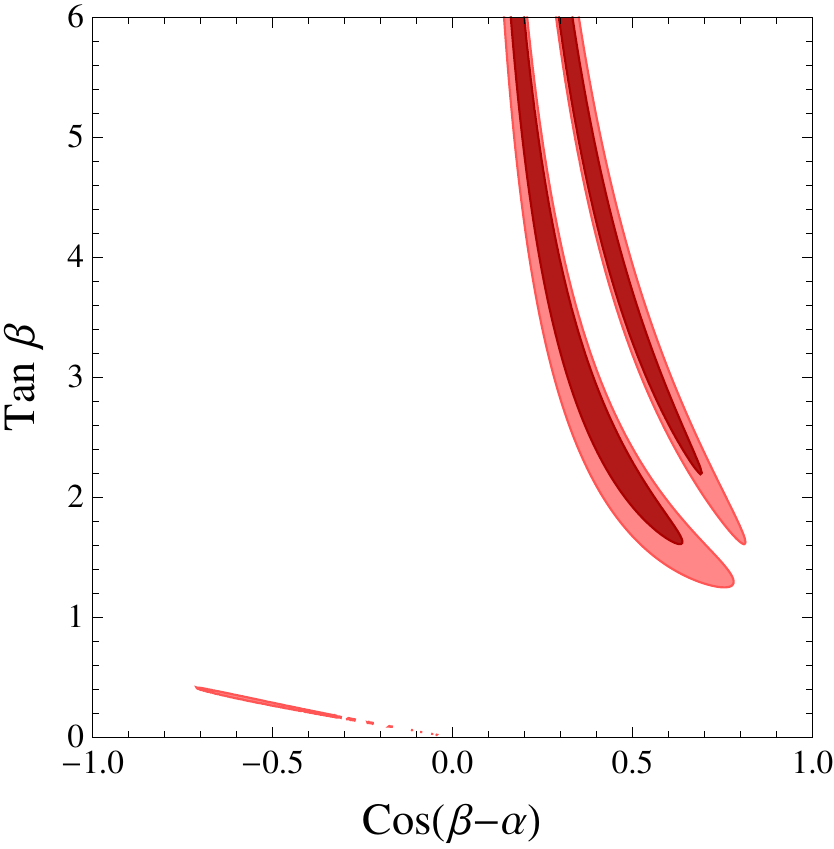}&
\includegraphics[width=.46\textwidth]{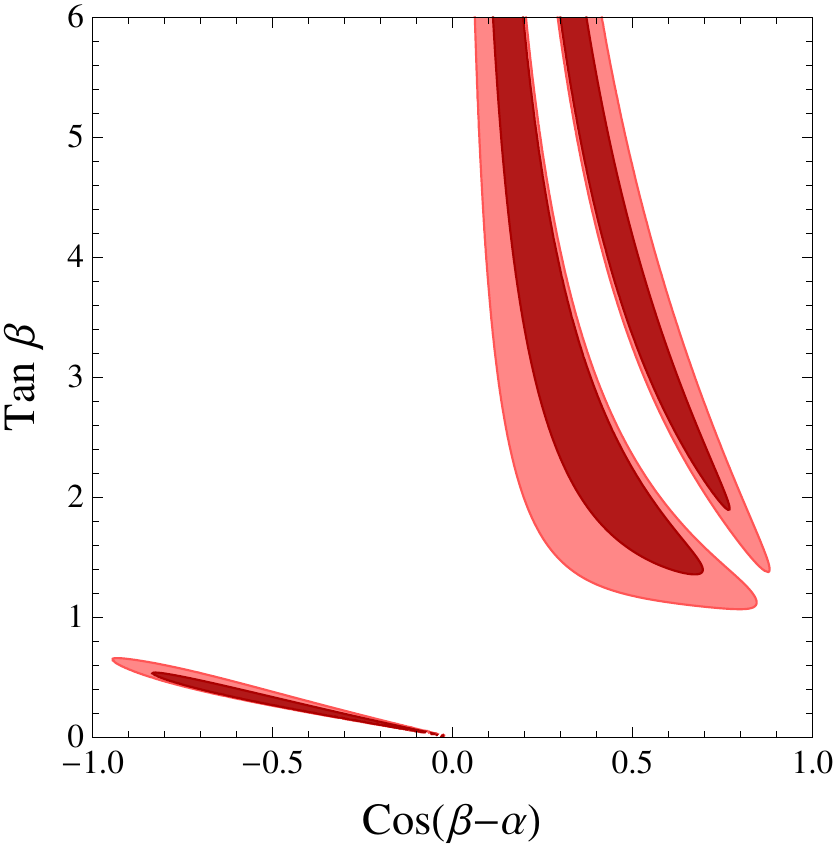}
\end{tabular}
\caption{Allowed 1$\sigma$ (dark red) and  2$\sigma$ (light red) regions, for a global fit to ATLAS and CMS 
data from measurements of Higgs boson decays in the left and right panel, respectively. The channels entering the fit 
are summarized in Table \ref{tab:Higgsconstraints} and errors are symmetrized. \label{fig:Higgsregions}}
\end{figure}
\begin{figure}[t!]
\centering
\begin{tabular}{cc}
\includegraphics[width=.46\textwidth]{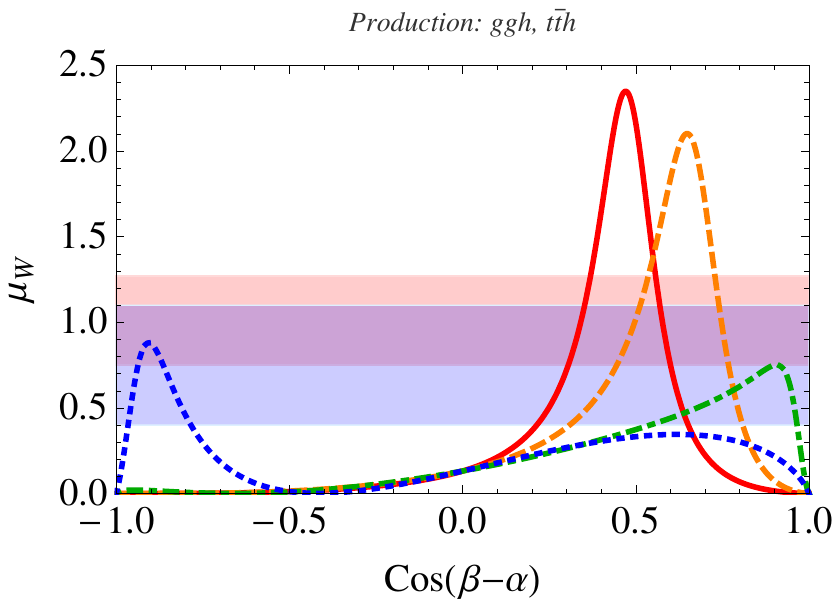}&
\includegraphics[width=.46\textwidth]{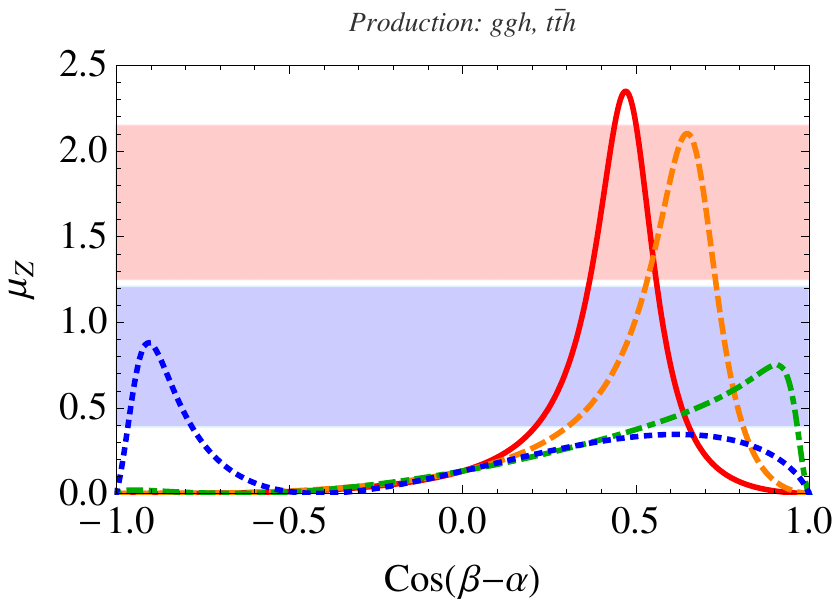}\\
\includegraphics[width=.455\textwidth]{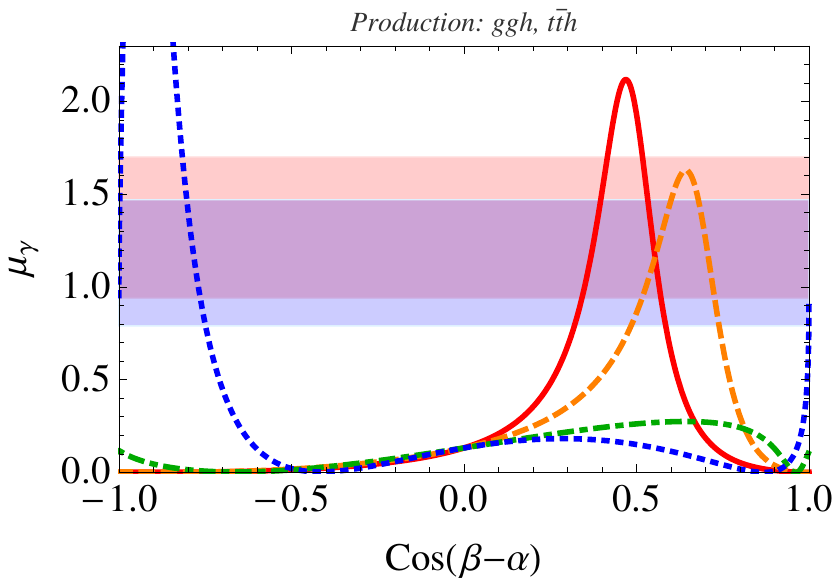}&
\includegraphics[width=.465\textwidth]{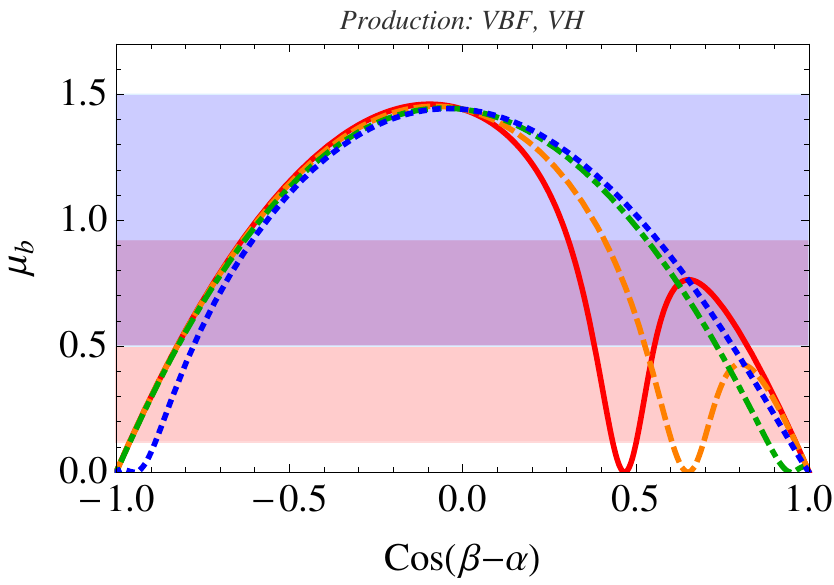}
\end{tabular}
\caption{ The upper panels show the signal strengths $\mu_W$ (left) and $\mu_Z$ (right) and the lower panel 
the signal strengths $\mu_\gamma$ (left) and $\mu_b$ (right) plotted against $c_{\beta-\alpha}$. The red (blue) 
band is the symmetrized $1 \sigma$ region of the corresponding ATLAS (CMS) measurement. Each plot shows 
curves for $t_\beta=3$ (solid red), $t_\beta=2$ (dashed orange), $t_\beta = 1$ (dot-dashed green) and $t_
\beta = 0.5$ (dotted blue).\label{fig:decays}
} 
\end{figure}

In Figure \ref{fig:Higgsregions} we show the result of a global $\chi^2$ fit based on the data collected in Table  
\ref{tab:Higgsconstraints}. Symmetrized errors are used for the fit. The left panel shows the plot for ATLAS and 
the right panel the plot for CMS. 
The two fit parameters are $c_{\beta-\alpha}$ and $t_\beta$. 
 The $1\sigma$ and $2\sigma$  regions consistent with the LHC data are shaded in dark and light red, 
respectively. It is clear, that the preferred parameter space is different from generic two Higgs doublet models, for which regions close to the alignment or decoupling limit $c_{\beta-\alpha}=0$ are favorable.   
\cite{Craig:2013hca, Haber:2013mia, Carena:2014nza}. In our case, $c_{\beta-\alpha}=0$ corresponds to the Babu-Nandi-Giudice-Lebedev model \cite{Babu:1999me, Giudice:2008uua}, which is clearly disfavored by the data. We observe, that while the allowed 
region for ATLAS is slightly smaller than in the case of CMS, both fits show a preference for values of $c_{\beta-
\alpha} > 0$ and $t_\beta \gtrsim 1$. The more constrained region of parameter space for ATLAS can be understood by the larger central values of $\mu_Z$, $\mu_W$ and $\mu_\gamma$ in the dominant gluon fusion channel, that are less compatible with larger values of $\kappa_b$, see Figure \ref{fig:kappabt}. The white area between the two branches in both fits can be explained by very small values of $\kappa_b$ for which all other branching fractions grow. 
Overall, the fermion couplings prefer a region in parameter space, where they approach their SM values, with the caveat that the value of the bottom Higgs coupling $\kappa_b$ has a negative sign with respect to the SM value in the upper right branch of the allowed red region. Note also that small values of $c_{\beta-\alpha}$ correspond to larger $t_\beta$ 
in the region preferred by the global fit as follows from equation \eqref{eq:kapb}.\\

In order to understand the features of the global fit,  we present the signal strengths of the relevant decay channels in Figure \ref{fig:decays}. In these plots, the red 
(blue) band is the $1 \sigma$ region of the corresponding ATLAS (CMS) measurement. Each plot shows the 
prediction of a particular signal strength for 
$\mu_W$, $\mu_Z$, $\mu_\gamma$ and $\mu_b$, depending on $c_{\beta-\alpha}$ for $t_\beta=3$ (solid red), $t_\beta=2$ (dashed orange), $t_
\beta = 1$ (dot-dashed green) and $t_\beta = 0.5$ (dotted blue). Excluding all but these four 
observables only marginally changes the global fits. For  $t_\beta\gtrsim 1$ all four measurements prefer values of $c_{\beta-\alpha} > 0 $. There is also an allowed region for $c_{\beta-\alpha} < 0 $ for values of $t_\beta <1$, however as will be shown later this region is phenomenologically less interesting.

We conclude, that the global fit to LHC Higgs measurements accommodates  $\tan\beta$ of $\mathcal{O}(1)$ for sizable values of $\cos(\beta-\alpha)$ away from the decoupling/alignment limit. 
This is a nontrivial result, given that $\tan\beta$ is already constrained to be of order one from the 
bound on the new physics scale. As we discuss below, values of $\tan\beta \lesssim 5$ are in agreement with flavor constraints as well as a possible UV completion scale in the TeV to a few TeV range.
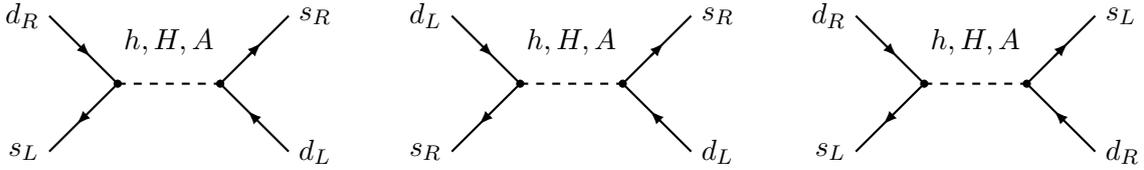
\begin{figure}[t]
\begin{align*}
  \begin{tikzpicture}[scale=.9, baseline=0]
\draw[thick,-latex] (-0.5,1) to (0.1,0.4);
\draw[thick] (0,0.5) to (0.5,0);
\draw[thick] (-0.5,-1) to (0,-0.5);
\draw[thick,latex-] (-0.1,-0.6) to (0.5,0);
\draw[dashed,decorate, thick] (0.5,0) to (2,0);
\node[above=.25cm] at (1.25,0){$h, H, A $};
\draw[thick,-latex] (2,0) to (2.6,.6);
\draw[thick] (2.4,.4) to (3,1);
 \draw[thick] (2,0) to (2.5,-.5);
\draw[thick,latex-] (2.4,-.4) to (3,-1);
\fill[color=black!] (0.5,0) circle (0.6mm);
\fill[color=black!] (2,0) circle (0.6mm);
\node[left] at (-.5,1){$d_R$};
\node[left] at (-.5,-1){$s_L$};
\node[right] at (3,-1){$d_L$};
\node[right] at (3,1){$s_R$};
 \end{tikzpicture}\qquad
    \begin{tikzpicture}[scale=.9, baseline=0]
\draw[thick,-latex] (-0.5,1) to (0.1,0.4);
\draw[thick] (0,0.5) to (0.5,0);
\draw[thick] (-0.5,-1) to (0,-0.5);
\draw[thick,latex-] (-0.1,-0.6) to (0.5,0);
\draw[dashed, thick] (0.5,0) to (2,0);
\node[above=.25cm] at (1.25,0){$h, H, A $};  
\draw[thick,-latex] (2,0) to (2.6,.6);
\draw[thick] (2.4,.4) to (3,1);
 \draw[thick] (2,0) to (2.5,-.5);
\draw[thick,latex-] (2.4,-.4) to (3,-1);
\fill[color=black!] (0.5,0) circle (0.6mm);
\fill[color=black!] (2,0) circle (0.6mm);
\node[left] at (-.5,1){$d_L$};
\node[left] at (-.5,-1){$s_R$};
\node[right] at (3,-1){$d_L$};
\node[right] at (3,1){$s_R$};
 \end{tikzpicture}\qquad
   \begin{tikzpicture}[scale=.9, baseline=0]
\draw[thick,-latex] (-0.5,1) to (0.1,0.4);
\draw[thick] (0,0.5) to (0.5,0);
\draw[thick] (-0.5,-1) to (0,-0.5);
\draw[thick,latex-] (-0.1,-0.6) to (0.5,0);
\draw[dashed, thick] (0.5,0) to (2,0);
\node[above=.25cm] at (1.25,0){$h, H, A $};  
\draw[thick,-latex] (2,0) to (2.6,.6);
\draw[thick] (2.4,.4) to (3,1);
 \draw[thick] (2,0) to (2.5,-.5);
\draw[thick,latex-] (2.4,-.4) to (3,-1);
\fill[color=black!] (0.5,0) circle (0.6mm);
\fill[color=black!] (2,0) circle (0.6mm);
\node[left] at (-.5,1){$d_R$};
\node[left] at (-.5,-1){$s_L$};
\node[right] at (3,-1){$d_R$};
\node[right] at (3,1){$s_L$};
 \end{tikzpicture}\end{align*}
 
  \caption{\label{fig:DF2treelvl} Tree-level contributions to 
  $\Delta S=2$ processes.}
  \end{figure}

\section{Constraints from Flavor Observables\label{sec:flavor}}

In addition to modifications of flavor-diagonal couplings, the misalignment of the mass and coupling matrices 
induces flavor changing couplings of the light Higgs $h$, the heavy neutral scalar $H$ and the pseudoscalar $A
$. These couplings generate FCNCs at tree-level, which are subject to strong constraints from neutral meson 
oscillations. In the following we calculate and analyze contributions to the relevant observables. We further estimate effects in $b \rightarrow s \gamma$ and 
give the prediction for the flavor-violating top decay $t\rightarrow h c$.

\subsection{Meson-Antimeson Mixing}

In the $K-\bar K$ system,  contributions from Higgs mediated FCNCs are captured by the effective Hamiltonian
\begin{align}
\Heff^{\Delta S=2}&=C_1^{sd} \,( \bar s_L\,\gamma_\mu \, d_L)^2+\tilde C_1^{sd} \,( \bar s_R\,\gamma_\mu \, 
d_R)^2 +C_2^{sd} \,( \bar s_R \, d_L)^2+\tilde C_2^{sd} \,( \bar s_L \, d_R)^2\notag\\
&+ C_4^{sd}\, ( \bar s_R \, d_L)\, ( \bar s_L \, d_R)\,+C_5^{sd}\, ( \bar s_L \,\gamma_\mu\, d_L)\, ( \bar s_R \,
\gamma^\mu d_R)\,+h.c. \, .\label{eq:heffdf2}
\end{align}
At tree-level, the corresponding Wilson coefficients can be read off from the diagrams in Figure 
\ref{fig:DF2treelvl}  \cite{Buras:2013rqa},
 \begin{align}\label{eq:wilsons}
C_2 ^{sd}&= 
-\frac{(g_{hds}^{\ast})^2}{m_h^2}\, -\frac{(g_{Hds}^\ast)^2}{M_H^2}\, -\frac{(g_{Ads}^\ast)^2}{M_A^2}\, \,,\notag\\
\tilde C_2 ^{sd}&=  -\frac{ g_{hsd}^2}{m_h^2}\,-\frac{g_{Hsd}^2}{M_H^2}\, -\frac{g_{Asd}^2}{M_A^2}\, \,,\notag\\
C_4 ^{sd}&=  -\frac{ g_{hsd}\,g_{hds}^\ast}{2m_h^2}\,-\frac{g_{Hsd}\,g_{Hds}^\ast}{2M_H^2}\, -\frac{g_{Asd}
\,g_{Ads}^\ast}{2M_A^2}\, .
 \end{align}
Similar expressions hold for $ B_s - \bar B_s$ mixing, with $ sd \rightarrow bs$, $ B_d - \bar B_d$ mixing, with $ sd \rightarrow bd$ and $ D - \bar D$ mixing  with $ sd \rightarrow uc$. 
Contributions from Higgs boson exchange are only suppressed by the weak scale, but the Froggatt-Nielsen mechanism induces 
an additional suppression of flavor off-diagonal couplings by the masses of the involved quarks as well as 
the expansion parameter $\varepsilon$. The relative size of the Wilson coefficients \eqref{eq:wilsons} depends therefore strongly on the explicit flavor structure. For the flavor charge assignment \eqref{eq:FCharges}, which is tailored to explain quark masses as well as CKM mixing angles, we collect the results in the left hand side of Table \ref{tab:Ccomp}.\\

In the case of $K - \bar K$ mixing, we find that the largest coefficient 
is $\tilde C_2$ with 
\begin{align}\label{eq:C2}
\tilde C_{2}^{sd} &= \,
-\frac{\tilde c_2^{sd}}{v^2}\,
\bigg\{\,\frac{f(\alpha,\beta)^2}{m_h^2}+\frac{F(\alpha,\beta)^2}{M_H^2}-\left(t_\beta+\frac{1}{t_\beta}\right)^2\frac{1}{M_A^2}\bigg\}\,\notag\\ 
 &\approx \frac{-10^{-15}}{\text{GeV}^2}\,  \bigg\{\,f(\alpha,\beta)^2+F(\alpha,\beta)^2\frac{m_h^2}{M_H^2}-\left(t_\beta+\frac{1}{t_\beta}\right)^2\frac{m_h^2}{M_A^2}\bigg\}\,, %
\end{align}
where we factored out the light Higgs mass in the second line, the trigonometric functions $f(\alpha, \beta)$ and $F(\alpha,\beta)$ are defined in \eqref{eq:trig}, and $\tilde c_2^{sd}$ is the flavor-dependent part of the Wilson coefficient given in Table \ref{tab:Ccomp}. The same expression holds for the Wilson coefficient $C_2^{sd}$, with the additional $\varepsilon^2$ suppression due to the replacement of $\tilde c_2^{sd}\rightarrow c_2^{sd}$. The flavor-dependent Wilson coefficient $c_4^{sd}$ is also suppressed by $\varepsilon$ with respect to $\tilde c_2^{sd}$, but the minus sign in the last line of \eqref{eq:C2} is replaced by a plus, which corresponds to a constructive interference of the different contributions,
\begin{align}\label{eq:C4}
 C_{4}^{sd} &= \,
-\frac{ c_4^{sd}}{v^2}\,
\bigg\{\,\frac{f(\alpha,\beta)^2}{m_h^2}+\frac{F(\alpha,\beta)^2}{M_H^2}+\left(t_\beta+\frac{1}{t_\beta}\right)^2\frac{1}{M_A^2}\bigg\}\,\notag\\ 
 &\approx \frac{-1.7\times 10^{-17}}{\text{GeV}^2}\,  \bigg\{\,f(\alpha,\beta)^2+F(\alpha,\beta)^2\frac{m_h^2}{M_H^2}+\left(t_\beta+\frac{1}{t_\beta}\right)^2\frac{m_h^2}{M_A^2}\bigg\}\,.%
\end{align}

\begin{figure}[t!]
\centering
\begin{tabular}{ccc}
\raisebox{1mm}{\includegraphics[width=.43\textwidth]{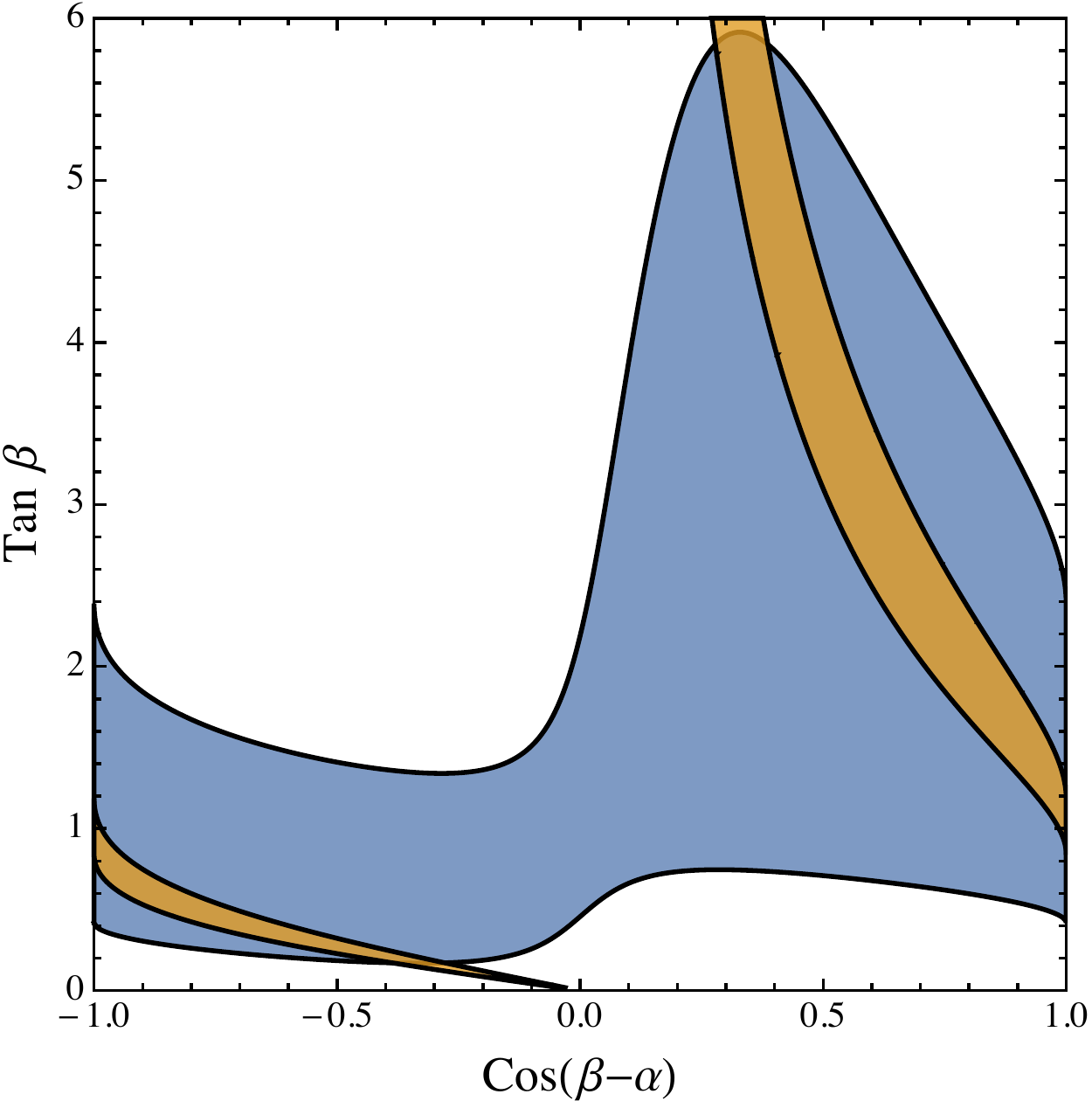}}&
\includegraphics[width=.43\textwidth]{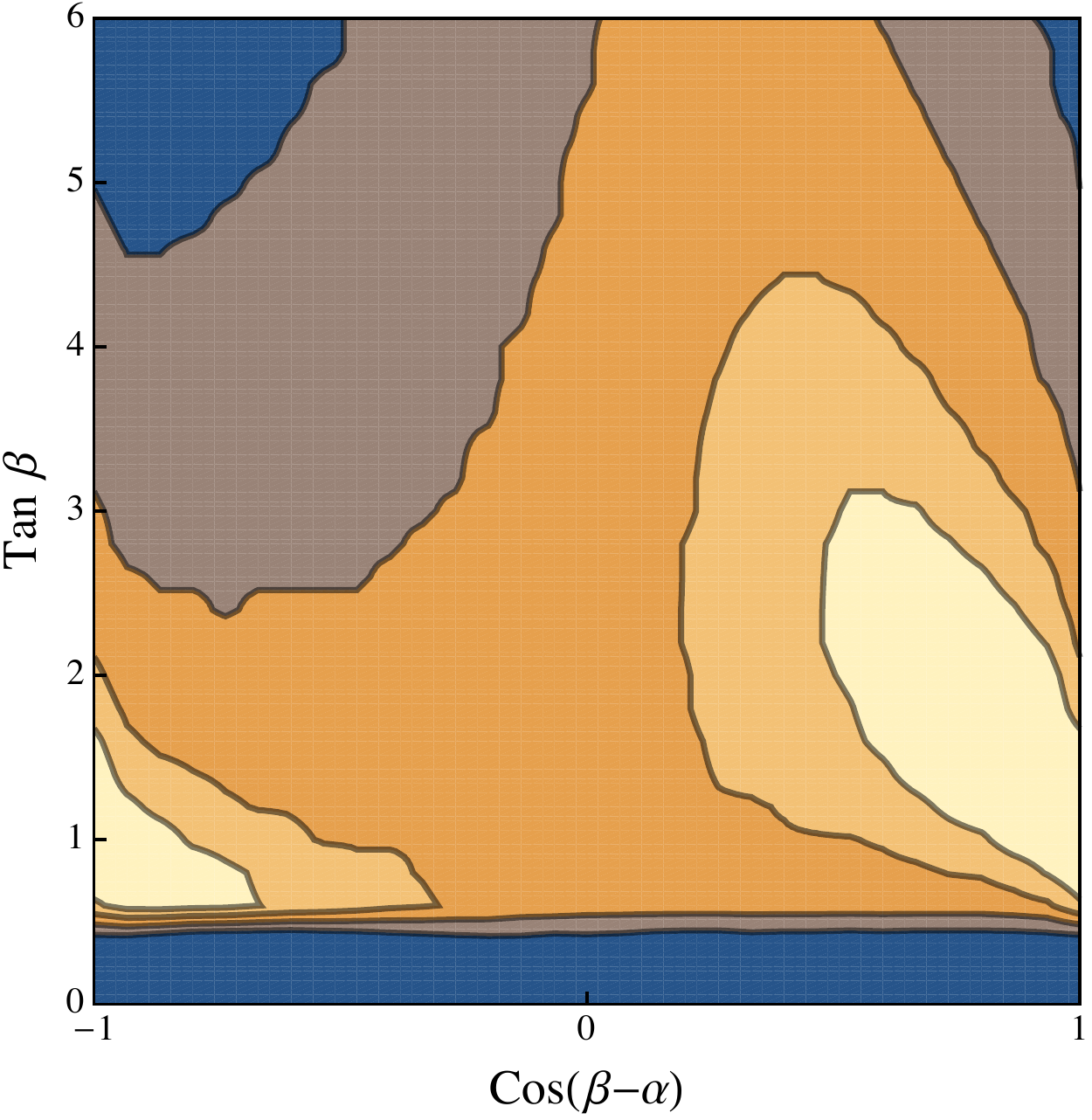}&
 \raisebox{.03\height} {\includegraphics[width=.044\textwidth]{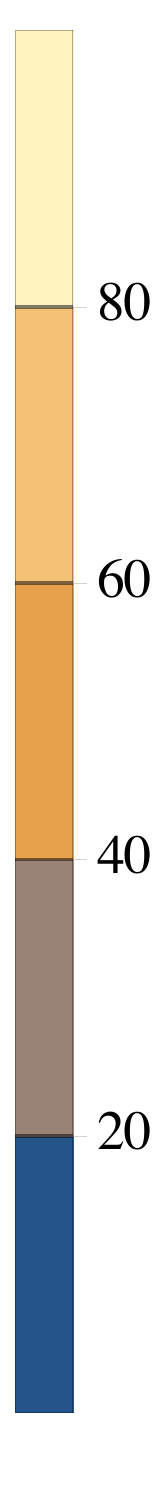}}\end{tabular}
\caption{The left panel shows the region in the $c_{\beta-\alpha}- t_\beta$ plane for which the tree-level contributions to $|\tilde C_2^{sd} | \leq 10^{-16}/\text{GeV}^2$ (orange) and the tree-level contributions to $ |C_4^{sd}|\leq 7\times 10^{-17}/\text{GeV}^2$ (blue). In the right panel we show regions of parameter space in which our sample points reproduce $C_{\varepsilon_K}$ within two standard 
deviations. The color coding indicates the percentage of points in agreement with the experimental constraint. In both plots,  the scalar masses are $M_A=M_H=M_{H^+}=500$ GeV. 
\label{fig:epsK}
}
\end{figure}

\begin{table}[t!]
\centering
\setlength\extrarowheight{3pt}
\begin{tabular}{|c| c c c|| ccc|}
\hline
&\multicolumn{3}{c||}{Scenario \eqref{eq:FCharges} } 
&\multicolumn{3}{  c|}{Scenario \eqref{eq:FCharges2} }\\[2pt]\hline 
$\Delta F=2$&$c_2^{ij}$& $\tilde c_2^{ij}$&$c_4^{ij}$
&$c_2^{ij}$&$\tilde c_2^{ij}$&$c_4^{ij}$\\\hline
$sd$&$\varepsilon^4\,m_s^2$&$\varepsilon^2\,m_s^2$&$\varepsilon^3\, m_s^2$
&$\varepsilon^2\, m_d^2$&$\varepsilon^2 \,m_s^2 $&$\varepsilon^2 \,m_dm_s$\\
$bd$&$\varepsilon^2\, m_b^2$&$\varepsilon^2 \,m_b^2 $&$\varepsilon^2\, m_b^2 $
&
          $\varepsilon^4\, m_d^2 $&$\varepsilon^4\, m_b^2 $&$\varepsilon^4 \,m_dm_b $\\
$bs$&$\varepsilon^2\, m_b^2 $&$\varepsilon^4\, m_b^2 $&$\varepsilon^3 \,m_b^2$
&
         $\varepsilon^2\, m_s^2 $&$\varepsilon^2\, m_b^2 $&$\varepsilon^2 \,m_sm_b $\\
$uc$&$\varepsilon^4\, m_c^2 $&$\varepsilon^4\, m_c^2 $&$\varepsilon^4 \,m_c^2$
&
          $\varepsilon^4\, m_c^2 $&$\varepsilon^4\, m_u^2 $&$\varepsilon^4 \,m_um_c $\\
          \hline
\end{tabular}
\caption{\label{tab:Ccomp} Flavor specific part of the  Wilson coefficients for meson-antimeson mixing in the case of the flavor charge assignments \eqref{eq:FCharges} with flavor structure \eqref{eq:struc} 
(left) and flavor charge assignments \eqref{eq:FCharges2} with flavor structure  \eqref{eq:struc2} (right) .}
\end{table}
The limit of exact cancellation in $C_2^{sd}$ and $\tilde C_2^{sd}$ and maximal interference in $C_4^{sd}$ corresponds to the $SU(2)_L$ symmetric limit, in which operators of the type $(\bar s_L d_R)^2$ are forbidden  \cite{Buras:2010mh}. In Table \ref{tab:wcref}, we present the current bounds on the Wilson coefficients at the electroweak scale for the different meson systems, based on \cite{Bevan:2014cya}. These bounds have been derived by assuming that new physics only contributes to a single Wilson coefficient and can therefore only be taken as a rough upper limit. For $K- \bar K$ mixing, the strongest constraint comes from the CP violating observable $\eps_K$, such that the bounds on the imaginary part of the Wilson coefficient is cited. Since we assume arbitrary phases, the estimate \eqref{eq:C2} holds for both real and imaginary parts of the Wilson coefficients. Comparing \eqref{eq:C2} with the bound in Table \ref{tab:wcref} shows that a partial cancellation in $\tilde C_2^{sd}$ is necessary in 
order to comply with the limit. For $M_A, M_H >m_h$, this corresponds to a preferred region in the $\cos (\beta-\alpha)-\tan \beta$ plane. In the left panel of Figure \ref{fig:epsK} we show the preferred region, for which $|\tilde C_{2}^{sd}| < 10^{-16}/\text{GeV}^2$ (shaded orange), assuming $M_A=M_H=500$ GeV. Contributions to $C_4^{sd}$ can be enhanced by the constructive interference between the scalar contributions. Also, the bound on $C_4^{sd}$ is particularly strong, because it is enhanced from Renormalization Group (RG) running as well as from the matrix element, that scales like $M_K^2/(m_s+m_d)^2\approx 14$, see Appendix \ref{app:RPG&running} for details. However, the additional suppression shown in Table \ref{tab:Ccomp} gives $C^{sd}_4 = \varepsilon \,\tilde C_2^{sd} $, such that a slight enhancement from interference effects is allowed. In the left panel of Figure \ref{fig:epsK} we show the region in the $\cos(\beta-\alpha) -\tan \beta $ plane for which $| C_{4}^{sd}| < 7\times 10^{-17}/\text{
GeV}^{2}
$ (shaded blue). 

\begin{table}[t!]\centering
\begin{tabular}{|c|cccc|}
\hline
$i$& $1$&$2$&$4$&$5$\\[1pt]\hline\\[-12pt]
$\IM C_i^{sd}$&$\lesssim 2\times 10^{-15}$&$\lesssim 1\times 10^{-16}$&$\lesssim 7\times 10^{-17}$&$
\lesssim 9\times 10^{-16}$\\[2pt]
$\IM C_i^{uc}$&$\lesssim 2\times 10^{-14}$&$\lesssim 2\times 10^{-14}$&$\lesssim 1\times 10^{-14}$&$
\lesssim 1\times 10^{-13}$\\[2pt]
$|C_i^{bd}|$&$\lesssim 1\times 10^{-12}$&$\lesssim 4\times 10^{-13}$&$\lesssim 6\times 10^{-13}$&$
\lesssim 1\times 10^{-12}$\\[2pt]
$|C_i^{bs}|$&$\lesssim 1\times 10^{-11}$&$\lesssim 2\times 10^{-12}$&$\lesssim 4\times 10^{-12}$&$
\lesssim 6\times 10^{-12}$\\[2pt]
\hline
\end{tabular}
\caption{Model-independent bounds on Wilson coefficients for meson-antimeson mixing evaluated at the electroweak scale in units of 
\mbox{GeV$^{-2}$}  \cite{Bevan:2014cya}, taking into account the running described in Appendix \ref{app:numbers}. The same bounds hold for the Wilson coefficients with flipped chirality $C_i \rightarrow \tilde 
C_i$\label{tab:wcref}. }\end{table}

\begin{figure}[tb]
\begin{align*}
& \begin{tikzpicture}[scale=.9, baseline=0]
 \draw[thick,-latex] (-0.5,1) to (0.1,1);
\draw[thick] (0,1) to (0.5,1);
\draw[thick] (-0.5,-1) to (0,-1);
\draw[thick,latex-] (-0.1,-1) to (0.5,-1);
\draw[thick] (0.5,-1) to (2,-1);
\draw[thick] (2,-1) to (1,-1);
\draw[thick,-latex] (2,-1) to (1.1,-1);
\draw[thick,-latex] (0.5,1) to (1.3,1);
\fill[color=black!] (0.5,1) circle (0.6mm);
\fill[color=black!] (0.5,-1) circle (0.6mm);
\draw[thick,-latex] (2,1) to (2.7,1);
\draw[thick] (2.4,1) to (3,1);
 \draw[thick] (2,-1) to (2.5,-1);
 \draw[thick,latex-] (2.4,-1) to (3,-1);
 \draw[thick, ] (2,1) to (0.5,1);
   \draw[ decoration={snake,amplitude=1.2}, thick, decorate] (.5,-1) to (0.5,1);
    \draw[ thick,dashed] (2,-1) to (2,1);
\node[left] at (-.5,1){$d$};
\node[left] at (-.5,-1){$s$};
\node[right] at (3,-1){$d$};
\node[right] at (3,1){$s$};
\fill[color=black!] (2,1) circle (0.6mm);
\fill[color=black!] (2,-1) circle (0.6mm);
\node[above] at (1.25,1.1){$u, c , t$};
\node[below] at (1.25,-1.1){$u, c, t$};
\node[] at (0,0.1){$W^-$};
\node[] at (2.5,0.1){$H^-$};
 \end{tikzpicture}\hspace{1cm}
 \begin{tikzpicture}[scale=.9, baseline=0]
\draw[thick,-latex] (-0.5,1) to (0.1,0.4);
\draw[thick] (0,0.5) to (0.5,0);
\draw[thick] (-0.5,-1) to (0,-0.5);
\draw[thick,latex-] (-0.1,-0.6) to (0.5,0);
 \draw[dashed, thick] (.5,0) .. controls (.6,.8) and (1.9,.8) .. (2,0);
  \draw[dashed, thick] (.5,0) .. controls (.6,-.8) and (1.9,-.8) .. (2,0);
\node[above=.25cm] at (1.25,.6){$h, H, A $};
\node[below=.25cm] at (1.25,-.6){$h, H, A $};
\draw[thick,-latex] (2,0) to (2.5,.5);
\draw[thick] (2.4,.4) to (3,1);
 \draw[thick] (2,0) to (2.5,-.5);
\draw[thick,latex-] (2.4,-.4) to (3,-1);
\fill[color=black!] (0.5,0) circle (0.6mm);
\fill[color=black!] (2,0) circle (0.6mm);
\node[left] at (-.5,1){$d$};
\node[left] at (-.5,-1){$s$};
\node[right] at (3,-1){$d$};
\node[right] at (3,1){$s$};
 \end{tikzpicture}\hspace{1cm}
 \begin{tikzpicture}[scale=.9, baseline=0]
\draw[thick,-latex] (-0.5,1) to (0.1,0.4);
\draw[thick] (0,0.5) to (0.5,0);
\draw[thick] (-0.5,-1) to (0,-0.5);
\draw[thick,latex-] (-0.1,-0.6) to (0.5,0);
\draw[dashed, thick] (0.5,0) to (2,0);
\node[] at (3.2,0){$H^+$};
\draw[thick,-latex] (2,0) to (2.4,.4);
\draw[thick] (2,0) to (3,1);
 \draw[thick] (2,0) to (2.5,-.5);
\draw[thick,latex-] (2.2,-.2) to (3,-1);
\draw[dashed, thick] (2.7,.7) to (2.7,-.7);
\fill[color=black!] (2.7,0.7) circle (0.6mm);
\fill[color=black!] (2.7,-.7) circle (0.6mm);
\fill[color=black!] (0.5,0) circle (0.6mm);
\fill[color=black!] (2,0) circle (0.6mm);
\node[left] at (-.5,1){$d$};
\node[left] at (-.5,-1){$s$};
\node[above] at (1.2,0.1){$h$};
\node[left] at (2.3,.6){$t$};
\node[left] at (2.3,-.6){$t$};
\node[right] at (3,-1){$d$};
\node[right] at (3,1){$s$};
 \end{tikzpicture}\\
& \begin{tikzpicture}[scale=.9, baseline=0]
 \draw[thick,-latex] (-0.5,1) to (0.1,1);
\draw[thick] (0,1) to (0.5,1);
\draw[thick] (-0.5,-1) to (0,-1);
\draw[thick,latex-] (-0.1,-1) to (0.5,-1);
\draw[thick] (2,-1) to (.5,-1);
\draw[thick,-latex] (2,-1) to (1.1,-1);
\draw[thick,-latex] (0.5,1) to (1.3,1);
\fill[color=black!] (0.5,1) circle (0.6mm);
\fill[color=black!] (0.5,-1) circle (0.6mm);
\draw[thick,-latex] (2,1) to (2.7,1);
\draw[thick] (2.4,1) to (3,1);
 \draw[thick] (2,-1) to (2.5,-1);
 \draw[thick,latex-] (2.4,-1) to (3,-1);
 \draw[thick, ] (2,1) to (0.5,1);
   \draw[ dashed,thick] (.5,-1) to (0.5,1);
    \draw[ thick,dashed] (2,-1) to (2,1);
\node[left] at (-.5,1){$d$};
\node[left] at (-.5,-1){$s$};
\node[right] at (3,-1){$d$};
\node[right] at (3,1){$s$};
\fill[color=black!] (2,1) circle (0.6mm);
\fill[color=black!] (2,-1) circle (0.6mm);
\node[above] at (1.25,1.1){$u, c , t$};
\node[below] at (1.25,-1.1){$u, c, t$};
\node[] at (0,0.1){$H^-$};
\node[] at (2.5,0.1){$H^-$};
 \end{tikzpicture}\hspace{1cm}
 \begin{tikzpicture}[scale=.9, baseline=0]
\draw[thick,-latex] (-0.5,1) to (0.1,0.4);
\draw[thick] (0,0.5) to (0.5,0);
\draw[thick] (-0.5,-1) to (0,-0.5);
\draw[thick,latex-] (-0.1,-0.6) to (0.5,0);
 \draw[dashed, thick] (.5,0) .. controls (.6,.8) and (1.9,.8) .. (2,0);
  \draw[dashed, thick] (.5,0) .. controls (.6,-.8) and (1.9,-.8) .. (2,0);
\node[above=.25cm] at (1.25,.6){$H^+$};
\node[below=.25cm] at (1.25,-.6){$H^- $};
\draw[thick,-latex] (2,0) to (2.5,.5);
\draw[thick] (2.4,.4) to (3,1);
 \draw[thick] (2,0) to (2.5,-.5);
\draw[thick,latex-] (2.4,-.4) to (3,-1);
\fill[color=black!] (0.5,0) circle (0.6mm);
\fill[color=black!] (2,0) circle (0.6mm);
\node[left] at (-.5,1){$d$};
\node[left] at (-.5,-1){$s$};
\node[right] at (3,-1){$d$};
\node[right] at (3,1){$s$};
 \end{tikzpicture}\hspace{1cm}
 \begin{tikzpicture}[scale=.9, baseline=0]
\draw[thick,-latex] (-0.5,1) to (0.1,0.4);
\draw[thick] (0,0.5) to (0.5,0);
\draw[thick] (-0.5,-1) to (0,-0.5);
\draw[thick,latex-] (-0.1,-0.6) to (0.5,0);
\draw[thick,dashed] (0.5,0) to (2,0);
\node[] at (3,0){$t$};
\draw[thick] (2.7,0.7) to (3,1);
\draw[thick,-latex] (2.8,0.8) to (3,1);
\draw[thick,dashed] (2,0) to (3,1);
 \draw[thick,dashed] (2,0) to (3,-1);
\draw[thick,latex-] (2.75,-.75) to (3,-1);
\draw[thick] (2.6,-.6) to (3,-1);
\draw[ thick] (2.7,.7) to (2.7,-.7);
\draw[ thick,-latex] (2.7,-.7) to (2.7,0.1);
\fill[color=black!] (2.7,0.7) circle (0.6mm);
\fill[color=black!] (2.7,-.7) circle (0.6mm);
\fill[color=black!] (0.5,0) circle (0.6mm);
\fill[color=black!] (2,0) circle (0.6mm);
\node[left] at (-.5,1){$d$};
\node[left] at (-.5,-1){$s$};
\node[above] at (1.,0.1){$h$};
\node[left] at (2.4,.6){$H^-$};
\node[left] at (2.4,-.6){$H^+$};
\node[right] at (3,-1){$d$};
\node[right] at (3,1){$s$};
 \end{tikzpicture}\\[2pt]
& \hspace{1.8cm}(a)\hspace{4.6cm}(b)\hspace{4.7cm}(c)
 \end{align*}
 \caption{\label{fig:DF2looplvl} Three types of one loop contributions to $\Delta S=2$ processes. }\end{figure}
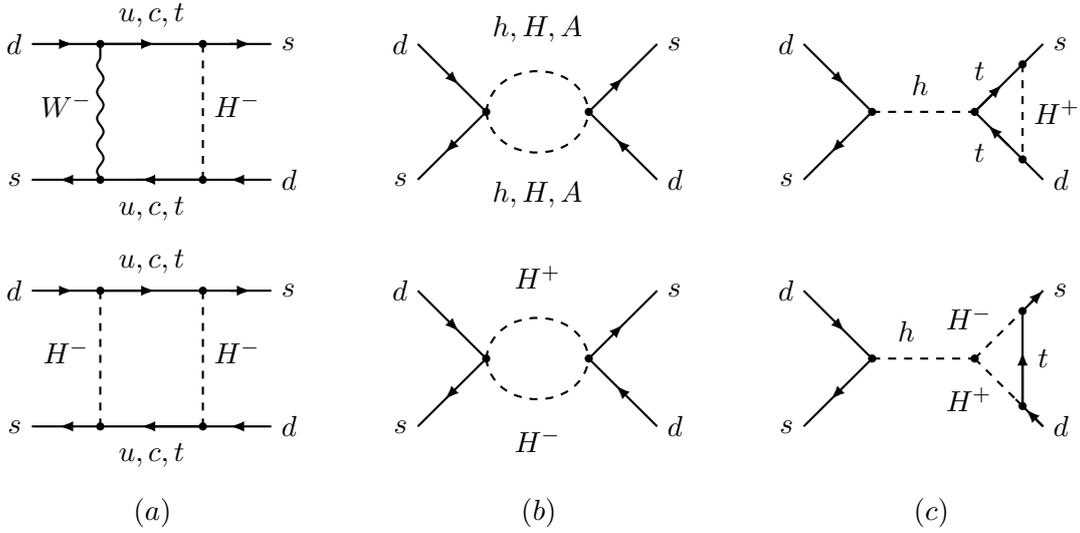

In addition to tree-level exchanges,  various one-loop contributions can potentially become large. The relevant 
diagrams are shown in Figure \ref{fig:DF2looplvl}.  The contributions from the box diagrams of type $(a)$ are 
completely analogous to 
the ones in a type II two Higgs doublet model, because the couplings of the charged Higgs 
\eqref{eq:Hpmcoupling} are indistinguishable between the two models. The leading contribution enters 
$C_1^{sd}$ and comes from the 
box with one charged Higgs \cite{Buras:1989ui}, a $W^\pm$ boson and top quarks running in the loop and one finds \begin{equation}
C_{1, \mathrm{box}}^{sd} \propto \frac{1}{16 \pi^2}\frac{1}{t_\beta^2}\left(\frac{m_t^2}{v^2}\,V_{ts}^\ast V_{td} 
\right)^2\frac{1}{M_{H^+}^2}\approx 9\times 10^{-16} \left(\frac{500\,\mathrm{GeV}}{M_{H^+}}\right)^2\mathrm{GeV}
^{-2}\,,
\end{equation}
where in the last equality we set $t_\beta = 1$.
For $t_\beta \lesssim 1$, this contribution is of the order of the largest tree-level contribution. We therefore require $t_\beta \gtrsim 1$ in order to be in compliance with experimental bounds in Table \ref{tab:wcref}.
In principle, there are also contributions from box diagrams to the other operators in $\eqref{eq:heffdf2}$ as well as box diagrams with neutral scalar exchange, but 
both are chirally suppressed by powers of light quark masses over the electroweak scale and turn out to be 
negligible. The loop diagrams labeled $(b)$ and $(c)$ in Figure \ref{fig:DF2looplvl} are also suppressed. Diagrams of type $(b)$ have the same coupling structure as the tree-level 
diagrams, but are additionally suppressed by a loop factor. Diagrams of type (c) are 
enhanced with respect to \eqref{eq:C4} by the light Higgs couplings to the top quark or charged scalars, but 
suppressed by CKM elements and a loop factor, such that we find for $C_4^{sd}$ \cite{Atwood:1996vj}
\begin{equation}
\frac{C_{ 4, \mathrm{penguin}}^{sd}}{C_{4, h}^{sd}}\approx \frac{1}{16\pi^2}\frac{m_t^2}{v^2}\frac{ V_{ts}^\ast\, 
V_{td}}{\varepsilon^2\, f (\alpha,\beta)}\approx 10^{-3}\,.
\end{equation}
The equivalent diagram with a charm quark in the loop is of the same order. \\

Having considered all different contributions we will map out the parameter space in the $\cos (\beta-\alpha) - \tan \beta$ plane in which the 
prediction for $\epsilon_K$ in our model agrees with the experimental bound within $2 \sigma$ in a numerical analysis.
 For this purpose we define 
\begin{equation}\label{eq:epsK}
C_{\eps_K}=\frac{\IM \,\langle  K^0| \Hfull^{\Delta S=2}| \bar K^0 \rangle 
}{\IM \,\langle K^0| \mathcal{H}_\mathrm{SM}^{\Delta S=2}| \bar K^0 \rangle 
}\,,
\end{equation}
where $\Hfull^{\Delta S=2} = \mathcal{H}_\mathrm{SM}^{\Delta S=2} +\Heff^{\Delta S=2} $ includes the 
Standard Model contribution. We compute the Wilson coefficients at the scale of the light Higgs and for
$M_H=M_A=M_{H^\pm}= 500$ GeV respectively, using the full expressions for the Wilson coefficients including 
tree-level and leading box diagrams. We collect the full analytic expressions of the latter in Appendix \ref{app:Boxes}. In the next step, the Wilson coefficients 
in \eqref{eq:heffdf2} are evolved down from the mass scale of the scalars to the scale $\mu= 2$ GeV  at which the hadronic 
matrix elements are evaluated using the RG equations in 
\cite{Ciuchini:1998ix}. The hadronic matrix elements are taken from \cite{Laiho:2009eu} and collected with the 
other numerical input in Appendix \ref{app:numbers}. We randomly generate a sample set of points of fundamental Yukawa couplings, defined in \eqref{eq:yuksector},  with $|y^{u,d}_{ij}|\in [0.5,1.5]$ and with arbitrary 
phases. We require the SM quark masses and Wolfenstein 
parameters to be reproduced within two standard deviations. More details to the procedure and input parameters can be found in Appendices \ref{app:RPG&running} and \ref{app:numbers}. At this stage, the mixing angles $\alpha$ and $\beta$ from the Higgs sector still remain free parameters and our sample set only fixes the fundamental Yukawas.  \\

In the right panel of Figure \ref{fig:epsK} we show the percentage of sample points which reproduce $C_{\eps_K}
^\mathrm{exp}$ within $2 \sigma$ in the $\cos(\beta-\alpha) - \tan \beta$ plane. 
We employ the value extracted from a fit to the CKM triangle by the UTfit group \cite{Bona:2007vi}, 
\begin{equation}
C_{\eps_K}^\mathrm{exp}= 1.05 ^{+0.36}_{-0.28}\quad @\, 95\%\, \text{CL}\,.
\end{equation} 
The result shows good agreement with the estimate of the separate contributions shown in the left panel of Figure \ref{fig:epsK}. The area for which $t_\beta 
<0.5$ is cut off, because of the one-loop contributions from charged Higgs exchange \cite{Jung:2010ik}. We find a large region of parameter space for which our model prediction is in agreement with the experimental bound without any tuning of parameters.

In the case of $B_{d}-\bar B_{d}$ and $B_{s}-\bar B_{s}$ mixing, the effective Lagrangian, as well as 
the tree-level contributions to the Wilson coefficients from scalar and pseudoscalar exchange can be read off 
from 
 \eqref{eq:heffdf2} and \eqref{eq:wilsons} with the replacements $s \leftrightarrow b$ and $d \leftrightarrow d, s$, 
respectively. The angle dependence of the Wilson coefficients is universal and therefore only the flavor dependent part changes from \eqref{eq:C2} and \eqref{eq:C4}, such that the parametric dependence presented in the left panel of Figure \ref{fig:epsK} also holds in the $B$ sector. For the Wilson coefficients it follows from Table \ref{tab:Ccomp},
\begin{align}
C^{bd}_4 &\approx C^{bd}_2 \approx \tilde C^{bd}_2 \approx  C_2^{bs} \propto \,\frac{m_b^2}{v^2}\,\frac{\varepsilon^2}{m_h^2} \approx \frac{2.5\times 10^{-12}}{\text{GeV}^2}\,, \\
\tilde C_2^{bs} &\propto \,\frac{m_b^2}{v^2}\,\frac{\varepsilon^4}{m_h^2}\approx \frac{7\times 10^{-16}}{\text{GeV}^2} \,,\qquad  C_4^{bs} \propto \,\frac{m_b^2}{v^2}\,\frac{\varepsilon^3}{m_h^2} \approx \frac{4\times 10^{-14}}{\text{GeV}^2}\,.
 \end{align}
The corresponding bounds in Table \ref{tab:wcref} imply, that $C_2^{bs}$ is at the border of the naive bound, while a much larger contribution to $C_4^{bs}$ is allowed. The contributions to $C_4^{bd}, C_2^{bd}$ and $\tilde C_2^{bd}$ are too large almost in the entire $\cos (\beta-\alpha) -\tan \beta $ plane, and therefore demand cancellations implying important restrictions for the permitted region of our parameter space. 

\begin{figure}[t!]
\centering
 {\begin{minipage}{6in}
 \hspace{-1.3cm} 
 \begin{tabular}{cccc}
\quad\includegraphics[width=.43\textwidth]{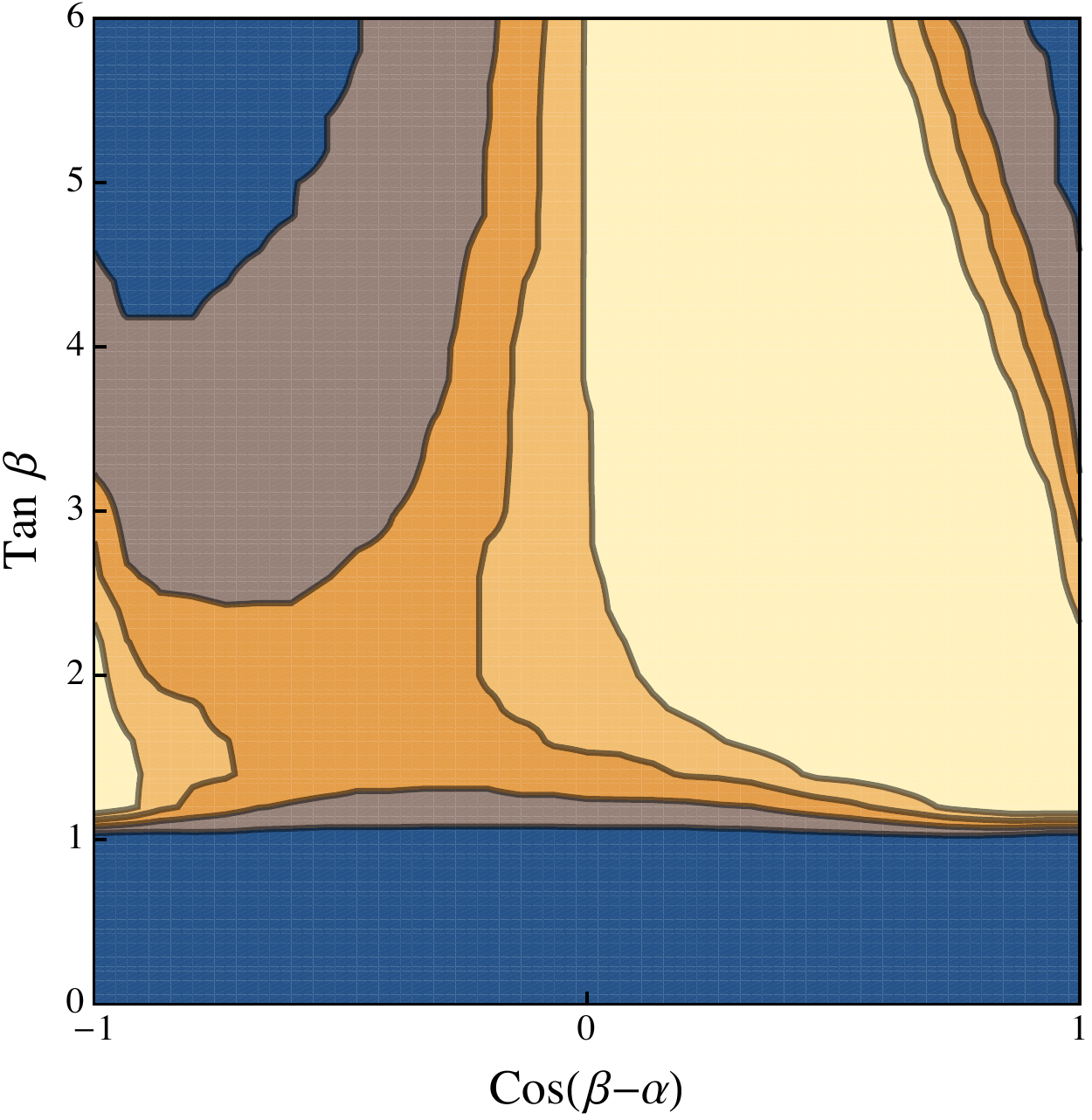}&
\raisebox{.04\height}{\includegraphics[width=.045\textwidth]{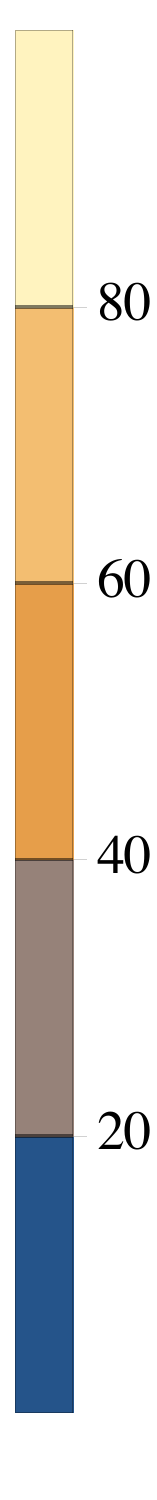}}&
\quad\includegraphics[width=.43\textwidth]{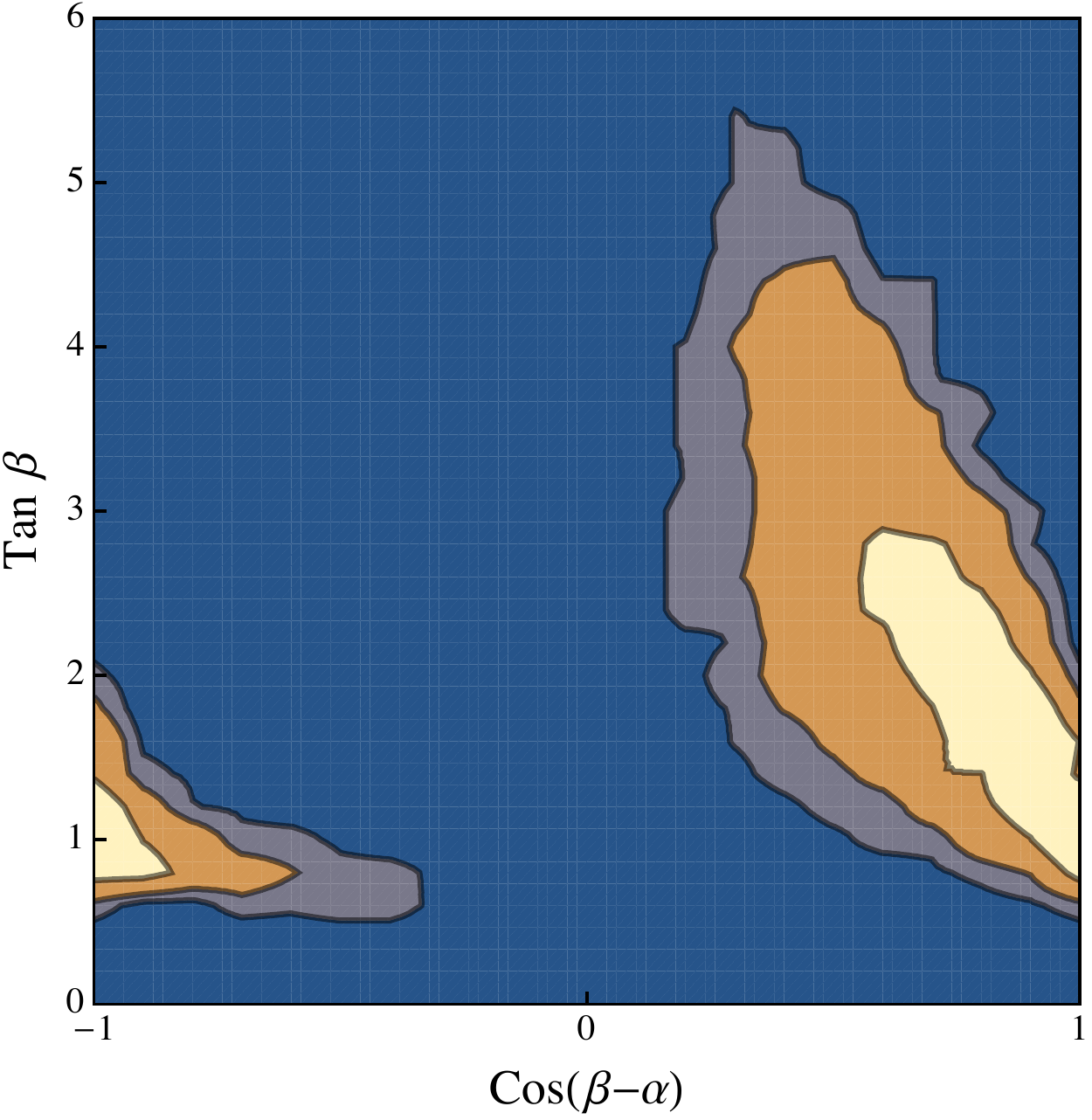}&
\raisebox{.05\height}
{\includegraphics[width=.044\textwidth]{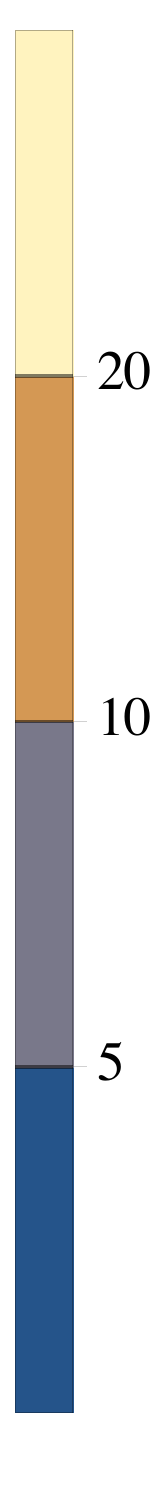}}
\end{tabular}
\end{minipage}}
\caption{In the left (right) panel we show regions of parameter space in which our sample points reproduce $C_{B_s} (C_{B_d})$ within two standard 
deviations. The color coding indicates the percentage of points in agreement with the experimental constraint.  In both plots,  the scalar masses are $M_A=M_H=M_{H^+}=500$ GeV.
\label{fig:BsBd}
}
\end{figure}
At the one-loop level, 
box diagrams generate the contributions
\begin{align}
C_{1, \mathrm{box}}^{bq} \propto \frac{1}{16 \pi^2}\frac{1}{t_\beta^2}\left(\frac{m_t^2}{v^2}\, V_{tb}^\ast V_{tq}
\right)^2\frac{1}{M_{H^+}^2}\approx \left(\frac{500\,\mathrm{GeV}}{M_{H^+}}\right)^2\begin{cases}
5\times 10^{-13}\,\mathrm{GeV}^{-2}\,,& q=d\,,\\[3pt]
1 \times 10^{-11}\, \mathrm{GeV}^{-2}\,, & q=s\,,
\end{cases}
\end{align}
for $\tan\beta=1$. In the $B_s - \bar B_s$ system for low $\tan \beta$, this contribution becomes larger than all tree-level contributions. Since the box is only sensitive to charged Higgs couplings, we expect comparable constraints as in a 
two Higgs doublet model of type II. In addition, since the contribution is independent of $\cos(\beta -\alpha)$, we expect a universal lower bound on $\tan \beta$, as observed in the left panel of Figure \ref{fig:BsBd}.
For both
the $B_{d,s} - \bar B_{d,s}$ system we also include the box diagram contributions to the other Wilson coefficients, which are suppressed by $m_b/m_W$. The corresponding expressions are collected in Appendix 
\ref{app:Boxes}.
 \\

\begin{figure}[t!]
\centering
\begin{tabular}{cc}
\includegraphics[width=.46\textwidth]{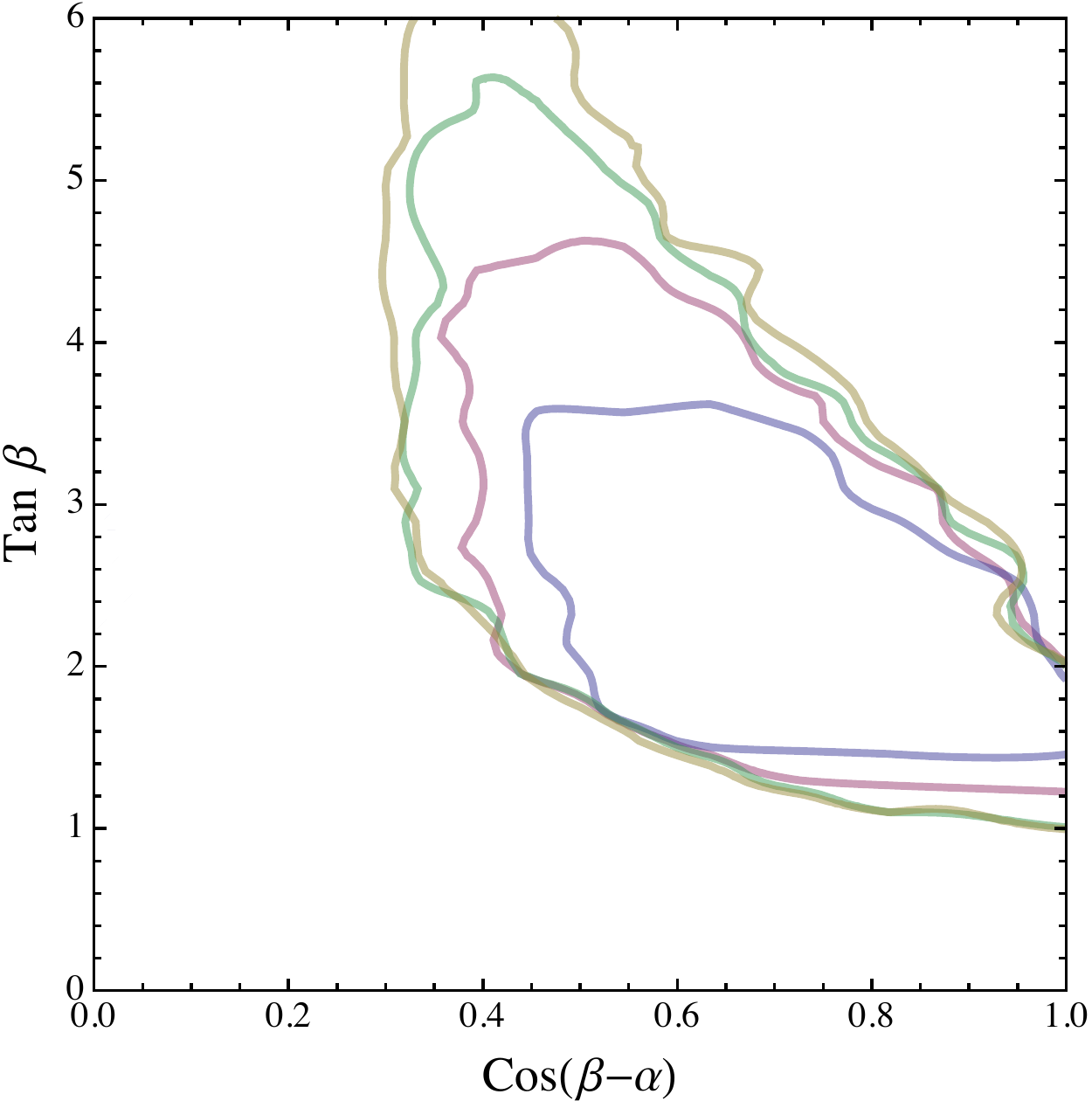}
\end{tabular}
\caption{Boundaries of the regions in which $10\%$ of our parameter points agree with the $C_{B_d}$ at the $95\%$ CL in the positive $c_{\beta-\alpha}$ plane. The different colors correspond to $M\equiv M_A = M_H = M_{H^+} = 400$ GeV (blue) $M =500$ GeV (purple),  $M= 600$ GeV (green), and $M=700$ GeV (light brown).\label{fig:10p}}
\end{figure}
Analogous to \eqref{eq:epsK},
we define 
 \begin{equation}\label{eq:CBq}
C_{B_q}e^{2i\,\phi_{B_q}}=\frac{ \,\langle B_q^0| \Hfull^{\Delta B=2}| \bar B^0_q \rangle 
}{\,\langle B^0_q| \mathcal{H}_\mathrm{SM}^{\Delta B=2}| \bar B^0_q \rangle 
}\,,
\end{equation}
such that $C_{B_q}=\Delta m_q/\Delta m_q^\mathrm{SM}$ measures new physics effects in the mass difference 
and new phases enter $\phi_{B_q}$. In the left (right) panel of Figure \ref{fig:BsBd}, we present the percentage of sample points in agreement with the experimental constraints at  $95\%$ CL for $C_{B_s}^\mathrm{exp}$ ($C_{B_d}^\mathrm{exp}$), based on the results obtained from the UTfit  group \cite{Bona:2007vi},
\begin{equation}
C_{B_s}^\mathrm{exp}= 1.052^{+0.178}_{-0.152}\quad @\, 95\%\, \text{CL}\,,\qquad C_{B_d}^\mathrm{exp}= 1.07^{+0.36}_{- 0.31}\quad @\, 95\%\, \text{CL}\,.
\end{equation}
In both plots we choose $M_H=M_A= M_{H^+}= 500$ GeV. As expected from our estimate above, in the $B_s-\bar B_s$ system, we find good agreement with the experimental bounds for a large region of parameter space. 
For the $B_d-\bar B_d$ system, we find only a small fraction of the parameter space in agreement with the experimental constraints. Since the new physics effects in all Wilson coefficients are too large, accidental cancellations in the fundamental Yukawa couplings are in effect in order to achieve agreement with data.
As a consequence, slightly tuned Yukawa couplings as well as rather heavy extra scalars $M_A\approx M_H\approx 500$ GeV are necessary in order to agree with the bounds from  $B_d -\bar B_d$ mixing. In the following, we will adopt the $10\%$ contour as the fine-tuning bound from flavor observables on the parameter space. Figure \ref{fig:10p} shows the corresponding contours in the positive $c_{\beta-\alpha}$ plane for $M\equiv M_A = M_H = M_{H^+}= 400$ GeV (blue) $M= 500$ GeV (purple),  $M= 600$ GeV (green), and $M= 700$ GeV (light brown). The bound for low $\tan\beta$ comes from the charged Higgs loops in $B_s - \bar B_s$ mixing. A future, more precise measurement of meson-antimeson mixing can reveal deviations from the SM prediction or further constrain the allowed parameter space, if no new physics effect is found.  \\

In $D-\bar D$ mixing, all tree-level contributions to the Wilson coefficients are strongly suppressed,
\begin{equation}
C^{uc}_4 \approx C^{uc}_2 \approx \tilde C^{uc}_2 \propto \,\frac{m_c^2}{v^2}\,\frac{\varepsilon^4}{m_h^2} \approx \frac{3.4\times 10^{-17}}{\text{GeV}^2}\, .\\
\end{equation}
In contrast to the down-sector however, the box diagram with neutral Higgs exchange is not suppressed by light quark masses, because the dominant contribution comes from the top in the loop \cite{Crivellin:2013wna}. The leading box contributions of the light Higgs to the coefficient $C_1^{uc}$ can therefore be larger than all tree-level effects
\begin{align}
C^{uc}_1&\approx -\frac{1}{128 \pi^2}\,\left(\frac{m_t}{v}\,\varepsilon\, f(\alpha,\beta)\right)^4\,D_2(m_t,m_h)\notag\\
&= -\frac{1}{128 \pi^2}\,\left(\frac{m_t}{v}\,\varepsilon\, f(\alpha,\beta)\right)^4\,\frac{m_h^4 - m_t^4 - 2 m_h^2 m_t^2 
\log\left(\frac{m_t^2}{m_h^2}\right)}{(m_h^2 - m_t^2)^3}\notag\\
&\approx-\frac{2\times 10^{-16}}{\mathrm{GeV}^2}\,,
\end{align}
for $f(\alpha,\beta)=1$, and the loop function defined in Appendix \ref{app:Boxes}. Boxes with heavy Higgs insertions are further suppressed.
However, the corresponding bound in Table \ref{tab:wcref} is orders of magnitude weaker than our estimate. The $D-\bar D$ system will therefore not induce further constraints.\\

In all the above analyses, we have concentrated on the solution for the flavor charges  \eqref{eq:FCharges}, but
the situation is quite different for the flavor charges given in \eqref{eq:FCharges2}. From \eqref{eq:struc2} it follows, that the contributions to the Wilson coefficients are highly suppressed, as is explicit in the flavor-dependent parts of the Wilson coefficients given on the right hand side of Table \ref{tab:Ccomp}. This shows, that although constraints from the $B_s-\bar B_s$ and $K -\bar K$ systems remain the same, the constraints from the $B_d -\bar B_d$ system can be very much relaxed due to the different charge assignment. Therefore, if only the hierarchies in the quark masses are explained by a Froggatt-Nielsen mechanism at the weak scale, but the CKM mixing angles have a different origin, bounds from meson-antimeson mixing are very mild and do not lead to any severe restrictions on the parameter space. \\

Rare Kaon and $B_{d,s}$ decays can in principle be subject to large corrections, but depend crucially on the 
implementation of the lepton sector, which will be discussed elsewhere. Processes 
in which the neutral scalars only enter at loop-level, such as $\Br(B_s \to X_s \gamma)$ are generically 
dominated by charged Higgs contributions, which are larger than the contributions from the neutral Higgs by a factor of
\begin{equation}
\frac{m_t\,V_{tb}\,V_{ts}^\ast}{ m_b\, f(\alpha, \beta )\, \varepsilon}\approx 
\mathcal{O}(10^2 - 10^3)\,,
\end{equation}
 for $f(\alpha,\beta) = 0.1 - 1$. We will therefore adopt the bounds from $\Br(B_s \to X_s \gamma)$ on the 
charged scalar mass in two Higgs doublet models for $\tan\beta \gtrsim 2$, considering values within a $3\sigma$ band in order to account for uncertainties of higher order corrections not included in the theoretical computation \cite{Hermann:2012fc, Misiak:2015xwa},

\begin{equation}
M_{H^\pm}\gtrsim  358 \,(480) \, \mathrm{ GeV} \quad  @ \, 99 \% (95 \%)\,\, \text{CL}\,.
\end{equation}

\subsection{Flavor Violating Top Decays}

We consider the flavor violating decays of the top quark $t \rightarrow h c$ and $t \rightarrow h u$. 
In contrast to the SM, in which flavor violating top quark decays are loop suppressed, in our model the top quark 
has tree-level couplings to the light Higgs and other up-type flavors. The corresponding 
branching ratios $\Br (t\rightarrow h\, c )\approx 3\times 10^{-15}$ and $\Br(t\rightarrow h\, u)\approx 2\times 
10^{-17}$ are tiny in the SM \cite{AguilarSaavedra:2004wm}. 
In our model the branching fraction of the top decaying to higgs and charm is given by \cite{Casagrande:2008hr}
 \begin{align}
\Br(t\rightarrow h\, c) = \,&\frac{2(m_t^2-m_h^2)^2 \, m_W^2}{g^2 (m_t^2-m_W^2)^2\,(m_t^2+2m_W^2)^2}
\left( |g_{hct}|^2+|g_{htc}|^2 +\frac{4m_tm_c}{m_t^2-m_h^2}\,\mathrm{Re}\left[ g_{hct}\,g_{htc}\right]\right)\,,
\end{align}
and similarly for $\Br(t\rightarrow h\, u)$ by replacing the appropriate flavor indices. Both branching ratios are parametrically of the same order, because the flavor off-diagonal couplings in equation $
\eqref{eq:hcoupling}$ yield 
$ g_{hct}\approx g_{hut}\propto m_t \varepsilon$.
  \begin{figure}[b!]
\centering
\includegraphics[width=0.46\textwidth]{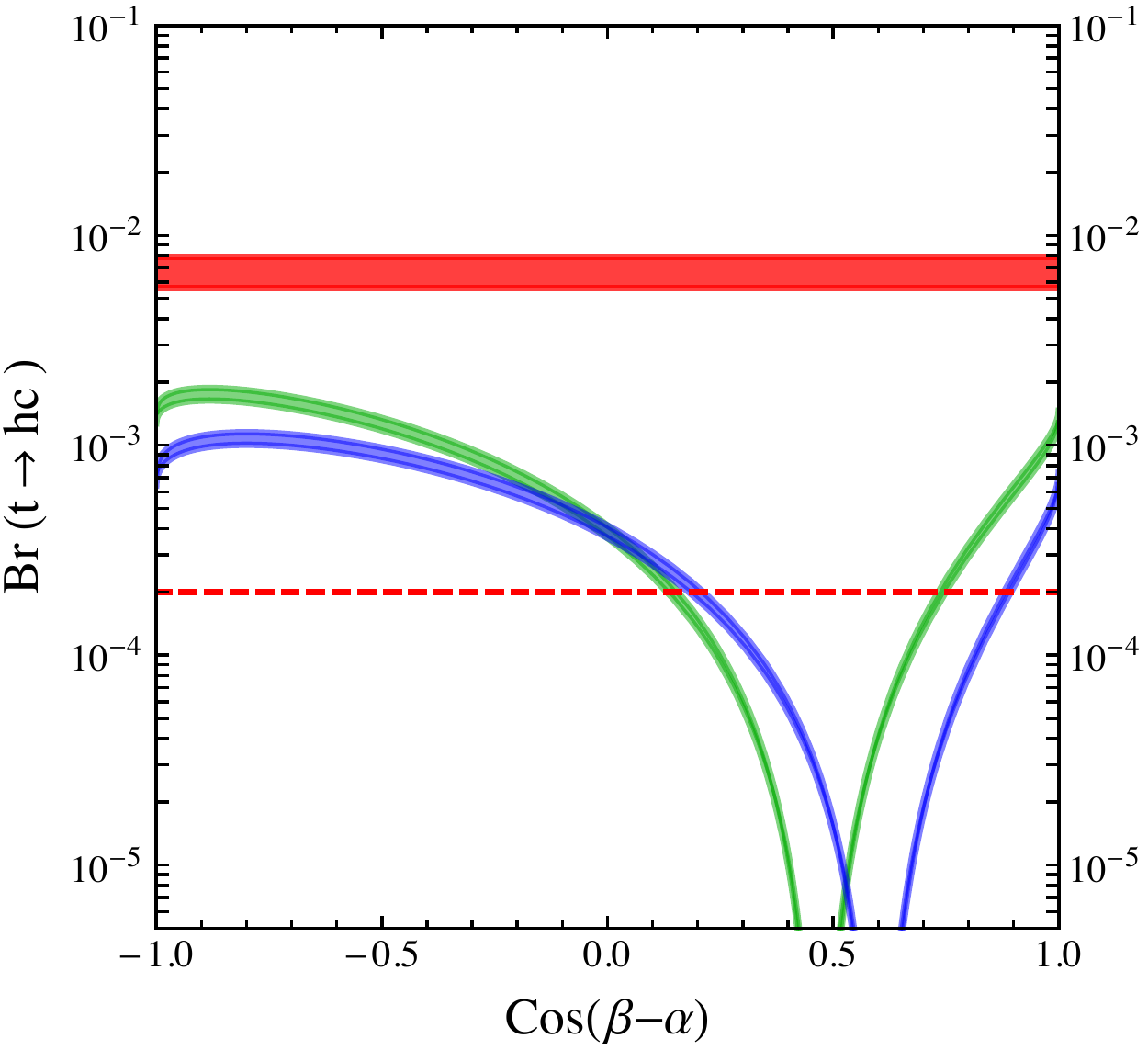}
\caption{The plot shows $\Br(t\rightarrow h c) $ vs. $\cos(\beta-\alpha)$ for $\tan\beta=3 (4)$ in blue (green) as well as the current 
exclusion limits for the 8 TeV LHC  (solid red) and projected limits at the high luminosity LHC (dashed red), respectively.}
\label{fig:thc}
\end{figure}
In Figure \ref{fig:thc} we show $\Br(t\rightarrow h\,c)$ plotted against $\cos (\beta-\alpha)$ for a range of parameter points and indicate the different predictions for 
$\tan \beta=3 (4)$ by a blue (green) band. The widths of these bands correspond to the range of values obtained by scanning over our sample set of random fundamental Yukawas.
The most recent limits are $\Br(t\rightarrow h c) < 0.56 \%$ from CMS  \cite{Aad:2014dya} and $\Br(t\rightarrow h c) < 0.79 \%$ from 
ATLAS \cite{Khachatryan:2014jya}
 and are shown in the plot as a red band. The projected exclusion limit for $3000 \,\text{fb}^{-1}$ at the high luminosity LHC $
\Br(t\rightarrow h c) < 2\times 10^{-4}$ 
\cite{Agashe:2013hma} is indicated by a dashed red line. The plot shows that this cross section can 
be even above $10^{-4}$ for negative values of $\cos (\beta-\alpha)$. However, 
the cross section drops for the same angles for which FCNCs become small, because the same trigonometric function governs flavor off-diagonal couplings between the light Higgs to up- and down-type 
quarks in equation \eqref{eq:hcoupling}.

 \section{Perturbativity, Unitarity, and Electroweak Precision Measurements}\label{sec:EWPT}

In this section we consider perturbativity bounds, as well as constraints from the unitarity of the S matrix and electroweak precision measurements on our model. 
The large scalar masses implied by flavor observables and the constrained scalar potential \eqref{eq:pot} result in potentially large quartic couplings. Mass splittings between the different scalar mass eigenstates can in addition generate sizable contributions to the oblique parameters $S, T$ and $U$.  We therefore scan over the allowed parameters, considering the various bounds described in \cite{Eriksson:2009ws}. This includes stability constraints on the Higgs potential, perturbativity bounds on the quartic scalar couplings, unitarity of the various scattering amplitudes involving scalars and the constraints from the oblique parameters. This calculation is not different from a generic two Higgs doublet model, since the oblique parameters only measure corrections to the gauge boson self-energies from loops of the new scalars, whose couplings are fixed by the kinetic terms \cite{Haber:2010bw, Celis:2013rcs}. 

\begin{figure}[t!]
 \centering
 \begin{tabular}{c c c}
\hspace{0.1 cm} \includegraphics[scale=.49
]{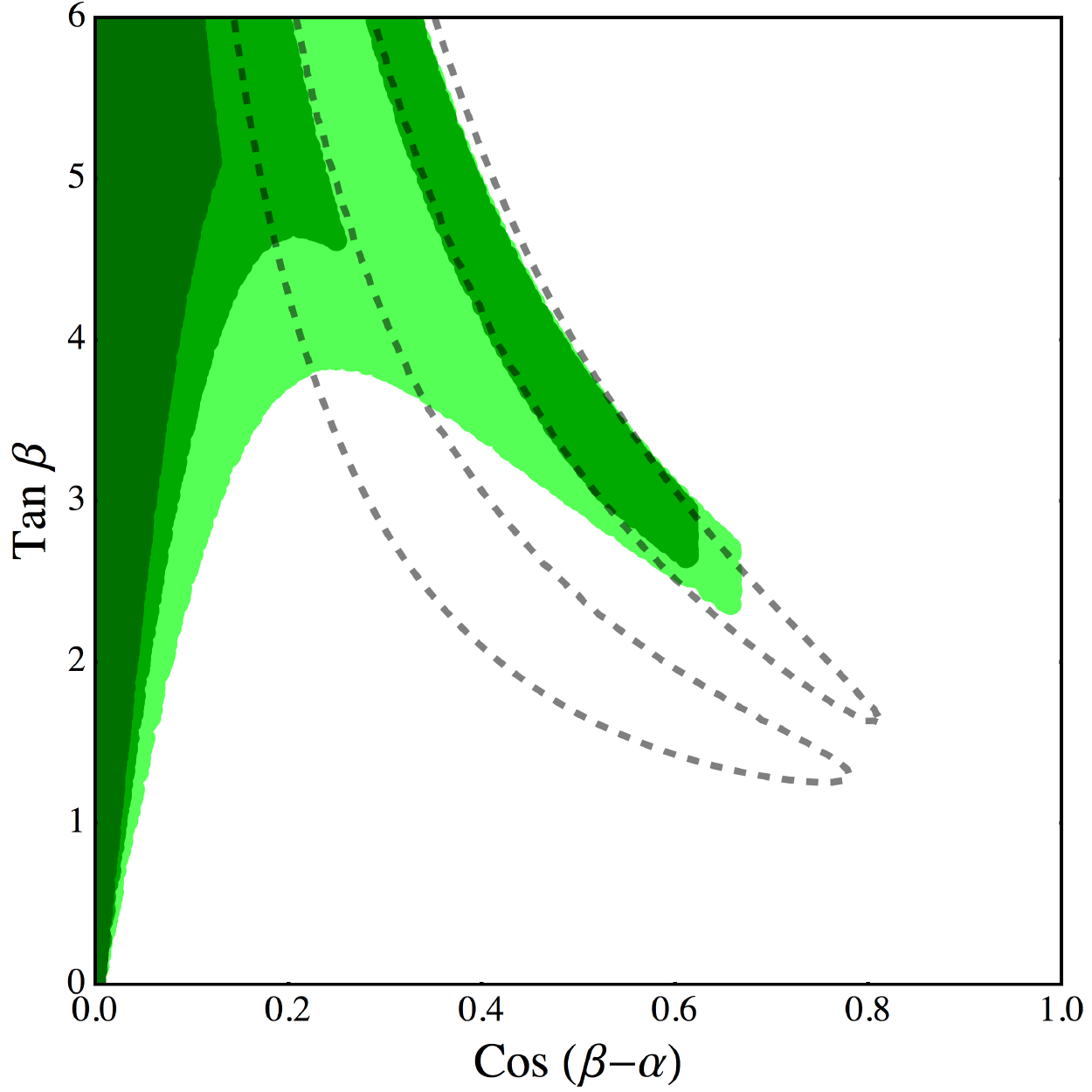} 
 &\hspace{.4cm}\includegraphics[scale=.49]{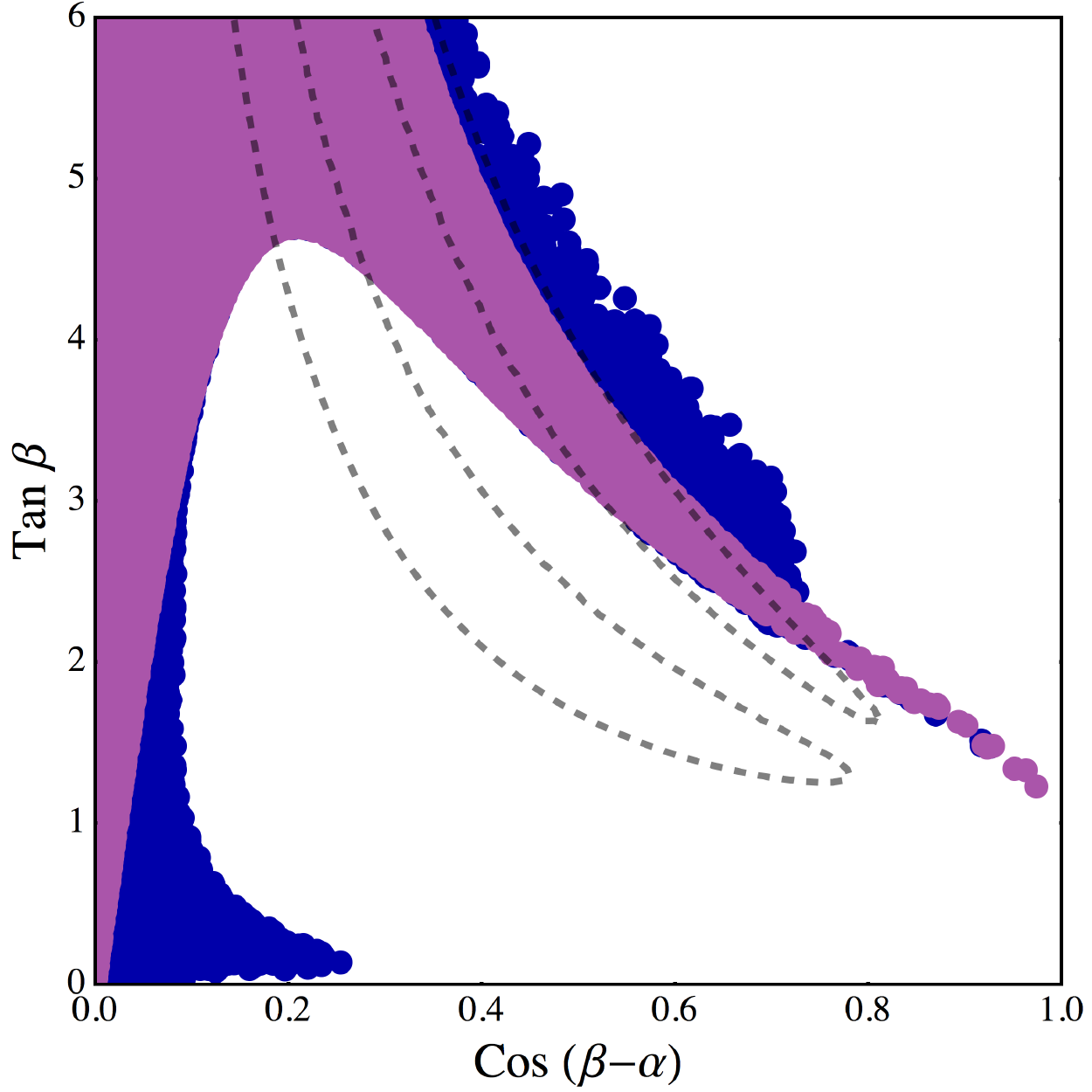}&\\
\includegraphics[scale=.5]{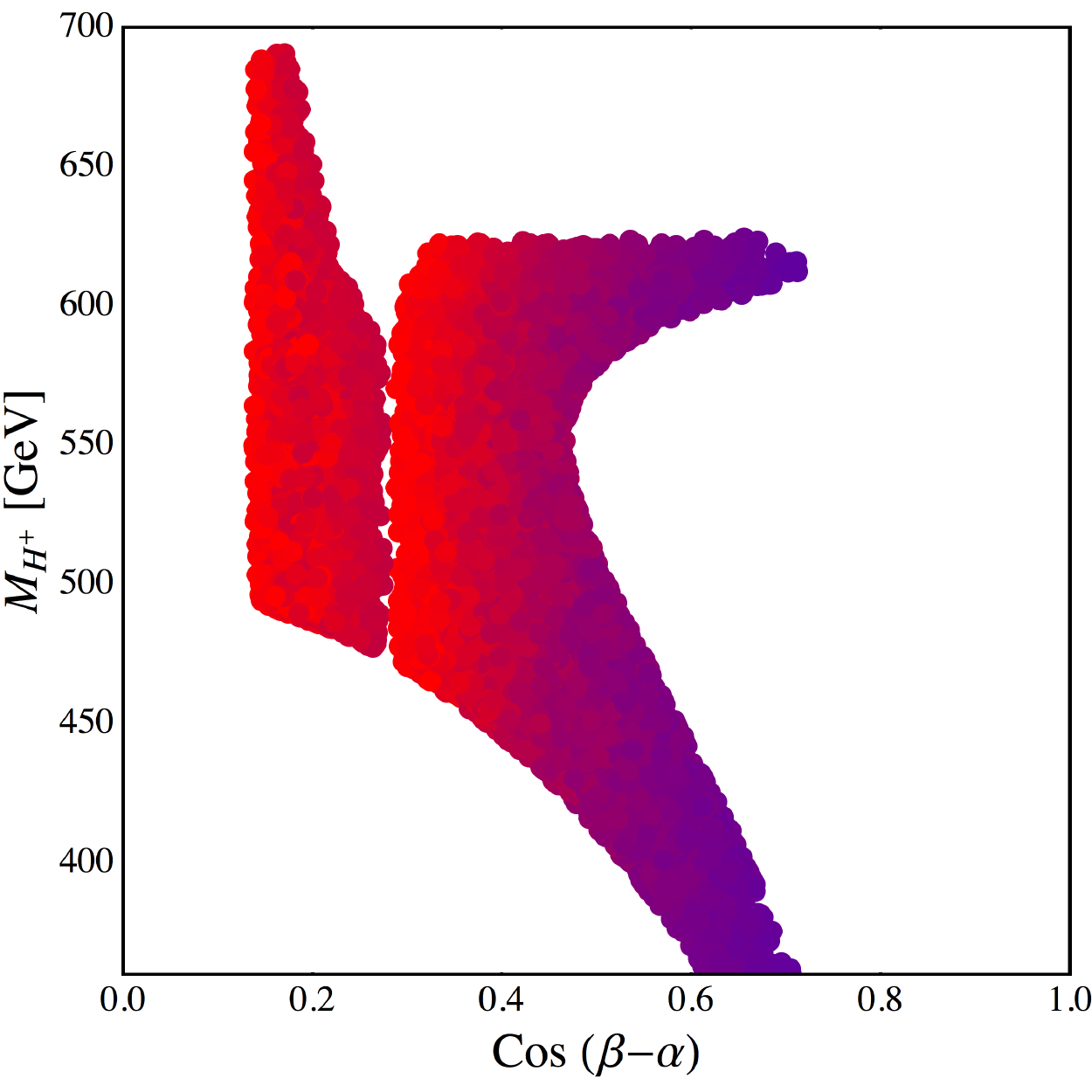}&\includegraphics[scale=.5]{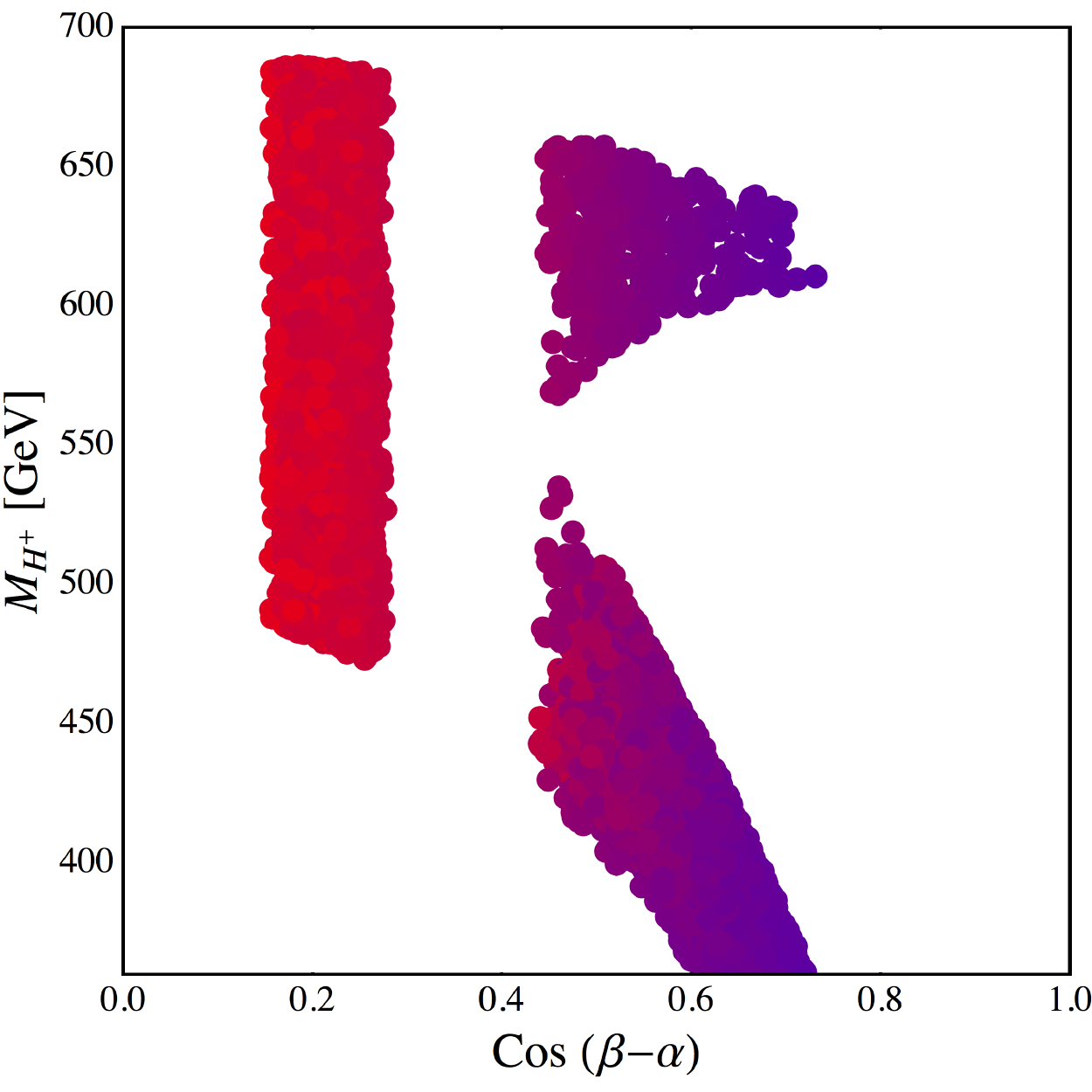}&\hspace{-2.5cm}\raisebox{.6cm}{ \includegraphics[width=.38\textwidth]{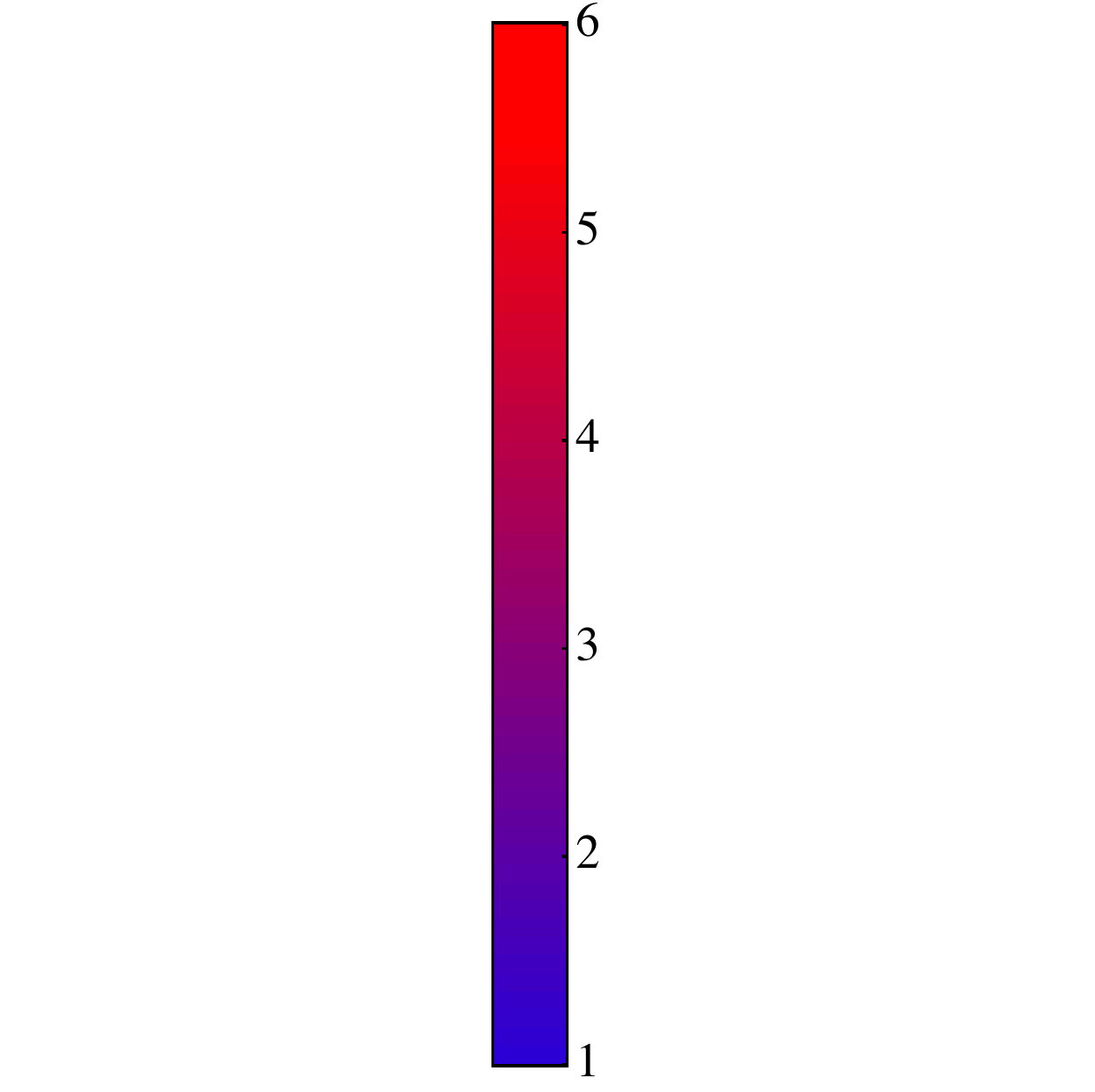}}
\end{tabular}
 \caption{
 \label{fig:S&T}
The upper left panel shows regions of parameter space in which the various constraints described in the text are fulfilled for scalar masses $M\equiv M_H=M_A=M_{H^+}= 500$ GeV (light green), $M=600$ GeV (green) and $M=700$ GeV (dark green).
The upper right panel shows the same plot for $M_{H^+} = 360 - 700$ GeV and $M_A = M_H = 600$ GeV ($M_A= 600$ GeV, $M_H= 550-650$ GeV) in purple (blue). The $2\sigma$ contours of the ATLAS fit to Higgs measurements is shown in dashed black. The lower panels show the parameter space in the $\cos (\beta-\alpha) - M_{H^+}$ plane in agreement with all bounds discussed in the text, including the $2\sigma$ global fit to ATLAS data. In the lower left (right) panel we assume $M_A = M_H = 600$ GeV  ($M_H = M_A \pm (10-20) $ GeV), with values of $\tan \beta $ indicated by the color coding bar on the right. } \end{figure}

The two plots in the upper panels of Figure \ref{fig:S&T} show the region in the positive $\cos(\beta-\alpha)-\tan \beta$ 
plane in which stability and perturbativity bounds are fulfilled, and the $S$ and $T$ parameters are at most $2 \sigma$ from the best fit point, corresponding to a global $\chi^2$ fit obtained by the Gfitter group \cite{Baak:2014ora}. The upper left panel illustrates the allowed regions for degenerate scalar masses of $M \equiv M_A=M_H=M_{H^\pm}=500$  GeV in light green,  $M=600$ GeV in green and  $M=700$ GeV in dark green. For masses $M=700$ GeV only values of $\cos(\beta-\alpha)\lesssim 0.2$ are allowed, approaching the decoupling limit. For masses $M=500$ GeV and $M=600$ GeV there is a region of parameter space in agreement with all constraints for values of $\cos(\beta-\alpha) > 0.2$, that partly overlaps the region preferred by the global fit to the SM Higgs signal strengths.
In the upper right panel, we show the same plot for masses $M_{H^+}= 360 - 700$ GeV and $M_A=M_H= 600$ GeV ($M_A= 600$ GeV, $M_H= 550-650$ GeV) in purple (dark blue). In both upper panels, we also superimpose the $2\sigma$ contours  (dashed lines) of the global Higgs fit using the ATLAS measurements of the signal strengths, that are the most stringent at present. Almost all of the right branch of the global Higgs fit can be populated for large scalar masses, while low values of $\cos(\beta-\alpha)<0.3$ are only allowed for $\tan\beta \gtrsim 4.5$. 

The lower left panel shows the region allowed by all constraints discussed above for which we further demand, that the ATLAS SM Higgs signal strengths measurements are reproduced within $2\sigma$ in the $\cos (\beta-\alpha) - M_{H^+}$ plane for $M_A=M_H = 600$ GeV. The value of $\tan \beta$ is indicated by the color coding. The tiny gap at $\cos (\beta-\alpha) \approx 0.3 $ is also visible in the upper left plot. For $\tan\beta \lesssim 4$ only degenerate masses $M_A=M_H= M_{H^+}$ or a sizable mass splitting of $M_{A}-M_{H^+}\gtrsim 100$ GeV are allowed. We show the same plot in the lower right panel, but with a moderate mass splitting between the neutral Higgs boson masses, $M_H = M_A \pm (10-20)$ GeV, while keeping $M_A=600$ GeV fixed.\footnote{If the mass splittings become larger than $|M_H-M_A| \gtrsim 30$ GeV, the full parameter space is excluded.} In that case the gap around $\cos(\beta-\alpha) \approx 0.3$ becomes much more prominent. 

Further, for some regions of the parameter space, one or more of the quartic couplings in the Higgs potential can become non-perturbative already at the TeV scale $\lambda_i(\mu = 1\text{TeV})\gtrsim 4 \pi $. We implement the one-loop beta functions for our model and match to the SM at  an approximate average scale of the Higgs boson masses in order to estimate the scale of strong coupling. In particular for larger values $\cos(\beta-\alpha)$ and larger and degenerate masses $M_A=M_H$, the cutoff scale becomes lower. Moreover, we find that for sizable mass splittings between the charged and neutral scalars, the scale of strong coupling is in the range of $2 - 5$ TeV. However, as mentioned in Section \ref{sec:setup} and in more detail in Section \ref{seq:tevcompletion} below, we expect the UV completion of our model to set in close to the TeV scale.

We conclude, that for fixed $M_A=600$ GeV, two qualitatively different choices of scalar masses are compatible with electroweak precision bounds, Higgs constraints and a low $\tan \beta$ as preferred by flavor constraints. Either the scalar masses are approximately degenerate $M_A\approx M_H\approx M_{H^+}$ or the charged scalar is considerably lighter than the neutral scalars $M_{A,H} - M_{H^+} \gtrsim 100$ GeV.  Of these possibilities, only for large mass splittings can the theory be valid up to several TeV and in the following we will concentrate on this setup. Note, that these restrictions would be slightly relaxed if we take the fit to the CMS measurements of the Higgs signal strengths as a constraint.\\

Another important electroweak precision observable is the  $Z b \bar b$  coupling. While the experimental value of the left-handed $Z 
b_L \bar b_L$ coupling is in good agreement with the SM prediction, there is a discrepancy between the measured 
right-handed  $Z b_R \bar b_R$  coupling and the SM prediction, see \emph{e.g.} \cite{Baak:2014ora, Buras:2010pz}. 
Higher order corrections with the neutral or charged scalars in the loop can in principle affect these couplings.

 The charged scalar contributions to the left-handed $Z 
b_L \bar b_L$ couplings in a two Higgs doublet model of type II can become sizable for low $\tan\beta$, inducing a bound of $t_\beta \gtrsim 0.5$ for masses of $M_{H^\pm}\approx 500$ GeV \cite{Jung:2010ik}, while corrections to the $Z b_R \bar b_R$ vertex are suppressed by $m_b/m_t$. In addition, the neutral scalar couplings to bottom quarks are very different from a 
generic two Higgs doublet model in a large range of parameter space.
 We define the couplings of the $Z$ boson to left-handed and right-handed bottom quarks by  
\begin{equation}
\Lag_{Zbb} = -\frac{e}{2 s_W c_W} Z_\mu \bar b \gamma^\mu \left(g^{L}(1 -\gamma_5) + g^{R} (1 -\gamma_5 )\right)\,b\,,
\end{equation}
with 
\begin{equation}
g^{L,R}= g^{L,R}_\mathrm{SM} +\delta g_{h}^{L,R}+ \delta g_{A,H}^{L,R}+\delta g_{H^\pm}^{L,R}\,.
\end{equation}
Here, $g_\mathrm{SM}^{L,R}$ are the SM couplings and we denote the corrections from neutral and charged Higgs exchange by $\delta g_{h}$,  $\delta g_{A,H}$ and $\delta g_{H^\pm}$, respectively. 
We estimate
\begin{equation}
\frac{\delta g_h^L}{\delta g_{H^\pm}^L} \propto \frac{M_{H^\pm}^2}{m_h^2}\,t_\beta^2\, \kappa_b^2\, 
\varepsilon^2\,,\qquad \frac{\delta g_h^R}{\delta g_{H^\pm}^R} \propto \frac{M_{H^\pm}^2}{m_h^2}\,
\frac{\kappa_b^2}{t_\beta^2}\,,
\end{equation} 
while contributions from the heavy neutral scalars are further suppressed by $\delta g_{h}/\delta g_{A,H}\approx  m_h^2/M_{A,H}^2$ and couple with  $\kappa_b^A$ and $\kappa_b^H$, as defined in the following section in equation \eqref{eq:kappas}.
Neutral Higgs contributions to $g^L$ are 
therefore at least an order of magnitude smaller than the charged Higgs contributions for the region preferred by the global Higgs fit, while corrections to the right-handed coupling $g^R$ are at most of a similar size. 
We numerically estimate the light neutral Higgs contributions following \cite{Haber:1999zh,Logan:1999if}. For $\kappa_b^2 =1$, we find for the right-handed coupling $\delta g_h^{R}\lesssim 10^{-6}\times g^{R}_\mathrm{SM}$,  and for the left-handed coupling $\delta g_h^{L}\lesssim 10^{-6}\times g^{L}_\mathrm{SM}$, which is many orders of magnitude too small in order to explain the anomalous $Z b_R b_R$ coupling. 
In order to improve the fit with respect to the SM, contributions of the order of $0.2\%$ to $g^L_\mathrm{SM}$ and $2\%-20\%$ to $g^R_\mathrm{SM}$ (depending on the sign) are necessary \cite{Batell:2012ca}. The neutral Higgs contributions to the $Z b \bar b$ vertex can therefore be safely neglected. It should be noted, that fermionic mixing effects in the UV completion of this model can affect both the oblique parameters and the $Zb\bar b$ vertex. These however depend sensitively on the exact realization of the UV completion, which is beyond the scope of this paper. 

\section{Collider Searches for Heavy Extra Scalars \label{sec:colliderboundsforscalars}}

Our model features heavy new scalars beyond the SM, namely the neutral scalar Higgs $H$, the pseudo-scalar 
$A$ and the charged Higgs $H^{\pm}$.  Their masses are bound to be less than $700$ GeV by perturbativity, and various flavor constraints set lower bounds on their masses as discussed in Section \ref{sec:flavor}.
In this section we consider the latest ATLAS and CMS bounds on new neutral and charged Higgs bosons.

\subsection{Couplings and Total Width of Heavy Scalars}
\begin{figure}[b!]
\centering
\begin{tabular}{cc}
\includegraphics[width=.46\textwidth]{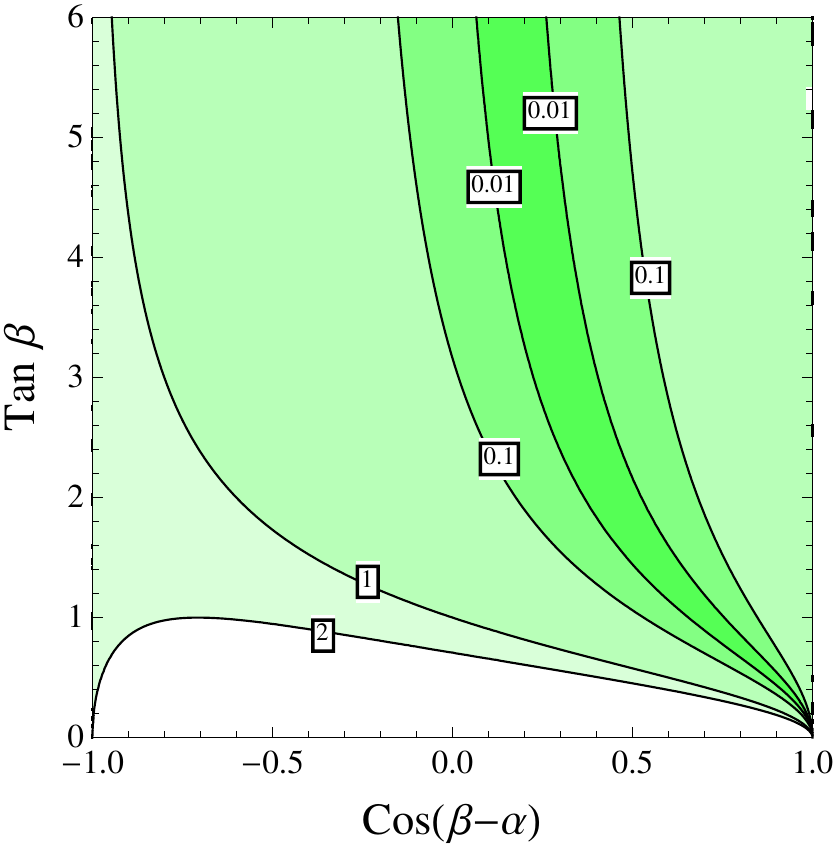}
\end{tabular}
\caption{Contours of $(\kappa_t^H)^2$ in the $\cos(\beta-\alpha)-\tan{\beta}$ plane. A suppression of the 
coupling with respect to the SM is achieved in the darker shaded area.   
\label{fig:kappatHsqu}}
\end{figure}
Similar to the case of the light scalar, the couplings of the heavy scalar $H$ and pseudoscalar $A$  to quarks - with the exception of the top quark - differ from the couplings in a two Higgs doublet model. The couplings to gauge bosons are instead the same as in a two Higgs doublet model. Specifically, the couplings of $H$ and $A$ to gauge bosons and third generation quarks normalized to the SM as in \eqref{eq:norm}, read   
\begin{align}\label{eq:kappas}
\kappa_{t}^H& =c_{\beta-\alpha}-\frac{s_{\beta-\alpha}}{t_\beta}\,,
\qquad \kappa_{b}^H=3c_{\beta-\alpha}+s_{\beta-\alpha}\left(2t_\beta-\frac{1}{t_\beta}\right)\,, \qquad 
\kappa_V^H=c_{\beta-\alpha}\,,\notag \\
\kappa_{t}^A&=\frac{1}{t_\beta}\,,
\qquad\qquad\qquad\,\,\, \kappa_{b}^A=2t_\beta+\frac{1}{t_\beta}\,,
\end{align}
where $t$, $b$ and $V$ denote the rescaling factor for top, bottom and vector boson couplings, respectively.
Since $(\kappa_{t}^{H})^2$ is relevant for the gluon fusion production of the heavy Higgs boson $H$, its parametric dependence is essential and we illustrate it in Figure 
\ref{fig:kappatHsqu}. Both flavor diagonal and flavor changing couplings of $H$ and $A$ involving the charm quark, are given by
\begin{align}\label{eq:kappaswithcharm}
\kappa_{c}^H& =3 c_{\beta-\alpha}+ s_{\beta-\alpha}\left(t_\beta - \frac{2}{t_\beta}\right)\,,
\qquad \kappa_{tc}^H=\left(2c_{\beta-\alpha}+s_{\beta-\alpha}\left(t_\beta-\frac{1}{t_\beta}\right)\right) \cdot \varepsilon\,, \notag \\
\kappa_{c}^A&=\frac{2}{t_\beta}+t_\beta\,,
\qquad\qquad\qquad\,\,\, \kappa_{tc}^A=\left(t_\beta+\frac{1}{t_\beta}\right) \cdot \varepsilon\,,
\end{align}
where $\kappa_{tc}^A$ and $\kappa_{tc}^H$ are defined according to equation \eqref{eq:AHtotc} below.
As discussed at the end of Section \ref{sec:setup}, we define the couplings to taus as
\begin{equation}
\kappa_{\tau}^H = \kappa_{b}^H\,, \qquad \kappa_{\tau}^A = \kappa_{b}^A\,.
\end{equation}
The couplings of the charged Higgs $H^+$ to fermions are the same as in a two Higgs doublet model of type II.
Similarly, all self-couplings between the scalars are the same as in a generic two Higgs doublet model. The coupling between the heavy scalar $H$ and the light Higgs $h$ is of particular interest for the following analysis and reads \cite{Gunion:2002zf,Craig:2013hca,Carena:2013ooa} 
\begin{align}
g_{Hhh} = \frac{c_{\beta-\alpha}}{v}\left[(3 M_A^2 - 2 m_h^2-M_H^2) \left( c_{2(\beta-\alpha)} -
\frac{s_{2(\beta-\alpha)}}{t_{2\beta}}\right) -M_A^2 \right]\,.
\label{eq:gHhh}
\end{align} 
Finally, the couplings between two Higgs bosons and one gauge boson read  \cite{Gunion:1989we} 
\begin{align}
 &g_{AhZ}= \frac{g}{2 \cos\theta_W} c_{\beta-\alpha}\,, \qquad   g_{AHZ}=\frac{g}{2 \cos\theta_W} s_{\beta-\alpha} \,, \qquad   g_{A H^+W^-}=\frac{g}{2} \,,\notag\\[4pt]
 & g_{h H^+W^-} =\frac{g}{2} c_{\beta-\alpha}\,, \qquad  g_{H H^+W^-}=\frac{g}{2} s_{\beta-\alpha}\,.
\end{align}
Further, we  define the total widths for $H$, $A$, and $H^+$, including all relevant and kinematically accessible decay channels (no off-shell decays are relevant in the regions we will consider)
\begin{align}
 \Gamma_{H} &= \Gamma (H \to WW)+\Gamma (H \to ZZ)+\Gamma (H \to hh)+\Gamma (H \to AZ) +\Gamma (H \to H^+W^-) \nn\\
&\quad +\Gamma (H \to  t\bar t)+\Gamma (H \to b\bar b)+\Gamma (H \to c\bar c)+\Gamma (H \to t\bar c) +\Gamma (H \to g\bar g)\nn \\
&\quad + \Gamma (H \to  \tau^+ \tau^-)\,,\notag\\[3pt]
  \Gamma_{A} &= \Gamma (A \to hZ)+\Gamma (A \to HZ)+\Gamma (A \to  H^+ W^-)+\Gamma (A \to  t\bar t)+\Gamma (A \to b\bar b) \nn \\
  &\quad +\Gamma (A \to c\bar c)+\Gamma (A \to t\bar c)+\Gamma (A \to g\bar g)+ \Gamma (A \to  \tau^+ \tau^-)\,,\nn\\[3pt]
\Gamma_{H^+} \!\!& = \Gamma (H^+ \to hW^+)+\Gamma (H^+ \to H W^+) + \Gamma (H^+ \to A W^+)+\Gamma (H^+ \to  t \bar b)\nn\\
 &\quad+\Gamma (H^+ \to  \tau \bar \nu) \,.
\label{eq:totalwidthHA}
\end{align}
\begin{figure}[t!]
\centering
\begin{tabular}{cc}
\quad \small{$M=600$ GeV} &\quad \small{$M=600$ GeV}\\
\includegraphics[width=.46\textwidth]{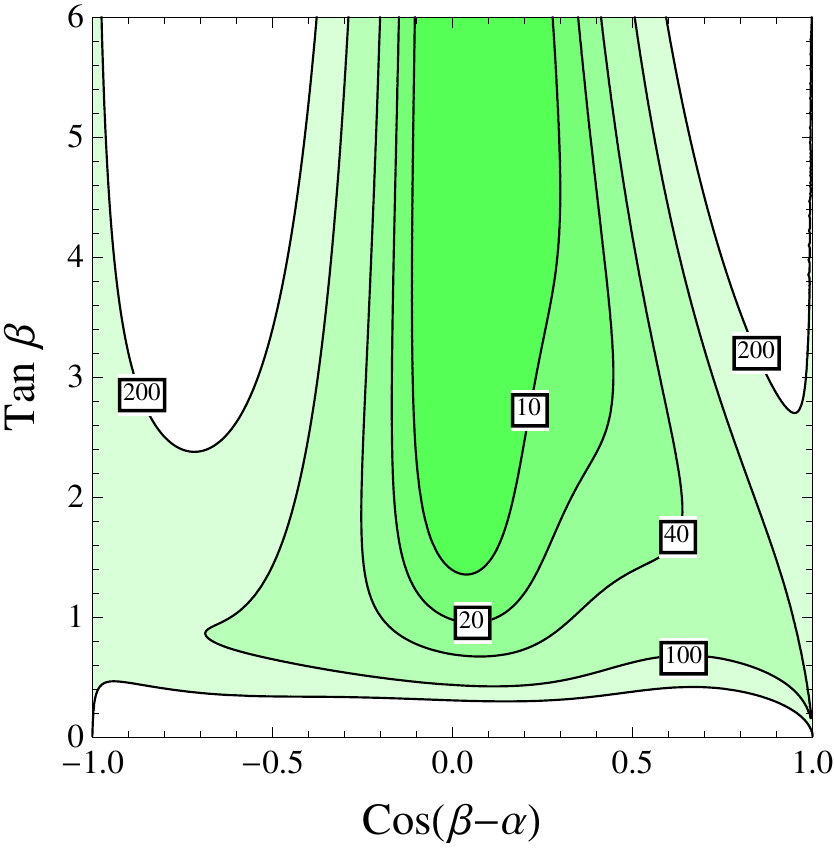}&
\includegraphics[width=.46\textwidth]{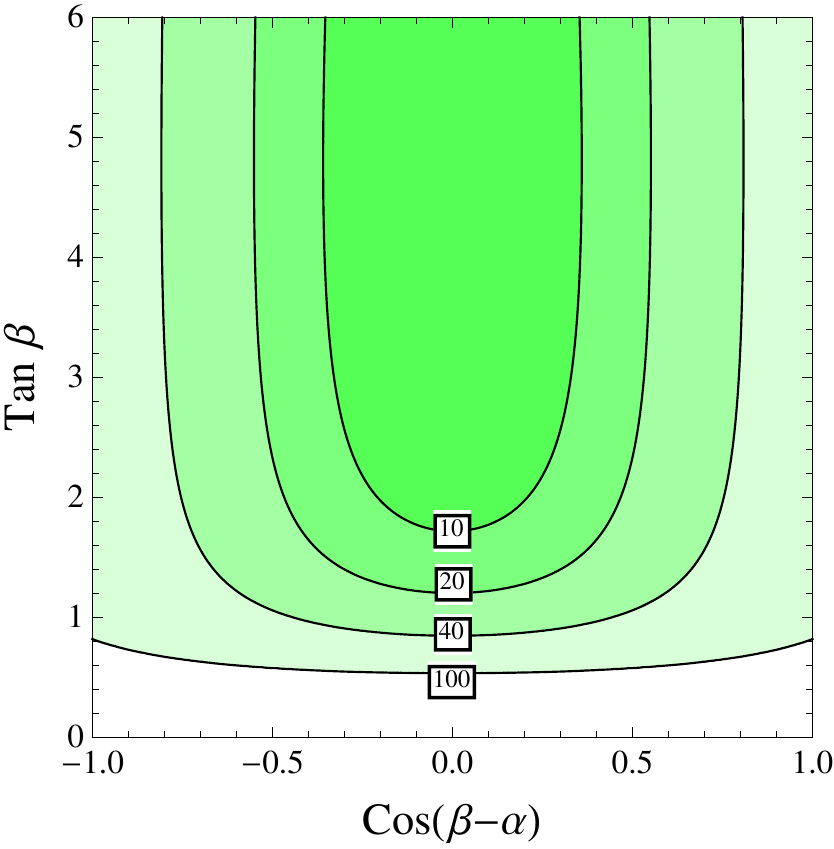}
\end{tabular}
\caption{The plot shows the parametric dependence of the total width for the heavy Higgs $H$ (left  panel) and total width for the pseudoscalar $A$ (right  panel) for $M=600$ GeV. The contours, labeled in GeV, show lines of constant width. 
\label{fig:totalwidth} }
\end{figure}
Note that, besides the usual decay channels the flavor violating channel $\Gamma (\Phi \to c\bar t)$ with $\Phi=H,A$ appears in \ref{eq:totalwidthHA}. This channel is characteristic for our model and we therefore give the partial width explicitely 
\begin{align}
\Gamma (\Phi \to c\bar t) &= \frac{3}{8 \pi} \left(\kappa_{tc}^\Phi\right)^2\,\frac{m_t^2}{v^2}  \,M_{\Phi} \, \sqrt{\lambda(1,\frac{m_t^2}{M_{\Phi}^2},\frac{m_c^2}{M_{\Phi}^2})} \left\{\begin{array}{ll} \left(\frac{(m_t-mc)^2}{M_A^2} -1 \right)& \quad \text{for} \quad M_{\Phi} =M_A \,,\\[5pt]
        \left(1-\frac{(m_t+mc)^2}{M_H^2} \right)& \quad \text{for} \quad M_{\Phi} =M_H\,,  \end{array}\right.
         \label{eq:AHtotc}
\end{align}
with
\begin{equation}
 \lambda(x,y,z)= x^2+y^2+z^2-2xy-2xz-2yz\,.
\end{equation}

The parametric dependence of the total width for the scalar (pseudoscalar) Higgs boson is illustrated in the left (right) panel of Figure \ref{fig:totalwidth}
for $M=M_A=M_H=M_{H^+}=600$ GeV. For large regions of parameter space the total width becomes large. In particular, for $\tan\beta > 1$ and $|\cos(\beta-\alpha)|> \ord(0.5)$  values of $\ord(100)$ GeV can be obtained, such that finite width effects need to be taken into account. The charged Higgs can also have a sizable branching ratio $\Br(H^+ \rightarrow h W^+)$, which can become the dominant decay channel for sufficiently large $\cos(\beta-\alpha)$.
In Appendix \ref{app:branchingratios} we show the branching ratios for all Higgs bosons for specific benchmark scenarios to be discussed later.
\begin{figure}[t!]
\centering
\begin{tabular}{cc}
\quad \small{$M=600$ GeV} &\quad \small{$M=600$ GeV}\\
\includegraphics[width=.46\textwidth]{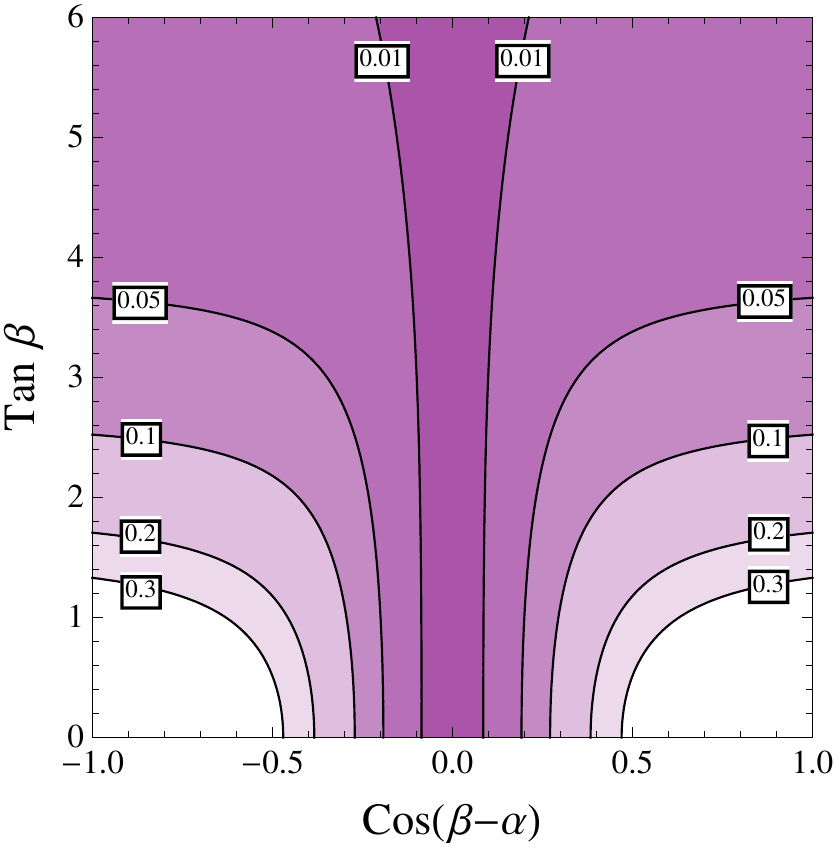}&
\includegraphics[width=.46\textwidth]{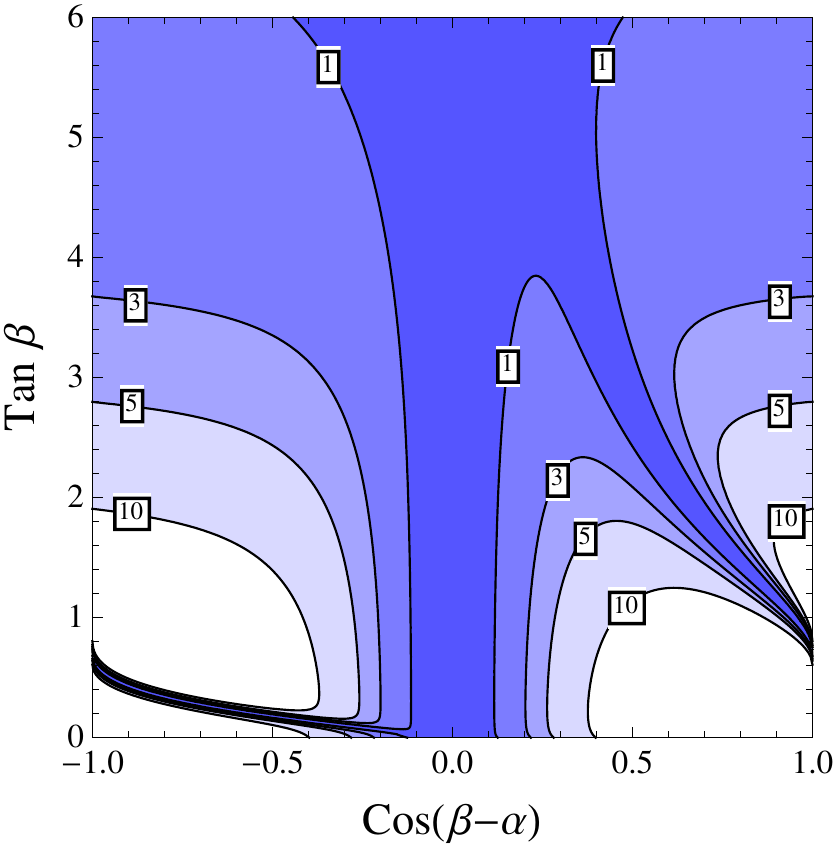}
\end{tabular}
\caption{We show contours of constant $\sigma (gg \to A) \times \Br(A \to hZ)$ in picobarn (left  panel) and $\sigma(gg \to A )\times \Br(A \to hZ \to \ell\ell b\bar b)$ in femtobarn (right  panel) for $8$ TeV $pp$ collisions and $M=600$ GeV. \label{fig:AtohZpred} }
\end{figure}

\subsection{Analysis of Production and Decay Channels}
In the following we study the impact of searches for heavy higgs bosons at ATLAS and CMS. To this end, we compute the production cross section and various decay rates for the heavy Higgs bosons. We generate the gluon-fusion production cross section at next-to-leading order (NLO) using HIGLU \cite{Spira:1995mt}, taking into account the contributions of the bottom quark loop and use the leading order expressions for the partial decay width with the appropriate couplings of our model \cite{Djouadi:1994mr,Djouadi:1995gv}. When relevant, we also consider the vector-boson fusion production cross section, using the values quoted in \cite{Dittmaier:2011ti, Heinemeyer:2013tqa}. For charged Higgs production we use the NLO results in \cite{Flechl:2014wfa}. In the following we will assume $M=M_A=M_H=M_{H^+}$, if not specified otherwise, and we discuss in detail the effects of a splitting between the neutral and charged Higgs boson masses.

One of the most interesting channels for the discovery of the pseudoscalar Higgs boson, involves the $A\rightarrow h Z$ decay, because the corresponding branching ratio becomes dominant for sizable values of $\cos(\beta-\alpha)$. There are several experimental studies constraining $\sigma (gg \to A) \times \Br(A \to hZ)$, with the light higgs further decaying into   bottom quarks \cite{Aad:2015wra,Khachatryan:2015lba}, tau leptons \cite{Aad:2015wra}, as well as multi-leptons \cite{Khachatryan:2014jya}.

\begin{figure}[t!]
\centering
\begin{tabular}{cc}
\quad CMS: \small{$M=(500,600)$ GeV} & \small{ ATLAS: $M=(500,600)$ GeV}\\
\includegraphics[width=.46\textwidth]{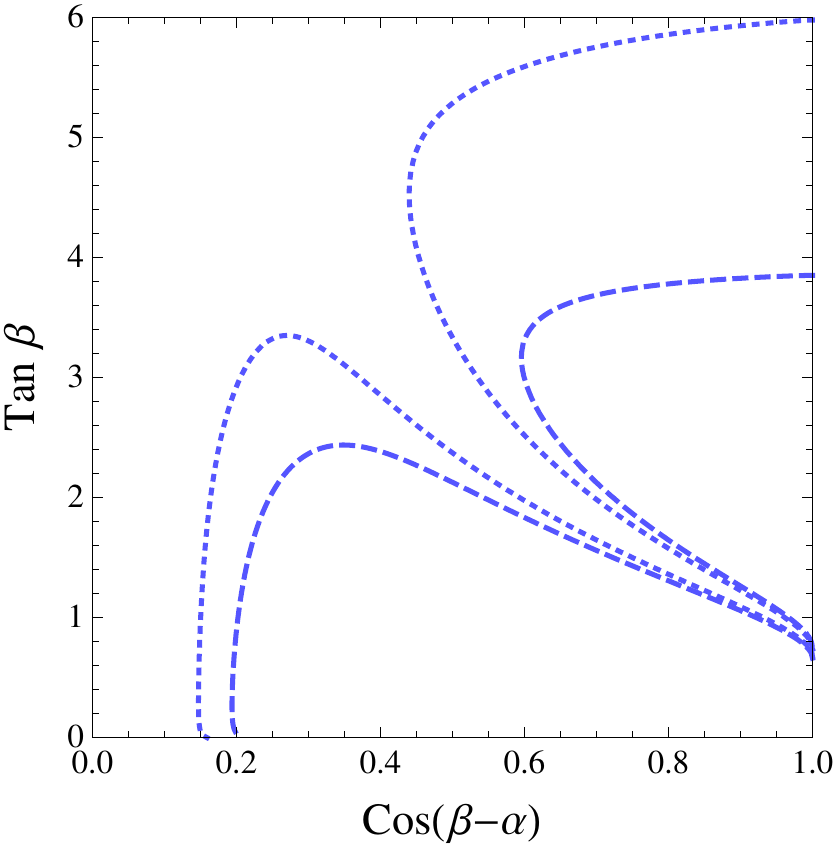}&
\includegraphics[width=.46\textwidth]{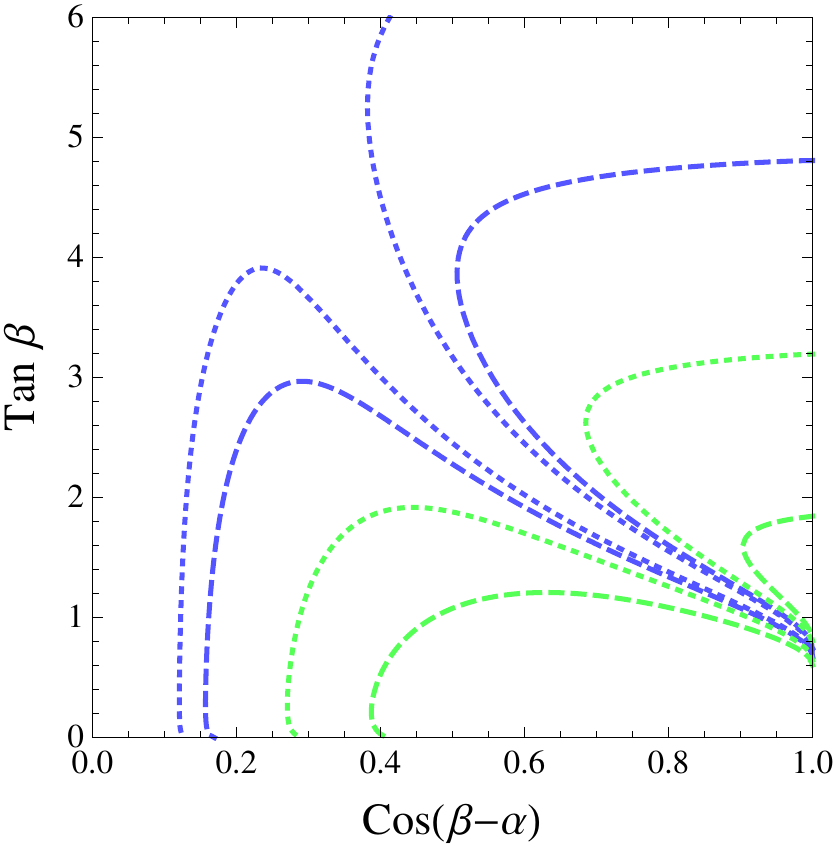}\\
\end{tabular}
\caption{In the left panel we show current exclusion bounds for $\sigma (gg \to A) \times \Br(A \to hZ \to \ell\ell b\bar b)$ based on the CMS data \cite{Khachatryan:2015lba}. In the right panel we show exclusion bounds for $\sigma (gg \to A) \times \Br(A \to hZ) \times \Br(h  \to b\bar b)$ (blue) and $\sigma (gg \to A) \times \Br(A \to hZ) \times \Br(h  \to \tau^+\tau^-)$ (green) based on ATLAS data \cite{Aad:2015wra}. In both plots we assume equal scalar masses, $M=500$ GeV (dotted) and $M=600$ GeV (dashed), and narrow-width approximation. The region below and to the right of the curves is excluded.}
\label{fig:AtohZtobbortautau} 
\end{figure}

The predictions of our model for both $\sigma (gg \to A) \times \Br(A \to hZ)$ and $\sigma(gg \to A )\times \Br(A \to hZ \to \ell^+\ell^- b\bar b)$ are presented in Figure \ref{fig:AtohZpred} in the left and right panels, respectively.  For the decay rate $\Gamma(h \to b \bar b)$, NLO corrections are sizable and therefore we include them in our analysis by setting
\begin{equation}
\Gamma(h \to b \bar b)= 0.57 \,\kappa_b^2\, \Gamma_{h}^{\mathrm{SM}}\,,
\end{equation}
where we use  $\Gamma_{h}
^{\mathrm{SM}}=4.07$ MeV \cite{Agashe:2014kda} and $\Br(Z \to \ell\ell)=6.729\%$ for $\ell^- = e^-, \mu^-$  \cite{Agashe:2014kda}.
In the left  panel of Figure \ref{fig:AtohZpred} we show the contours of $\sigma (gg \to A) \times \Br(A \to hZ)$ in picobarn for $8$ TeV 
proton-proton ($pp$) collisions in the $\cos (\beta-\alpha) - \tan\beta$ plane for $M=600$ GeV. The shape of the contours follows naturally from the fact that the branching ratio scales as $ \cos(\beta-\alpha)^2$, while the production cross section depends only on $\tan\beta$. This is no different than in a generic two Higgs doublet model \cite{Craig:2013hca, Khachatryan:2014jya}, but it is particularly relevant in our model, since it cannot live close to the decoupling limit, as discussed in Section \ref{seq:Higgsprod&decay}.
The experimental exclusion bounds from  \cite{Khachatryan:2014jya} constrain $\sigma (gg \to A) \times \Br(A \to hZ)$ considering a multi-lepton final state, but the study is only performed for pseudoscalars with masses up to $M_A <360$ GeV. 

In the right  panel of Figure \ref{fig:AtohZpred} we show the contours of  $\sigma (gg \to A) \times \Br(A \to hZ \to \ell^+\ell^- b\bar b)$ in  femtobarn  for $8$ TeV $pp$  collisions in the $\cos(\beta -\alpha)-\tan\beta$ plane and  $M=600$ GeV.  Two branches with suppressed values for $\sigma (gg \to A) \times \Br(A \to hZ \to \ell\ell b\bar b)$ appear. The first branch is the decoupling or alignment limit, where $g_{AhZ}$ vanishes. The second branch is given by the region for which the coupling of the light Higgs $h$ to bottom quarks becomes small.

\begin{figure}[t!]
\centering
\begin{tabular}{cc}
\quad  \small{$M_A=M_H=600$ GeV, $M_{H^+}=(400,600)$ GeV}\\
\includegraphics[width=.45\textwidth]{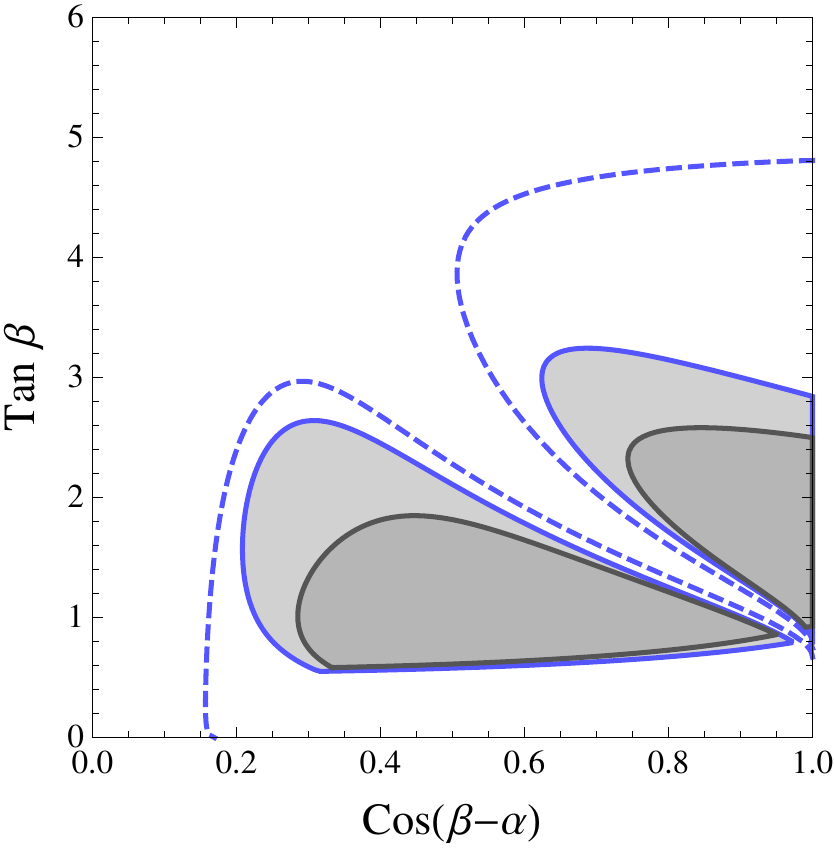}&
\includegraphics[width=.46\textwidth]{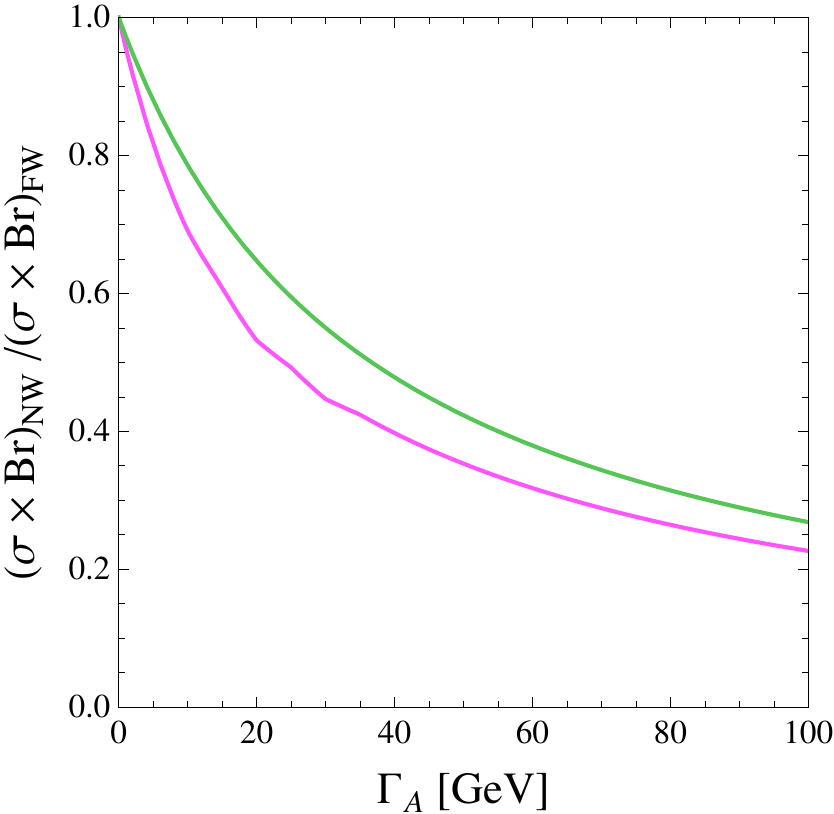}
\end{tabular}
\caption{In the left panel, we show the exclusion contour for $\sigma (gg \to A) \times \Br(A \to hZ) \times \Br(h  \to b\bar b)$ for $M=600$ GeV in the narrow width (NW) approximation (dashed blue) and taking into account finite width (FW) effects (solid blue). The black contour additionally shows mass splitting effects, assuming $M_A=M_H=600$ GeV and $M_{H^+}=400$ GeV. The shaded region inside each contour depicts the excluded area. The right panel shows the rescaling factor due to FW effects with respect to the NW approximation extrapolated from the CMS analysis \cite{Khachatryan:2015lba}, for $M_A=500\, (600)$ GeV in pink (green).} 
\label{fig:finitewidthandsplittings} 
\end{figure}

We consider the measurement of  $\sigma (gg \to A) \times \Br(A \to hZ) \times \Br(h \to b\bar b)$ by ATLAS \cite{Aad:2015wra} and 
the measurement of  $\sigma (gg \to A) \times \Br(A \to hZ \to \ell^+\ell^- b\bar b)$ by CMS \cite{Khachatryan:2015lba} with $\ell^- = e^-, \mu^-$. Both experiments give their bounds assuming narrow width approximation for the heavy scalar. In Figure \ref{fig:AtohZtobbortautau} we compare these bounds (blue curves) from both for equal masses of the heavy scalars with $M=500$ GeV (dotted) and $M=600$ GeV (dashed).
For both mass choices the ATLAS measurement gives a stronger bound. For $M=500$ GeV, substantial regions of the model parameter space are ruled out, however for $M=600$ GeV the model is considerably less constrained.
The right panel of Figure \ref{fig:AtohZtobbortautau} also shows the ATLAS bounds \cite{Aad:2015wra} of $\sigma (gg \to A) \times \Br(A \to hZ)$ with the light Higgs $h$  decaying further to tau leptons. The corresponding bounds are shown as green contours for $M=500$ GeV (dotted) and $M=600$ GeV (dashed).
These constraints are substantially weaker than the corresponding bounds for the $h\rightarrow b \bar b$ decay.

In the following we consider the impact of finite width effects on the previous bounds.  In the right panel of Figure \ref{fig:finitewidthandsplittings}, we show the rescaling factor for the cross section times branching ratio due to finite width effects, extrapolated frm the CMS analysis \cite{Khachatryan:2015lba}, for $M_A=500\, (600)$ GeV in pink (green). In the left panel of Figure \ref{fig:finitewidthandsplittings} we first show for comparison the exclusion bound from ATLAS data for $M=600$ GeV in the narrow width approximation. 
Under the assumption that the scaling effects for ATLAS and CMS are similar and assuming sensitivity up to a total width of $\Gamma_A \simeq 100$ GeV, we consider finite width effects for each point in the $\cos(\beta -\alpha) - \tan\beta$ plane and reinterpret the ATLAS results (solid, blue line in the left panel of Figure \ref{fig:finitewidthandsplittings}). Although finite width effects significantly weaken the exclusion bound, this channel remains the most promising discovery channel at the LHC run II.
The bound is further relaxed in the case of a mass splitting, $M_A \gg M_{H^+} (M_H)$, such that the decay channels $A \rightarrow H^+ W^- (H Z)$ open up. Our discussion in Section \ref{sec:EWPT} showed that such a mass splitting is only allowed between the pseudoscalar and the charged Higgs boson. We present the bound for $\sigma (gg \to A) \times \Br(A \to hZ) \times \Br(h  \to b\bar b)$ including finite width effects for $M_A=M_H=600$ GeV and $M_{H^+}=400$ GeV in the left panel of Figure \ref{fig:finitewidthandsplittings} (black).

\begin{figure}[t!]
\centering
\begin{tabular}{c}
\quad \small{$M=600$ GeV}\\
\includegraphics[width=.46\textwidth]{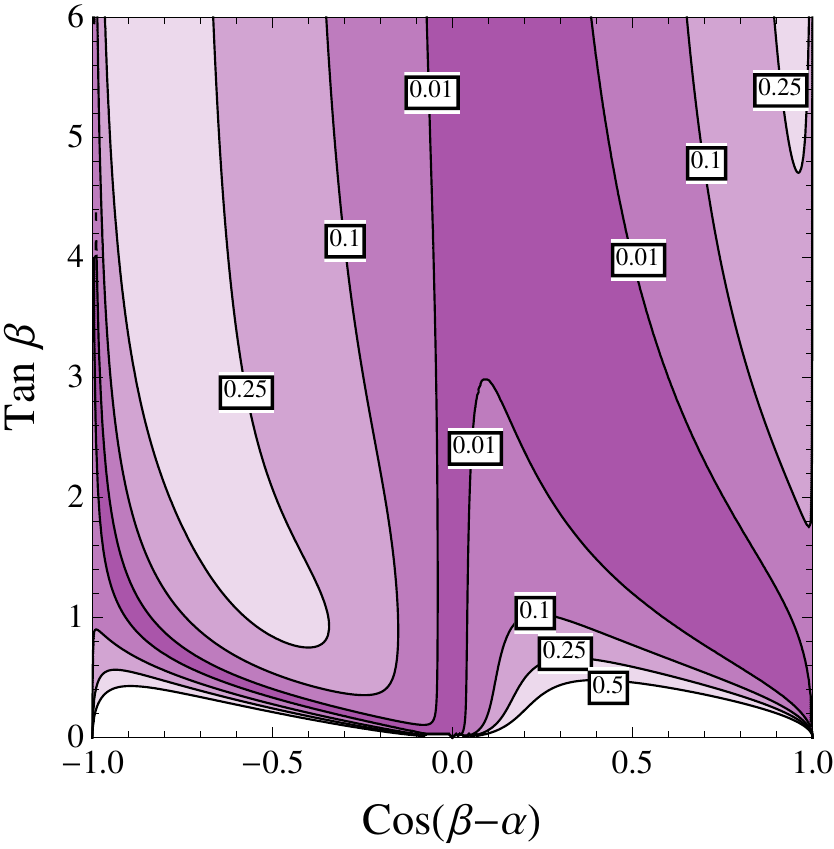}
\end{tabular}
\caption{Model predictions for the contours of $\sigma (gg \to H) \times \Br(H \to hh)$ in picobarn for $8$ TeV $pp$ collisions and heavy scalar masses $M=600$ GeV.
\label{fig:Hpred} }
\end{figure}

In the following we will consider the experimental bounds from searches for the neutral CP-even Higgs boson $H$. There are two channels of particular interest, the CP even scalar decaying into light Higgs bosons $H \rightarrow hh$ and the CP even scalar decaying to vector bosons $H \rightarrow VV$ with $V=W,Z$.

In Figure \ref{fig:Hpred}  we present predictions for $\sigma (gg \to H ) \times \Br(H \to hh)$ in picobarn for $8$ TeV $pp$ collisions in the $\cos (\beta -\alpha) - \tan\beta$ plane for $M=600$ GeV. From \eqref{eq:gHhh} we observe that the self coupling $g_{Hhh}$ is proportional to $\cos(\beta-\alpha)$ and has an explicit $M_A$ dependence. For $\cos (\beta-\alpha) \geq 0$ we observe two branches of contours with suppressed $\sigma  \times \Br$. The first branch approaches zero at $\cos (\beta-\alpha)=0$, and for the second branch both the coupling $g_{Hhh}$ and the production cross section become small.  
Predictions for $\sigma (gg \to H ) \times \Br(H \to hh)$ are comparable to the ones in a generic two Higgs doublet model of type II \cite{Craig:2013hca}.
Similar to the pseudoscalar case, the experimental exclusion bounds for $\sigma (gg \to H ) \times \Br(H \to hh)$   \cite{Khachatryan:2014jya} are only available up to $M_H < 360$ GeV. However for the CP even Higgs, the model predictions seem to be much below the present experimental sensitivity.

The most important search channel for the heavy CP even neutral Higgs boson $H$ is the inclusive production with subsequent decay of $H\rightarrow VV$ with $V=W,Z$. In our specific model, in particular, there is an interesting region of parameter space in which the vector boson fusion production is competitive with the gluon fusion production due to the behavior of $\kappa_{t}^H$. Normalized to the corresponding SM Higgs production and decay processes for a SM Higgs of mass $M_H$, we have for gluon fusion and vector boson fusion production processes, respectively,
\begin{align}
\frac{\sigma (gg \to H) \times \Br(H \to VV)}{(\sigma (gg \to H)  \times \Br(H \to VV))_{\text{SM}}}
&= ({\kappa_t^H})^2 \left(1+ \xi_b^H \frac{\kappa_b^H}{\kappa_t^H}\right)^2 
\left(\kappa_{V}^H\right)^2 \frac{\Gamma_H^{\text{SM}}}{\Gamma_H}\,,\label{eq:HVVggh}\\
\frac{\sigma (pp \to qqH) \times \Br(H \to VV)}{(\sigma (pp \to qqH)  \times \Br(H \to VV))_{\text{SM}}}
&= \left(\kappa_{V}^H\right)^4 \frac{\Gamma_H^{\text{SM}}}{\Gamma_H}\,,
\label{eq:HVVVBF}
\end{align}
where $\xi_b^H$ denotes the correction from a bottom quark in gluon fusion with respect to the leading top contribution. We take the SM total width $\Gamma_H^{\text{SM}}$ for a heavy Higgs of mass $M_H$ from the LHC Higgs Cross Section Working Group \cite{Dittmaier:2011ti,Dittmaier:2012vm,Heinemeyer:2013tqa}.

\begin{figure}[t!]
\centering
\begin{tabular}{cc}
\quad \small{$M=600$ GeV} &\quad \small{$M=600$ GeV}\\
\includegraphics[width=.46\textwidth]{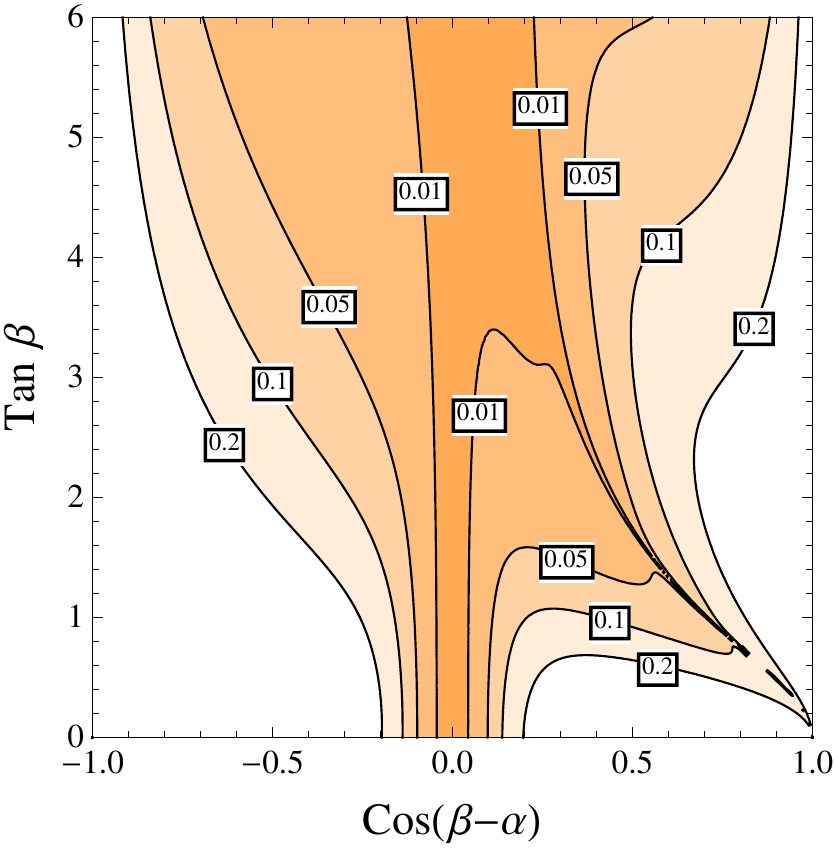}&
\includegraphics[width=.46\textwidth]{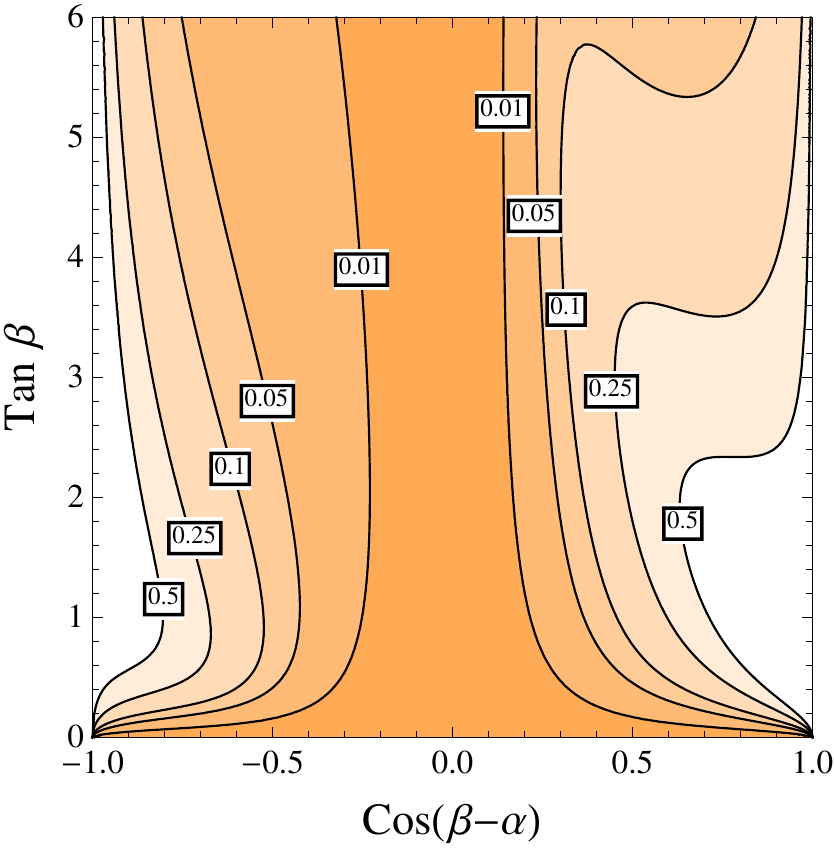}
\end{tabular}
\caption{Contours of $\sigma (pp \to H+X)  \times \Br(H \to VV)/(\sigma (pp \to H+X) \times \Br(H \to VV))_{\text{SM}}$ (right  panel) and $\sigma (pp \to qq H)  \times \Br(H \to VV)/(\sigma (pp \to qq H) \times \Br(H \to VV))_{\text{SM}}$ (left panel).  The heavy scalar masses are set to $M=600$ GeV.
\label{fig:HVVpred} }
\end{figure}

In Figure \ref{fig:HVVpred} we present theoretical predictions for contours of inclusive heavy neutral CP even Higgs production (left panel) and vector boson fusion production (right panel) with subsequent decay into $H \to VV$, using \eqref{eq:HVVggh} and \eqref{eq:HVVVBF}, for $M=M_A=M_H=M_{H^+}=600$ GeV. The vector boson fusion is governed by $\kappa_{V}^H$ and becomes strongly suppressed for small $\cos(\beta-\alpha)$. The gluon fusion production mode in \eqref{eq:HVVggh} is suppressed for small values of $\kappa_t^H$ or for small $\kappa_V^H$ and this effect shows in the inclusive production mode above. We observe that for small $\kappa_t^H$, both production cross sections become competitive. The theory prediction for these two observables differs from a two Higgs doublet model of type II only by the different scaling of the width $\Gamma_{H}$ and the contribution of the bottom quark to gluon fusion, which is small for $\tan\beta \sim \ord(1)$.

\begin{figure}[t!]
\centering
\begin{tabular}{cc}
\quad \small{$M=(500,600)$ GeV}\\
\includegraphics[width=.46\textwidth]{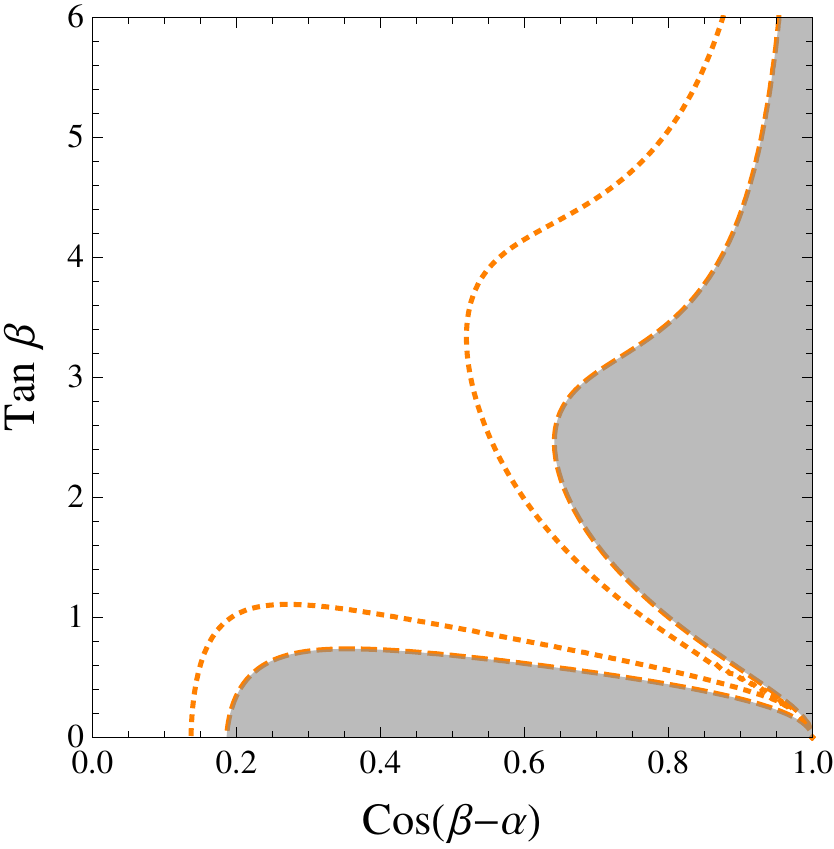}
\\
\end{tabular}
\caption{
Exclusion bounds for $\sigma (pp \to H +X)   \times \Br(H \to VV)/(\sigma (pp \to H+X) \times \Br(H \to VV))_{\text{SM}}$ of CMS \cite{Khachatryan:2015cwa} for  $M=500$ GeV (dotted) and $M=600$ GeV (dashed).
\label{fig:HtoVVexcl} }
\end{figure}

The CMS collaboration has reported updated results from an inclusive search for a heavy Higgs decaying into $W^+W^-$ and $ZZ$ in the range of $M_H= 145 - 1000$ GeV \cite{Khachatryan:2015cwa}. They consider both fully leptonic and semileptonic final states.  
In Figure \ref{fig:HtoVVexcl} we illustrate those bounds for $M=M_A=M_H=M_{H^+}$ with $M=500$ GeV (dotted) and $M=600$ GeV (solid). We observe that this search mode is competitive with the bounds obtained from the $A \to hZ$ channel. We note that for the neutral CP even Higgs analysis no finite width effects have been taken into account, although we expect sizable finite width effects in a large region of parameter space, compare the left panel of Figure \ref{fig:totalwidth} .

The CMS collaboration also performed an analysis for a heavy neutral Higgs boson decaying into $W^+W^-$ in vector boson fusion production channel in the mass range  $M_H= 110 - 600$ GeV \cite{CMS:2013yea}. The observed signal significance is close to the SM prediction for a Higgs of $M_H = 300 - 600$ GeV,  and hence from the right panel of Figure \ref{fig:HVVpred} it follows that there is no sensitivity to the preferred parameter region from this search.
\\

Searches for heavy charged Higgses have been performed by both ATLAS and CMS collaborations. In particular, they searched for production modes in association with a single top, $\sigma (bg \rightarrow H^-  t)$, or top and bottom quarks, $\sigma (gg \rightarrow H^-  t \bar b)$, with subsequent decays into third generation fermions: $H^- \rightarrow \bar t b$ and $H^- \rightarrow \tau \nu_\tau$ \cite{CMS:2014cdp,CMS:2014pea,Aad:2014kga}. The most recent limits are   
\begin{align}
 &\Br(H^- \to \tau \nu ) < \,0.153\, \mathrm{pb} - 0.026 \,\mathrm{pb}\qquad \,\text{for}\quad \, M_{H^+} = 300 - 600\, \mathrm{GeV}\,,\\
 &\Br(H^- \to t \bar b ) <\, 6\,\mathrm{pb} - 4\, \mathrm{pb}\qquad \qquad \qquad  \text{for}\,\quad M_{H^+} = 300 - 600 \,\mathrm{GeV}\,,
\end{align}
 assuming  $\Br(H^- \to \tau \nu )  =100\%$ and $\Br(H^- \to t\bar b )  =100\%$, respectively. These values are below the expected production cross section,  $\sigma(pp \rightarrow H^- t (b) ) = 70 \,\mathrm{fb} - 6 \,\mathrm{fb} $ for $M_{H^+} = 300\, \mathrm {GeV}- 600 \,\mathrm{GeV}$ and $\tan\beta \approx 2$ (lower values of the production cross section occur for $2<\tan\beta <6$) \cite{Flechl:2014wfa}. A heavy charged Higgs boson is therefore not constrained for the parameter region of interest, through current direct search limits. \\

For a heavy charged Higgs $M_{H^+}\approx 360 - 400$ GeV,$\cos (\beta - \alpha)\gtrsim 0.3 (0.2)$ and $\tan \beta = 2 (4)$, the decay channel $H^+ \rightarrow h W^+$ dominates over $H^+ \rightarrow t \bar b$. The branching ratio can become as large as $\Br (H^+ \rightarrow h W^+) \approx 85 \% $ for $\tan\beta = 2.5$, $\cos(\beta-\alpha) = 0.6$. For a lighter charged Higgs, this is slightly less pronounced and we find  $\Br (H^+ \rightarrow h W^+) \approx 70 \% $ for  $\tan\beta = 2.5$, $\cos(\beta-\alpha) = 0.6$ and $M_{H^+}= 400$ GeV.

 \section{Origin of the Effective Yukawa Couplings \label{seq:tevcompletion}}
In this section we present an example of the origin of the effective Yukawa couplings at the TeV scale for the bottom quark sector. Similar considerations can explain the generation of the other effective light quark Yukawa couplings in our model. A complete description of the UV completion is beyond the scope of this paper.  

A possible completion of the Froggatt Nielsen model may introduce new colored vector-like fermions or additional scalar doublets \cite{Calibbi:2012at}, whose masses determine 
the suppression scale $\Lambda$ in the expansion parameter \eqref{eq:defeps}.
Since in our model the flavor breaking scale is identified with the electroweak scale and the expansion parameter is fixed by the ratio of bottom and top quark masses $\varepsilon= m_b/m_t$,  the 
UV scale is constrained to be of the order of $\Lambda\sim 1$ TeV.

The relevant operators 
 that would provide a UV completion for the bottom Yukawa interactions are  
\begin{align} 
  \Lag_\mathrm{UV}&=y_{1} \,\overline{b}_L\, H_d\, \eta_R+ y_{2}\,\overline \eta_R\,  H_u\, \psi_L+ 
y_{3}\,\overline \psi_L\,H_d\, b_R\
 + M_\eta \bar \eta_L \eta_R + M_\psi \bar \psi_L \psi_R\,,
\end{align}
such that after integration of the heavy fields the effective Lagrangian is given by
\begin{equation}
\Lag_{EFT}=Y_b^\mathrm{eff}\, \overline b_L\, H_d\,b_R \,,
\end{equation}
with

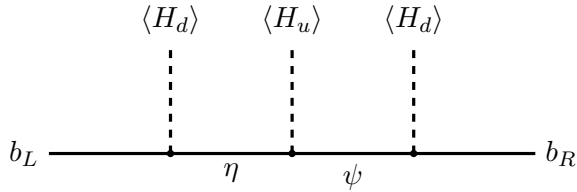
\begin{figure}[t!]
\centering
\begin{tikzpicture}[scale=.8, baseline=-0.5]
\draw[very thick] (0,0) to (2,0);
\node[left] at (0,0){$b_L$};
\draw[very thick,dashed] (2,0) to (2,1.8);
\fill[color=black!] (2,0) circle (0.6mm);
\node[above] at (2,1.8){$\langle H_d\rangle $} ;
\draw[very thick] (2,0) to (4,0);
\draw[very thick,dashed] (4,0) to (4,1.8);
\fill[color=black!] (4,0) circle (0.6mm);
\node[above] at (4,1.8){$\langle H_u\rangle $} ;
\draw[very thick] (4,0) to (6,0);
\node[below] at (3,0){$\eta$};
\node[below] at (5,0){$\psi$};
\draw[very thick,dashed] (6,0) to (6,1.8);
\draw[very thick] (6,0) to (8,0);
\fill[color=black!] (6,0) circle (0.6mm);
\node[above] at (6,1.8){$\langle H_d\rangle $} ;
\node[right] at (8,0){$b_R$};
 \end{tikzpicture}
 \caption{Diagram in the full theory, which generates the  Yukawa coupling between the Higgs and the bottom 
quarks after integrating out the heavy vector-like fermions $\psi$, $\eta$.} 
\label{fig:diagYuk}
\end{figure}

\begin{figure}[b!]
\centering
\rotatebox{90}{\phantom{MMMMMMMMMM}
$\overbar M$  [GeV]}\!\!\!\!
\includegraphics[scale=0.65]{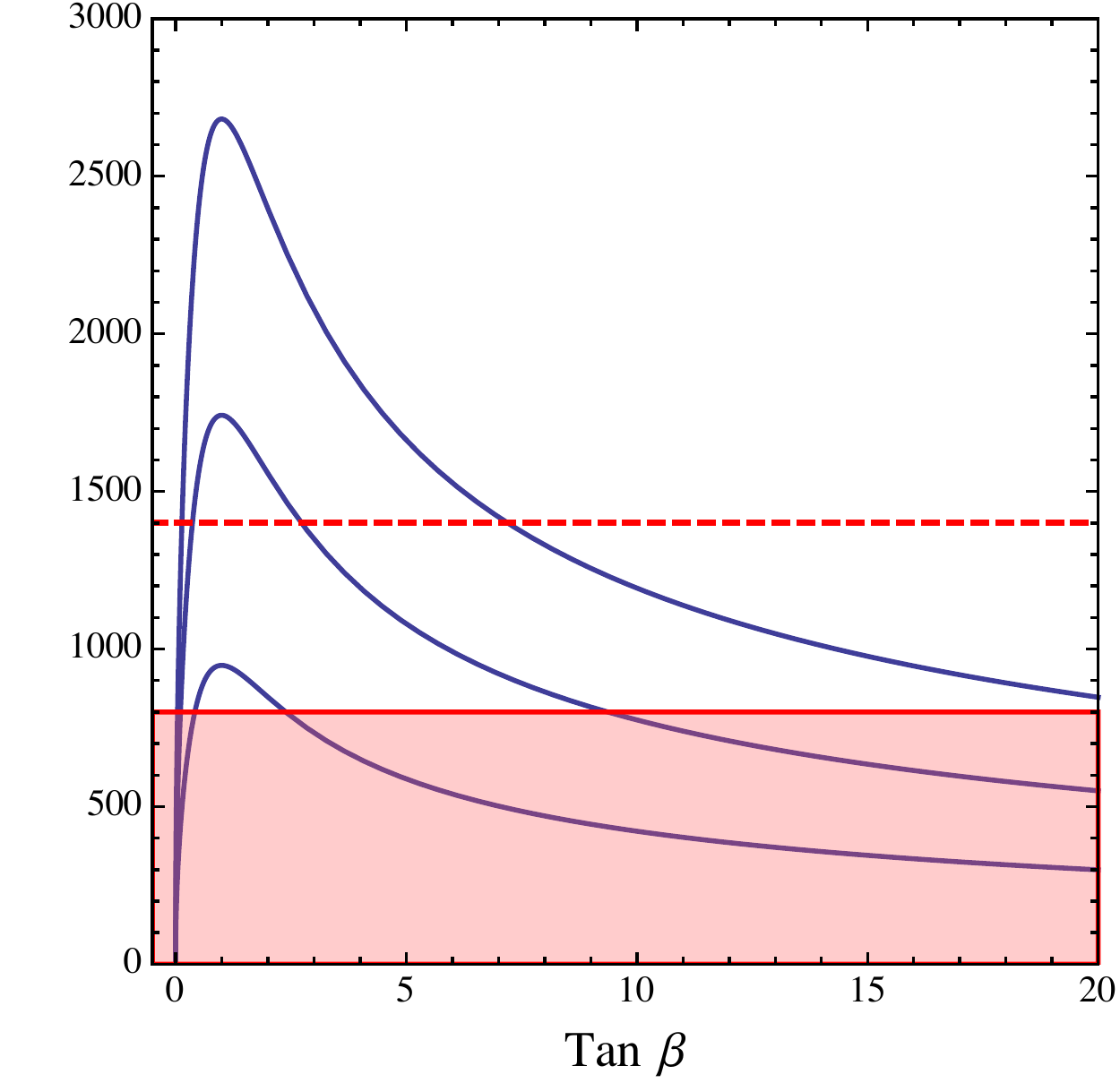}
\caption{\label{fig:UVmass} Masses of the new fermions in the UV completion depending on $\tan \beta$ and for three different values of the average Yukawa coupling $\bar y = 1,1.5,2$ (from bottom to top). Fermion masses below the solid red line are excluded by current LHC bounds, while the dashed red line shows the expected exclusion reach for the 14 TeV run of the LHC.}
\end{figure}

\begin{equation}\label{eq:YeffTeVcom}
Y_{b}^\mathrm{eff} \equiv \varepsilon\, y^d\,=
\frac{y_{1}\,y_{2}\,y_{3}}{M_{\eta}\,M_{\psi} }\,\frac{v_u v_d }{2}\,.
\end{equation}
The corresponding diagram is given in Figure \ref{fig:diagYuk} in which the new vector-like fermions carry quantum numbers 
\begin{equation}
\eta_{L,R}\sim (\mathbf{3}, \mathbf{1}, -1/3 , 2)\,, \qquad \psi_{L,R}\sim (\mathbf{3}, \mathbf{2}, 1/6 ,1)\,,
\end{equation}
 with respect to the groups $\big(SU(3)_C, SU(2)_L, U(1)_Y, U(1)_F\big)$.
 
From \eqref{eq:YeffTeVcom} is follows that for fixed $y_1=y_2=y_3=1$ 
and $y^d \in [0.5,1.5]$ this 
predicts the masses $M_\eta=M_\psi\approx \Lambda = 1 $ TeV. 
It is evident that slightly larger fundamental Yukawa couplings $y_1$, $y_2$ and $y_3$, 
allow for heavier vector-like fermions, while any $\tan\beta \gg 1$ or $\tan\beta \ll 1$ lead to lower mass scales. 
In the spirit of avoiding hierarchies between the fundamental couplings, including the top Yukawa coupling, we shall consider the ratio $y_i/y_t \sim \ord(1)$ with $i=1,2,3$. This constrains the masses of the vector-like fermions to be at most of the order of a few TeV. In particular, we define a generic mass $\overbar{M}\equiv \sqrt{M_\eta M_\psi}$, and an average fundamental Yukawa coupling $\bar y = (y_1 y_2 y_3)^{1/3}$, such that
\begin{align}
\overbar{M}^2 &= \frac{\bar y^3}{ y^d} \, \frac{v^2}{2\varepsilon} \, \frac{t_\beta}{1+t_\beta^2} \,.
\end{align}

In Figure \ref{fig:UVmass}, we show the expected masses of the new fermions 
for varying $\tan\beta$ and fixed $y^d=1$, for three different values of average Yukawa couplings $\bar y = 1,1.5,2$ (from bottom to top). These predictions for the expected masses remain the same for $\bar y = 1$ and change at most by 15\% (25\%) for $\bar y = 1.5\, (2)$ for the first generation quarks and at most $10\%$ (20\%) for second generation quarks.

The solid and dashed red lines in Figure \ref{fig:UVmass} indicate the present and projected experimental bounds from searches for pair produced heavy quarks at the LHC.  
These searches have been performed both by  ATLAS  
and CMS, and exclude vector resonances with masses of $600 - 800$ GeV  \cite{Aad:2014efa,CMS:2013una, Chatrchyan:2013uxa}, depending on the decay mode, with some channels already probing top partners $T$ up to $900$ GeV for  $\Br(T \rightarrow W^+ b) = 100 \%$ \cite{CMS:2014dka}. The next run of the LHC has a projected reach of $\overbar{M}\gtrsim 1.2\, (1.4)$ TeV for $20 \, \mathrm{fb}^{-1} (100 \,\mathrm{fb}^{-1})$ and $\Br(T \rightarrow W^+ b) =  50\%$ \cite{Matsedonskyi:2014mna}. 
Searches for heavy vector-like quarks in single production have also been considered \cite{Atre:2008iu,Atre:2011ae,VanOnsem:2014dba} and could be much more effective as a discovery channel for sufficiently heavy vector-like quarks compared to the previously mentioned pair production searches. However, the LHC reach in the single production channel depends very strongly on the model parameters which define the couplings of the heavy quarks to SM quarks. A reinterpretation of any of the existing LHC bounds in single heavy quark production channels would demand a detailed study of production cross sections and decay branching ratios for a specific UV completion.
Similarly, a specific UV completion would be subject to constraints from electroweak precision measurements as well as from flavor physics  \cite{Calibbi:2012at}. The latter have been addressed in some detail in the original Giudice-Lebedev paper \cite{Giudice:2008uua}.

\section{Benchmark Scenarios}\label{sec:bench}

The global fit to Higgs signal strength measurements discussed in Section \ref{sec:Higgscouplings} universally constrains the allowed parameter space to two branches within $\cos (\beta -\alpha) = 0.35 - 0.8$. Smaller values of $\cos (\beta -\alpha) <0.35$ are in principle possible for $\tan\beta >5$, but such large values of $\tan\beta$ are in tension with flavor observables. Electroweak precision observables and collider searches for the extra scalars provide additional constraints that narrow the parameter space significantly. In the following we examine the allowed window in the $\cos(\beta-\alpha)-\tan\beta$ plane and specify three benchmark points, that highlight the interesting features for the phenomenology of this model, and for which we give a detailed list of couplings, production cross sections and decay widths. \\

\begin{figure}[t!]
\begin{center}
\begin{tabular}{cc}
\quad \small{$M_A=M_H=600$ GeV, $M_{H^+}=450$ GeV}&\quad\,\, \small{$M_A=M_H=500$ GeV, $M_{H^+}=360$ GeV}\\
\includegraphics[width=.47\textwidth]{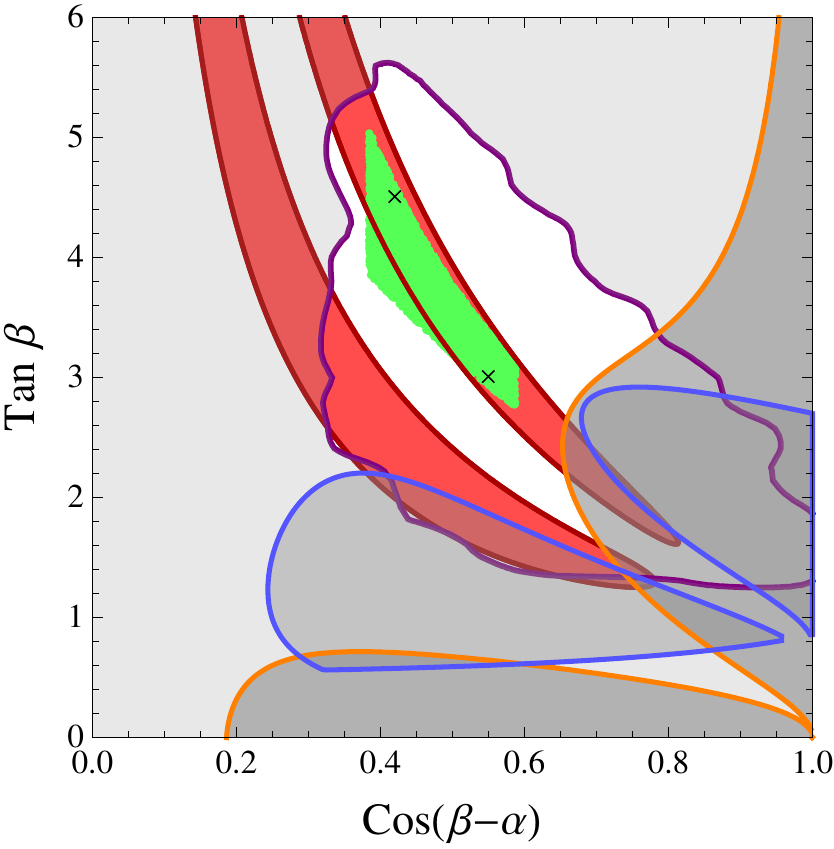}&\quad \includegraphics[width=.47\textwidth]{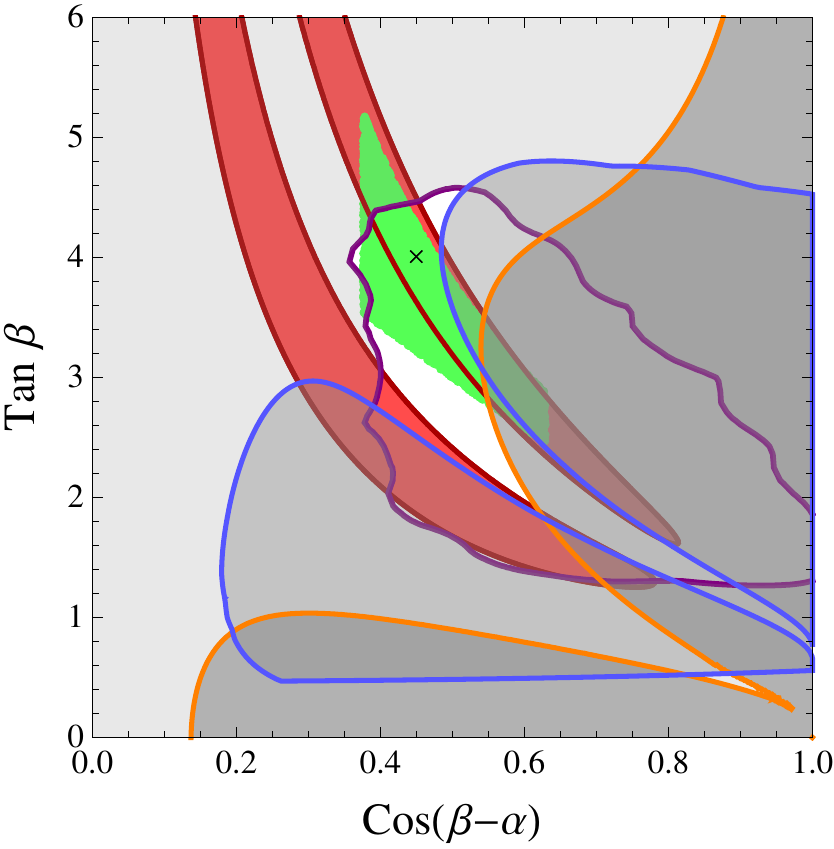}
\end{tabular}
\end{center}
\caption{\label{fig:bench} Summary plots showing constraints from flavor observables (purple contour) and direct collider searches for $A \rightarrow hZ \rightarrow \ell^+\ell^- b \bar b$ (blue contour) as well as $H \rightarrow W^+W^-/ZZ$ (orange contour), where the gray shaded area shows exclusion. The red shaded region is allowed at the $95\%$ CL from the global fit to Higgs signal strength measurements to ATLAS data. The green area highlights the allowed region from electroweak precision observables, perturbativity and unitarity constraints. The panels correspond to $M_A=M_H=600$ GeV and $M_{H^+}=450$ GeV (left), and $M_A=M_H=500$ GeV and $M_{H^+}=360$ GeV (right). The black crosses in both panels indicate the benchmark scenarios. }

\end{figure}

As a result of the discussion in Sections  \ref{sec:EWPT}, the combination of constraints from flavor physics, electroweak precision observables, unitarity and perturbativity lead to a constrained region of mass values for the additional Higgs bosons
$M_A \approx M_H\approx 500-600$ GeV and $M_{H^+} \approx 360-500$ GeV. Perturbativity puts an upper bound of $600$ GeV on the neutral Higgs masses and requires a splitting between the neutral and charged Higgs masses of $M_{A,H} - M_{H^+}\gtrsim 100 $ GeV. 
Electroweak precision measurements exclude the left branch of the global fit to Higgs coupling measurements for values of $\tan\beta \lesssim 4.5$. In addition, tree-level contributions to meson-antimeson mixing put an upper bound of $\tan\beta \lesssim 5$, while loop contributions from charged Higgs exchange result in a lower bound $\tan\beta \gtrsim 1.5$. Collider searches for the two heavy neutral Higgs bosons further constrain the allowed parameter space and probe the right branch of the global Higgs fit for $\cos (\beta-\alpha) = \mathcal{O}( 0.5)$ and $\tan \beta \lesssim 3$. As a result, there is a specific window of allowed masses as well as values of $\cos(\beta-\alpha)$ and $\tan\beta$, which translates into a precise prediction for searches for the extra scalars and constrain the possible deviations in the SM Higgs couplings. In Figure \ref{fig:bench}, we illustrate this window by showing the $95\%$ CL region of the global fit to ATLAS
Higgs signal strengths measurements (red shaded area), the region preferred by electroweak precision constraints (shaded green) and the bound induced from flavor constraints (solid purple contour), as shown in Figure \ref{fig:10p}. Further, we superimpose the bounds derived from the ATLAS and CMS measurements of $\sigma (gg \rightarrow A)\times \Br(A\rightarrow hZ \rightarrow b \bar b \ell^+ \ell^- )$ (solid blue) and $\sigma (pp\rightarrow H+X)\times \Br(H\rightarrow  VV )$ (solid orange). In the left panel we assume scalar masses of $M_H=M_A = 600 $ GeV, $M_{H^+}=450$ GeV, and in the right panel $M_H=M_A = 500 $ GeV, $M_{H^+}=360$ GeV. The gray shaded area is excluded, the overlap of the light green and red regions is allowed.

\begin{table}
\begin{center}
    \begin{tabular}    {l}          
\textbf{Benchmark 1} $:M_A=M_H= 600$ GeV, $M_{H^+}= 450$\,GeV\,,\\
\phantom{\textbf{Benchmark}} \textbf{1a}\,  $\cos(\beta-\alpha)= 0.55 \,,\quad \tan\beta = 3,$\\
\phantom{\textbf{Benchmark}} \textbf{1b}\,  $\cos(\beta-\alpha)= 0.42\,,\quad \tan\beta = 4.5,$\\[3mm]

Light Higgs Couplings:\\[3mm]

 \textbf{1a}\,  $\kappa_t= 1.02 \,,\quad \kappa_V= 0.84 \,, \quad \kappa_b = \kappa_\tau = -0.61  \,, \quad \kappa_c=1.22 \,, \quad \kappa_s = -0.41$\\
  \textbf{1b}\,  $\kappa_t= 1.00 \,,\quad \kappa_V= 0.91 \,, \quad \kappa_b = \kappa_\tau = -0.96  \,, \quad \kappa_c=1.02 \,, \quad \kappa_s = -0.95$\\[5pt]
 
 Higgs Signal Strength:\\[3mm]
 
\,\begin{tabular}{l  | c|c|c|c  }
  \textbf{1a}&    $\mu_V$ & $\mu_\gamma$ &  $\mu_b$&$\mu_c$      \\\hline
      $\sigma_{gg\rightarrow h}$&                                
      $1.38$&$1.21 $&$0.74$&$2.95$\\\hline
      $ \sigma_{t\bar t \rightarrow h}$&$1.33$ &$1.17$ &$0.71$&$2.84$\\\hline
      $\sigma_{VBF}, \sigma_{VH}$       & $0.89$&$0.78 $       &$0.48$ &$1.91$\\                               \end{tabular}\,\quad 
 \,\begin{tabular}{l  | c|c|c|c  }
  \textbf{1b}&    $\mu_V$ & $\mu_\gamma$ &  $\mu_b$&$\mu_c$      \\\hline
      $\sigma_{gg\rightarrow h}$&                                
      $0.96$&$0.91 $&$1.09$&$1.22$\\\hline
            $ \sigma_{t\bar t \rightarrow h}$ &$0.90$ &$0.85$ &$1.02$&$1.14$\\\hline
      $\sigma_{VBF}, \sigma_{VH}$       & $0.74$&$0.70 $       &$0.84$ &$0.94$\\                               \end{tabular}
      
\\\\[-2mm]
 Heavy Scalar Production Cross Sections for 1a (1b):\\[3mm]
      
\textbf{8 TeV:}  \,\, $ \sigma(gg \rightarrow A) = 78 (36) $ fb\,, \quad $ \sigma(gg \rightarrow H) =  32(21)$ fb\,, \quad\\
 \quad \qquad \quad \,\,$ \sigma(pp \rightarrow H^- t (b)) = 9 (4)$ fb\,,\\
\textbf{14 TeV:}  $ \sigma(gg \rightarrow A) = 361 (157) $ fb\,, \quad $ \sigma(gg \rightarrow H) =  166(97)$ fb\,, \quad\\
\quad \qquad \quad \,\,$ \sigma(pp \rightarrow H^- t (b)) = 63 (25)$ fb\,,\\[3mm]

      Heavy Scalar Decay Modes:\\[5pt]
     
               \begin{tabular}{c|   c|c   }
             $A$& \multicolumn{2}{c}{$\Gamma_i/\Gamma_A$}\\\hline
            & \textbf{1a}& \textbf{1b}\\\hline
              $Zh$  &$70.2 \%$ & $62\%$ \\
              $W^-H^+$& $14.4\%$& $21.8\%$\\
              $b\bar{b}$& $1.6\% $& $5.2\%$ \\
              $t\bar{t}$& $12.9\% $&$8.7\%$\\
            $\tau^+\tau^-  $ & $0.2\% $&$0.7\%$\\
            $t \bar{c}$ & $0.4\% $&$1.1\%$
              \end{tabular}
            \quad
          
                  \begin{tabular}{c|   c |c }
             $H$& \multicolumn{2}{c}{$\Gamma_i/\Gamma_H$}\\\hline
            & \textbf{1a}& \textbf{1b}\\\hline
              $WW$  &$52.9\%$  &$43\%$ \\
              $ZZ  $& $25.6\% $ &$20.9\%$\\
              $hh$& $9.2\%$&$16.9\%$\\
               $W^-H^+$& $6.8\%$& $11.2\%$\\
            $t\bar t$ & $3.9\%$&$3.5\%$
              \end{tabular}
              
             \quad
               \begin{tabular}{c|   c|c  }
                $H^+$& \multicolumn{2}{c}{$\Gamma_i/\Gamma_{H^+}$}\\\hline
            & \textbf{1a}& \textbf{1b}\\\hline
              $hW$  &$78.7\%$&$81.5\%$  \\
              $t \bar b  $& $21.2\% $&$18.2\%$ \\
              $\tau \nu$& $0.048 \%$&$0.33\%$\\
              \end{tabular}\\\\

Total Width for 1a (1b):\\[5pt]
    
$\Gamma_h= 2.22 \,(3.71)$ MeV\,,\quad $\Gamma_A=24.6\,(16.3)$ GeV\,,\quad  $\Gamma_H = 36.4 \,(26.1)$ GeV\,,
\\ $\Gamma_{H^+}=10.2 \,(5.8)$ GeV\,. \\

   \end{tabular}
   \caption{\label{tab:B1}    Values for the Higgs signal strength, heavy scalar production cross sections for the dominant channels at the LHC, partial and total widths for the benchmarks \textbf{1a} and \textbf{1b}.  }
   \end{center}
   \end{table}

   \begin{table}[t!]
   \begin{center}
   \begin{tabular}    {l}         

\textbf{Benchmark 2} $:M_A=M_H= 500$ GeV, $M_{H^+}= 360$\,GeV\,,\\
\phantom{\textbf{Benchmark} \textbf{1a}}\,  $\cos(\beta-\alpha)= 0.45 \,,\quad \tan\beta = 4,$\\[3mm]

Light Higgs Couplings:\\[3mm]

  \textbf{1b}\,  $\kappa_t= 1.01 \,,\quad \kappa_V= 0.9 \,, \quad \kappa_b = \kappa_\tau = -0.81  \,, \quad \kappa_c=1.1 \,, \quad \kappa_s = -0.71$\\[5pt]
 
 Higgs Signal Strength:\\[3mm]

\hspace{2cm}\,\begin{tabular}{l  | c|c|c|c  }
  \textbf{2}&    $\mu_V$ & $\mu_\gamma$ &  $\mu_b$&$\mu_c$      \\\hline
      $\sigma_{gg\rightarrow h}$&                                
      $1.15$&$1.07 $&$0.94$&$1.76$\\\hline
            $ \sigma_{t\bar t \rightarrow h}$&$1.09$&$1.02$&$0.90$&$1.67$\\\hline
      $\sigma_{VBF}, \sigma_{VH}$       & $0.86$&$0.80 $       &$0.71$ &$1.32$\\                               \end{tabular}\,\quad \\\\\\[-2mm]

 Heavy Scalar Production Cross Sections:\\[3mm]
      
 \textbf{8 TeV}:\,\, $ \sigma(gg \rightarrow A) = 130 $ fb\,, \quad $ \sigma(gg \rightarrow H) =  53$ fb\,, \quad $ \sigma(pp \rightarrow H^- t (b)) = 12$ fb\,,\\
  \textbf{14 TeV}: $ \sigma(gg \rightarrow A) = 546 $ fb\,, \quad $ \sigma(gg \rightarrow H) =  224$ fb\,, \quad $ \sigma(pp \rightarrow H^- t (b)) = 66$ fb\,,\\[3pt]

      Heavy Scalar Decay Modes:\\[5pt]

   \hspace{2cm}           
                    \begin{tabular}{c|   c   }
             $A$& $\Gamma_i/\Gamma_A$\\\hline
              $Zh$  &$56.6 \%$ \\
              $W^-H^+$& $23.3\%$\\
              $b\bar{b}$& $5.3\% $ \\
              $t\bar{t}$& $12.4\% $\\
            $\tau^+\tau^-  $ & $0.66\% $\\
            $t \bar c$ &$1.1\%$
              \end{tabular}
            \quad
          
                  \begin{tabular}{c|   c  }
             $H$& $\Gamma_i/\Gamma_H$\\\hline
                 $WW$  &$45.4\%$  \\
              $ZZ  $& $21.8\% $ \\
              $hh$& $11.5\%$\\
               $W^-H^+$& $12.6\%$\\
            $t\bar t$ & $3.65\%$
              \end{tabular}
              
             \quad
               \begin{tabular}{c|   c  }
                $H^+$& $\Gamma_i/\Gamma_{H^+}$\\\hline
               $hW$  &$71.8\%$  \\
              $t \bar b  $& $27.8\% $\\
              $\tau \nu$& $0.4 \%$\\
              \end{tabular}\\\\

Total Width:\\[5pt]
    
$\Gamma_h = 3$ MeV\,,\quad 
$\Gamma_A=10.7$ GeV\,,\quad     $\Gamma_H = 15.7 $ GeV\,,\quad  $\Gamma_{H^+}=3$ GeV\,.

   \end{tabular}
   \end{center}
   \caption{\label{tab:B2}    Values for the Higgs signal strength, heavy scalar production cross sections for the dominant channels at the LHC, partial and total widths for the benchmark \textbf{2}. }

   \end{table}

Comparing the two plots in Figure \ref{fig:bench}, bounds from flavor physics as well as collider constraints become weaker for larger masses.
The area in agreement with electroweak precision bounds is slightly larger for smaller mass splittings, but similar for the two examples given in Figure \ref{fig:bench}. The right boundary of the right branch of the global Higgs fit is close to the contour of $\kappa_b=-1$, for which the Higgs coupling to bottom quarks has the same size, but opposite sign compared to the SM one. The left boundary of the right branch is close to $\kappa_b=-0.5$. For all of the allowed parameter space, we can therefore infer $-1\lesssim\kappa_b\lesssim-0.5$. 
 In addition to the sign and the reduction of the Higgs bottom coupling, we find a universal enhancement of the Higgs charm couplings. Both can in principle be probed by 
measurements of exclusive radiative Higgs boson decays, which can test the sign of $\kappa_b$ at the 14 TeV LHC, and establish possible departures from the SM Higgs charm couplings of the order of $20\%$ at a prospective 100 TeV collider \cite{Bodwin:2013gca,Koenig:2015pha}. In the presence of a Higgs portal to dark matter, such corrections to the Higgs couplings to quarks could significantly modify the direct detection cross section \cite{Bishara:2015cha}.

In Table \ref{tab:B1} and \ref{tab:B2} we give the values for the Higgs couplings, signal strengths, production cross sections and branching ratios for three representative benchmark points indicated by black crosses in Figure \ref{fig:bench}.  Typical values of $\cos(\beta-\alpha)\approx 0.4 -0.55$ and $\tan\beta \approx 3 -4.5$ are considered. In all cases, $\kappa_t \approx 1$, implying a gluon fusion production rate of order of the SM one. \\

Benchmarks \textbf{1a} and \textbf{1b} allow for larger values $M_{A,H} \approx 600$ GeV and a charged Higgs mass $M_{H^+}\approx 450$ GeV, close to the $2\sigma$ bound derived from the experimental $b \rightarrow s \gamma$ measurement in a type II two Higgs doublet model with $\tan\beta > 2$. 

In Benchmark \textbf{1a}, the tree-level gauge boson and down type fermion third generation couplings are suppressed by factors of order $20\%$ and $40\%$, respectively, while the Higgs coupling to charm is enhanced by about $20\%$. The sizable suppression of $\kappa_b$ yields a suppression of the branching ratio into gauge bosons and hence of the corresponding signal strength of those channels. The charm signal strength instead, is increased by a factor $\sim 2- 3$ (depending on the production mode) due to the combined effects of an enhancement in $\kappa_c$ and a suppression in $\kappa_b$ and $\kappa_V$. All other vector boson fusion and VH production channels are suppressed with respect to the SM, in particular the $h\rightarrow b \bar b$ search mode. 

In Benchmark \textbf{1b} all tree-level  fermion and gauge Higgs couplings are within less than $5 -10\%$ of the SM expectations, hence the signal strengths in gluon fusion production are also within $5 -10\%$ of the SM ones, with the exception of a $20\%$ enhancement in $\mu_c$. All vector boson fusion/VH production channels are suppressed with a maximal suppression of $25 - 30\%$ in the case of light Higgs decaying into gauge bosons.

Benchmark \textbf{2} allows for the smallest possible values of $M_{H^+} = 360$ GeV compatible with the $3\sigma$ bounds derived from the experimental $b \rightarrow s \gamma$ measurement in a type II 2HDM with $\tan\beta > 2$. 
Benchmark \textbf{2} has a similar tendency in the couplings of gauge bosons and fermions to the light Higgs boson and hence in the corresponding signal strengths as in Benchmark \textbf{1a}, but with percentual effects in the deviations from SM predictions that are a factor $2-3 $ smaller. In addition to improving signal strength measurements, the ongoing run of the LHC will probe these benchmarks by direct searches for the additional Higgs bosons. All three benchmark scenarios will be primarily tested by the search for $A\rightarrow Z h $ and $H \rightarrow VV$, that have branching ratios of $55\% - 75\%$, depending on the scenario.
 In the case of $H \rightarrow VV$, the inclusive and vector boson fusion production modes will play a complementary, relevant role. In addition to these discovery channels, other interesting search modes such as 
$A, H \rightarrow W^+ H^-$, $H \rightarrow hh$, $A \rightarrow t \bar {t}$, $H^+ \rightarrow h W^+$, and $H^+ \rightarrow t \bar{b} $
would yield additional valuable information about this model. The mass splitting between neutral and charged scalars give rise to an additional decay chain, that can potentially allow to discover the charged Higgs even for masses of $M_{H^+}\approx 360-400$ GeV, in particular for the subsequent decay of $H^+ \rightarrow W^+ h$. Although challenging due to the small branching ratio, a novel channel in these scenarios is $A \rightarrow t \bar c$.

Predictions for particular observables can be computed from the information provided in Table \ref{tab:B1} and Table \ref{tab:B2}. Finite width effects play a relevant role and in the case of $A\rightarrow h Z$ we have compiled them in the right panel in Figure \ref{fig:finitewidthandsplittings}. 

Finally, improved measurements of flavor observables, in particular in the neutral $B_d$ system could additionally constrain the parameter space significantly.

\section{Conclusion \label{seq:conclusions}}

In this article we propose an explanation to the hierarchies in fermion masses and mixings based on a Froggatt-Nielsen mechanism, in which two Higgs doublets play the role of the flavon. Therefore, the underlying 
flavor symmetry is broken at the electroweak scale. The flavor charges are fixed to reproduce the SM quark mass hierarchies and CKM mixing angles up to rescalings,  that have no effect on any physical quantity. As a result, this two Higgs doublet flavor model can be described by few effective parameters, the masses of the extra scalars $M_H$, $M_A$, $M_{H^+}$, $\cos(\beta-\alpha)$ and $\tan\beta$. This allows us to present our main findings in the $\cos(\beta-\alpha)-\tan\beta$ plane for fixed mass values, as shown in Figure \ref{fig:bench}. 

Modified interactions between the SM-like Higgs $h$ and quarks are characteristic for our two Higgs doublet flavor model, leading to strong constraints from Higgs signal strength measurements. 
The results of our Higgs global fit to ATLAS and CMS data constrain possible deviations of the couplings of the light Higgs to fermions and gauge bosons with respect to the SM ones, and select sizable values of $\cos(\beta-\alpha)\approx \mathcal{O}(0.5)$. This implies a suppression of the tree-level couplings of the Higgs to gauge bosons, which is proportional to $\sin(\beta-\alpha)$ as in any two Higgs doublet model and therefore a suppressed vector boson fusion production rate with respect to the SM. The alignment/decoupling limit $\cos(\beta-\alpha)=0$ is excluded for all values of $\tan \beta$, since in this limit  our model approaches the Babu-Nandi-Giudice-Lebedev model for which there is a factor of three enhancement for coupling of the light Higgs to bottom quarks. The Higgs global fit allows for two branches in the $\cos(\beta-\alpha)-\tan\beta$ plane (red shaded areas in Figure \ref{fig:bench}) with opposite sign of the bottom Yukawa coupling. However, other constraints end up 
singling out 
the branch with values of the SM normalized light Higgs-bottom Yukawa coupling between $-0.5$ and $ -1$. On this branch the light Higgs-top Yukawa coupling is close to its SM value, implying gluon fusion signal strengths of $\mathcal{O}(1)$. Furthermore, on this branch, the coupling of the light Higgs to charm quarks is universally enhanced by up to $30\%$,  leading to a possible enhancement of the Higgs to charm signal strength by a factor of three. Both the negative sign of the bottom Higgs coupling as well as the enhanced Higgs to charm signal strength can in principle be measured at a high luminosity/energy collider through exclusive Higgs decays with a final state photon, such as $h \rightarrow \Upsilon \gamma$ and $h \rightarrow J/\psi\, \gamma$. 

Flavor changing neutral currents arise at tree-level, mediated by the light Higgs as well as the extra neutral 
scalars. Remarkably, light Higgs FCNCs become automatically small for the branch of the global Higgs fit with negative light Higgs-bottom Yukawa coupling. While the masses of the extra neutral scalars are constrained to be larger than $500$ GeV in this region, we need a mild fine-tuning of $\mathcal{O}(10\%)$ in the Yukawa couplings in order not to exceed the strongest constraint from $B_d-\bar B_d$ mixing (shown as purple contour in Figure \ref{fig:bench}). These tree-level FCNCs result in an upper bound of $\tan\beta \lesssim 5.5$.
Moreover, contributions from box diagrams with charged Higgs exchange can 
compete with the tree-level diagrams for low $\tan\beta$ and exclude values of $\tan\beta \lesssim 1$. Thus the interplay of tree-level and loop contributions in flavor observables predicts  $5.5 \gtrsim \tan\beta  \gtrsim 1$. Interestingly, if we discard the explanation of the CKM angles by the two Higgs doublet flavor model, we find almost no constraints from flavor observables in the region preferred by the global Higgs fit. As in any two Higgs doublet model, charged Higgs exchanges also induce FCNCs through penguin diagrams, for example $b \rightarrow s \gamma$, which imply a lower bound on the charged Higgs mass of $360$ GeV for $\tan\beta \gtrsim 2$.

The two Higgs doublet flavor model offers exciting possibilities for direct collider searches for the additional Higgs bosons. Electroweak precision observables, perturbativity and unitarity constraints choose a preferred range of masses and mass splittings for the new heavy scalars. In particular, almost degenerate values for the CP-odd and CP-even Higgs boson masses and sizable splitting between the neutral and charged Higgs masses are strongly favoured. This opens the opportunity of new decay channels, $A \rightarrow H^+ W^-$ and $H \rightarrow H^+ W^-$ in addition to the regular decay channels $H \rightarrow W^+ W^-/ ZZ$, $A \rightarrow h Z$,  that are importantly enhanced in the $\cos (\beta-\alpha) \approx \mathcal{O}(0.5) $ region. The latter are the leading discovery modes for these scalars (present bounds are shown by blue and orange contours in Figure \ref{fig:bench}). Furthermore, the $\cos (\beta-\alpha)$ dependence of the $H \,W^+W^-$, $H ZZ$ couplings are of particular relevance because the 
vector boson fusion production mode can compensate for the suppression of the gluon fusion production mode of the CP even Higgs in the relevant regions of parameter space. Direct searches for a charged Higgs boson are not sensitive for masses compatible with the flavor constraints, however future searches via Higgs decay chains with the subsequent decay $H^+ \rightarrow W^+ h$ may be promising. The other possible decay of heavy Higgs bosons to the SM Higgs is in the channel $H \rightarrow hh$  with branching ratios of order $10\%$. 

The fact that the flavor symmetry is broken at the electroweak scale predicts a UV completion in the few TeV range, as well as a low value of $\tan \beta$ in agreement with flavor constraints. The necessity of new physics at the TeV scale provides an additional motivation for the search for new vector-like fermions at the run II of LHC.

We conclude, that in the two Higgs flavor model constraints from flavor observables, Higgs precision measurements, direct heavy Higgs searches, and precision electroweak observables, as well as unitarity and perturbativity restrictions on the theory, can be fulfilled simultaneously. We propose three benchmark scenarios in this region, that highlight different characteristics of the two Higgs doublet flavor model (black crosses in Figure \ref{fig:bench}). In Table \ref{tab:B1} and \ref{tab:B2} we provide all the relevant information to compute production cross sections and decay rates for these benchmark scenarios and test the two Higgs doublet flavor model at the run II of LHC.

\section*{Acknowledgments}

We thank Prateek Agrawal, Wolfgang Altmannshofer, Andrzej Buras, Thorsten Feldmann, Elisabetta Furlan, Joerg Jaeckel, Matthias Neubert, Tilman Plehn, Raoul R\"ontsch and Carlos Wagner for useful comments and discussions. We specially thank  Zhen Liu for very helpful comments about the Higgs boson phenomenology and Mikolaj Misiak for private discussions on the bound of charged Higgs masses from $\Br(B_s \to X_s \gamma)$ in two Higgs doublet models. KG was supported by the Deutsche Forschungsgemeinschaft (DFG), grant number GE 2541/1-1.  MB acknowledges the support of the Alexander von Humboldt 
Foundation. Fermilab is operated by Fermi Research Alliance, LLC under Contract No. DE-AC02-07CH11359 with the United States Department of Energy.

\begin{appendix}

\section{The Higgs Potential \label{app:Higgspotential}}

In this appendix we consider the scalar potential and related topics.

The fact that $H_uH_d$ carries a flavor charge strongly constrains the scalar potential. We need a (soft) source of flavor breaking in order to generate a $b$-term.  We consider this additional source of flavor
breaking to be irrelevant for the texture of the Yukawa couplings. 

The potential reads then 
\begin{align}\label{eq:pot}
 V(H_u,H_d)= &\,\mu_u^2\,H_u^\dagger\,H_u + \mu_d^2\,H_d^\dagger\,H_d- 
\left[b\,H_u\, H_d+h.c.\right]\\
&+\frac{\lambda_1}{2}\,(H_u^\dagger\,H_u)^2+\frac{\lambda_2}{2}(H_d^\dagger\,
H_d)^2
+\lambda_3\,(H_u^\dagger\,H_u)(H_d^\dagger\,H_d)+\lambda_4(H_u^\dagger\,
H_d)(H_d^\dagger\,H_u)\notag\,,
\end{align}
in which $H_uH_d \equiv H_u^T (i \sigma_2) H_d$.
Note that the potential is the same as in a generic CP-conserving two Higgs doublet model, see for example \cite{Gunion:2002zf,Carena:2013ooa}, whith $\lambda_5=\lambda_6=\lambda_7=0$.

In order to diagonalize the potential, we introduce the neutral scalar mass eigenstates, 
\begin{align}\label{eq:hmasses}
\begin{pmatrix}
  h\\
  H
 \end{pmatrix} = \begin{pmatrix}
  c_\alpha& -s_\alpha\\
  s_\alpha& c_\alpha  \end{pmatrix}\, \begin{pmatrix}
  \mathrm{Re}\, H_u^0\\
  \mathrm{Re}\, H_d^0
 \end{pmatrix}
\,,
\end{align}
with the mixing angles $c_\alpha=\cos\alpha$ and $s_\alpha=\sin\alpha$ as well as the pseudo-scalar mass 
eigenstates
\begin{align}\label{eq:Amasses}
 \begin{pmatrix}
  \pi^0\\
  A
 \end{pmatrix} = \begin{pmatrix}
  s_\beta&- c_\beta\\
  c_\beta& s_\beta
 \end{pmatrix}\,
  \begin{pmatrix}
  \mathrm{Im}\, H_u^0\\
  \mathrm{Im}\, H_d^0
 \end{pmatrix}\,,
\end{align}
and similarly for the charged mass eigenstates $H^\pm$, 
\begin{align}\label{eq:Hpmmasses}
 \begin{pmatrix}
  \pi^-\\
  H^-
 \end{pmatrix} = \begin{pmatrix}
  s_\beta&- c_\beta\\
  c_\beta& s_\beta
 \end{pmatrix}\,
  \begin{pmatrix}
  H_u^-\\
  H_d^{+\ast}
 \end{pmatrix}\,.
\end{align}
Performing these rotations the explicit formulas for masses of scalar fields can be obtained, see for 
details for example \cite{Gunion:2002zf,Carena:2013ooa}. 

Finally from the scalar potential we obtain all couplings between the scalars  \cite{Gunion:2002zf,Craig:2013hca,Carena:2013ooa}. In particular, relevant for our analysis are the coupling between the heavy scalar $H$ and the light Higgs $h$ given in equation \eqref{eq:gHhh},
and the coupling of the light Higgs $h$ to two charged Higgses $H^{\pm}$ 
\begin{equation}
g_{hH^+H^-} = \frac{1}{v} \left[(2 M_A^2 - 2 M_{H^\pm}^2-m_h^2) s_{\beta-\alpha}+ 2 (M_A^2 -m_h^2) \frac{c_{2\beta}c_{\beta-\alpha}}{s_{2\beta}}\right]\,.
\label{eq:ghHpHm}
\end{equation}

\section{Box Diagrams and Loop Functions}
 \label{app:Boxes}
In this appendix, we collect the contributions to the Wilson 
coefficients \eqref{eq:heffdf2} from box diagrams and the relevant loop functions \cite{Buras:1989ui, Buras:2001mb,Jung:2010ik}. For $K-\bar K$ mixing we have the following Wilson coefficients:
\begin{align}
C^{sd}_{1\,,\mathrm{box}}=\frac{G_F^2\,m_W^2}{16\pi^2}\, \left(
(\lambda^{t}_{sd})^2 \,C^t_\mathrm{1\,, box} + (\lambda^{c}_{sd})^2\,C^c_\mathrm{1\,,box}+ 2 \lambda^{t }_{sd}\,
\lambda^{c}_{sd}\, C^{ct}_\mathrm{1\,,box}\right)\,,
\end{align}
with $\lambda_t=V_{td} V_{ts}^\ast$, $\lambda_c=V_{cd} V_{cs}^\ast$ and
\begin{align}
C^t_\mathrm{1,box}=&\left(4 x_t+x_t^2\right)\,m_W^2\, D_2(m_t,m_W)-8\,x_t^2 m_W^4\,D_0(m_t,m_W)\notag\\
&+\frac{2x_t^2}{t_\beta^2}\left[m_W^2\,D_2(m_t,m_W,M_{H^\pm})-4\, m_W^4\,D_0(m_t,m_W,M_{H^\pm})\right]\notag\\
&+\frac{x_t^2}{t_\beta^4}\,m_W^2 \,D_2(m_t,M_{H^\pm})\,,\\[2pt]
C^c_\mathrm{1,box}=&\left(4 x_c+x_c^2\right)\,m_W^2\, D_2(m_c,m_W)-8\,x_c^2 m_W^4\,D_0(m_c,m_W)\,,
\\[2pt]
C^{ct}_\mathrm{1,box}=&\left(4 x_{ct}+x_{ct}^2\right)\,m_W^2\, D_2(m_c,m_t,m_W)-8\,x_{ct}
^2\,m_W^4\,D_0(m_c,m_t,m_W)\,\notag \\
&+\frac{2x_{ct}^2}{t_\beta^2}\left[m_W^2 D_2(m_c,m_t,m_W,M_{H^\pm})-4m_W^4\, D_0(m_c,m_t,m_W,M_{H^\pm})
\right]\notag\\
&+\frac{x_{ct}^2}{t_\beta^4}\,m_W^2\,D_2(m_cm_t,M_{H^\pm})\,,
\end{align}
in which $x_t=m_t^2/m_W^2, x_c=m_c^2/m_W^2 $ and $x_{ct}=m_c\,m_t/m_W^2$.
For $B_{d,s}-\bar B_{d,s}$ mixing, we have 
\begin{align}
C^{bq}_{1\,,\mathrm{box}}&=\frac{G_F^2\,m_W^2}{16\pi^2}\, 
(\lambda_{bq}^{t})^2 \,C_{1\,,\mathrm{box}}^t\,,\qquad \qquad\quad\,\,
\tilde C^{bq}_{1\,,\mathrm{box}}=\frac{G_F^2\,m_W^2}{16\pi^2}\, \frac{ m_q^2 m_b^2}{m_W^4}\, \, 
(\lambda_{bq}^{t})^2\tilde C_{1\, ,\mathrm{box}}\,,\\
C^{bq}_{2\,,\mathrm{box}}&=\frac{G_F^2\,m_W^2}{16\pi^2}\, \frac{4 m_q^2 }{m_W^2}\,\, 
(\lambda_{bq}^{t})^2 \,C_{2\, ,\mathrm{box}}\,,\qquad\quad\,
\tilde C^{bq}_{2\,,\mathrm{box}}=\frac{G_F^2\,m_W^2}{16\pi^2}\, \frac{4 m_b^2 }{m_W^2}\, \, 
(\lambda_{bq}^{t})^2\,C_{2\, ,\mathrm{box}}\,,\\
C^{bq}_{4\,,\mathrm{box}}&=\frac{G_F^2\,m_W^2}{16\pi^2}\, \frac{2 m_q m_b}{m_W^2}\, \, 
(\lambda_{bq}^{t})^2\,C_{4\, ,\mathrm{box}} \,,\qquad
C^{bq}_{5\,,\mathrm{box}}=\frac{G_F^2\,m_W^2}{16\pi^2}\, \frac{ m_q m_b}{m_W^2}\,\, 
(\lambda_{bq}^{t})^2 \,C_{5\, ,\mathrm{box}}\,,
\end{align}
with $\lambda_t=V_{tb}^\ast V_{tq}$ and $(q=s,d)$ and
\begin{align}
C_{1\, ,\mathrm{box}}&=m_W^2\left[t_\beta^4\frac{m_t^2}{M_{H^\pm}^2}\,D_2(m_t,M_{H^\pm})+t_\beta^2\, \bar 
D_2(m_t,m_W,M_{H^\pm}) \right]\,,\\
C_{2\, ,\mathrm{box}}&=x_t^2\,m_W^4\left[D_0(m_t,M_{H^\pm})+2D_0(m_t,m_W,M_{H^\pm})\right]\,,\\
C_{4\, ,\mathrm{box}}&= x_t^2\, m_W^4\Big[4\,D_0(m_t, M_{H^\pm})+\Big(t_\beta^2+\frac{1}{t_\beta^2}\Big)
\,D_2(m_t,m_W,M_{H^\pm})\Big]\notag\\
&\phantom{=}-4\,t_\beta^2\,x_t\,m_W^2\bar D_2 (m_t,m_W,M_{H^\pm})\,,\\
C_{5\, ,\mathrm{box}}&= x_t^2\, m_W^2\left[D_2(m_t, M_{H^\pm})+2D_2(m_t,m_W,M_{H^\pm})\right]\,.
\end{align}
The loop functions are given by
\begin{align}
D_0(m_1,m_2,M_1,M_2)=&\frac{m_2^2\log\big(\frac{m_2^2}{m_1^2}\big)}{(m_2^2-m_1^2)(m_2^2-M_1^2)
(m_2^2-M_2^2)}\notag\\
&+\frac{M_1^2\log\big(\frac{M_1^2}{m_1^2}\big)}{(M_1^2-m_1^2)(M_1^2-m_2^2)(M_1^2-M_2^2)}\notag\\
&+\frac{M_2^2\log\big(\frac{M_2^2}{m_1^2}\big)}{(M_2^2-m_1^2)(M_2^2-m_2^2)(M_2^2-M_1^2)}\,,\\
D_2(m_1,m_2,M_1,M_2)=&\frac{m_2^4\log\big(\frac{m_2^2}{m_1^2}\big)}{(m_2^2-m_1^2)(m_2^2-M_1^2)(m_2^2-
M_2^2)}\notag\\
&+\frac{M_1^4\log\big(\frac{M_1^2}{m_1^2}\big)}{(M_1^2-m_1^2)(M_1^2-m_2^2)(M_1^2-M_2^2)}\notag\\
&+\frac{M_2^4\log\big(\frac{M_2^2}{m_1^2}\big)}{(M_2^2-m_1^2)(M_2^2-m_2^2)(M_2^2-M_1^2)}\,,
\end{align}
and for $i=1,2$
\begin{align}
D_i(m_1,M_1,M_2)&=\lim_{m_2\rightarrow m_1}\, D_i(m_1,m_2,M_1,M_2)\,,\\
D_i(m_1,M_1)&=\lim_{M_2\rightarrow M_1}\, D_i(m_1,M_1,M_2)\,,\\
\bar D_2(m_1,M_1,M_2)&=D_2(m_1,M_1,M_2)-D_2(0,M_1,M_2)\,.
\end{align}
	
\section{Random Parameter Generation and Running}
\label{app:RPG&running}
In order to find sample parameter points, we generate random fundamental Yukawa couplings with $y_{ij}^{u,d}= |y_{ij}^{u,d}|
\,e^{i\phi^{u,d}_{ij}}$ and  $|y_{ij}^{u,d}|\in [0.5,1.5]$ and $\phi^{u,d}_{ij}\in [0,2 \pi]$. The effective Yukawa 
couplings \eqref{eq:effuse} have to reproduce the quark masses and Wolfenstein parameters in Table 
\ref{tab:bmws} in Appendix \ref{app:numbers}. To this end we perform a $\chi^2$ fit, with symmetrized $2 \sigma$ errors and  require $
\chi^2 < 10$. 

In order to obtain the new contributions to $K-\bar K$ and $B_{s,d}-\bar 
B_{s,d}$ mixing we compute the Wilson coefficients with these effective Yukawas, including the tree-level \eqref{eq:wilsons} and one loop Wilson coefficients given by Appendix \ref{app:Boxes}. These Wilson coefficients are at the high scale $\mu=m_h$ and $\mu=M_H, M_{H^\pm}, M_A$, respectively.
The next step is running the Wilson coefficients from the electroweak scale to the scale at which the matrix elements are 
evaluated, $\mu=2$ GeV in the case of $K-\bar K$ mixing and $\mu=m_b$ in the case of $B_{s,d}-\bar 
B_{s,d}$ mixing. For  $K-\bar K$ mixing  we use \cite{Bona:2007vi}
\begin{equation}\label{eq:running}
\langle \bar K | \mathcal{H}^{\Delta S=2}_\mathrm{eff} |  K\rangle_i= \sum_{j=1}^5 \sum_{r=1}^5\,
\left(b_j^{(r,i)}+ \eta\, c_j^{(r,i)}\right)\,\eta^{a_j}\, C_i^{sd}(\mu)\,B_i^K\, \langle \bar K | Q_r^{sd} | K\rangle\,,
\end{equation}
in which $\eta = \alpha_s(\mu)/\alpha_s(m_t)$, $a_i, b_j^{(r,i)}$ and $ c_j^{(r,i)}$ are "magic numbers" collected in 
\cite{Ciuchini:1998ix} and $B_K^i$ are the $B$ parameters collected in Table \ref{tab:KKmix}. The matrix elements are given by
\begin{align}\label{eq:matrixelements}
\langle \bar K | Q_1^{sd} | K\rangle&=\frac{1}{3}M_K\, f_k^2\,, \nonumber\\
\langle  \bar K | Q_r^{sd} | K\rangle&=N_r\,\left(\frac{M_K}{m_d+m_s}\right)^2\, M_K f_k^2\,,
\end{align}
with $N_r= (-5/24, 1/24, 1/4, 1/12)$ for $r=(2,3,4,5)$ and $M_K$ and $m_d+m_s$ again given in Table 
\ref{tab:KKmix}. For $B_{d,s}-\bar B_{d,s}$ mixing, \eqref{eq:running} and \eqref{eq:matrixelements} hold 
with the obvious replacements. The corresponding "magic numbers" can be found in \cite{Bona:2007vi}, and all other 
parameters in Table \ref{tab:BBmix}.

{\section{Numerical Input} \label{app:numbers}
In this Appendix we collect the numerical input used throughout this paper.

\begin{table}[!ht]
 {\begin{minipage}{6in}
\centering
\begin{tabular}{c p{2cm} c}
\begin{tabular}{cll}
\hline
\multicolumn{3}{l}{Quark Masses in GeV\hfill\cite{Xing:2007fb}}\\\hline
$m_u(m_Z)$& 0.00127&$\pm 0.0005$ \\
$m_d(m_Z)$& 0.0029&$\pm 0.0012$\\
$m_s(m_Z)$& 0.055&$\pm 0.016$\\
$m_c(m_Z)$& 0.619&$\pm 0.084$\\
$m_b(m_Z)$& 2.89&$\pm 0.09$\\
$m_t(m_Z)$& 171.7&$\pm 3.0$\\\hline
\end{tabular}
&&
 \raisebox{.3\height} {\begin{tabular}{cll}
\hline
\multicolumn{3}{l}{Wolfenstein Parameters \hfill \cite{Charles:2004jd} }\\\hline
$\lambda$& 0.22551& $\pm 0.00091$ \\
$A$& 0.813&$\pm 0.035$ \\
$\bar \eta$& 0.342&$\pm 0.024$ \\
$\bar \rho$& 0.149&$\pm 0.033$\\\hline
\end{tabular}}
\end{tabular}
\end{minipage}}
\caption{Quark masses and Wolfenstein parameters at the electroweak scale. Errors are symmetrized. \label{tab:bmws}}
\end{table}

\begin{table}[!ht]
\centering
\begin{tabular}{cl}
\hline
\multicolumn{2}{l}{Couplings and Boson Masses \, \hfill\cite{Agashe:2014kda,Mohr:2012tt}}\\\hline
$\alpha_e(m_Z)$&$ 1/127.9$ \\
$\alpha_s(m_Z)$& $0.1185\pm0.0006$ \\
$m_Z$& $91.1876\pm 0.0021$ GeV\\
$m_W$& $80.385\pm 0.015$ GeV\\
$G_F$& $1.16638 \cdot 10^{-5}$ GeV$^{-2}$
\\\hline
\end{tabular}
\caption{Gauge boson masses and couplings. }
\end{table}

\begin{table}[!ht]
\centering

\begin{tabular}{cl}
\multicolumn{2}{l}{Parameters in $K-\bar K$ mixing \hfill\cite{Bae:2013tca,Agashe:2014kda}}\\\hline
$B_1^K$& $0.537\pm 0.007\pm 0.024$ \\
$B_2^K$& $0.620\pm 0.004\pm 0.031$ \\
$B_3^K$& $0.433\pm 0.003\pm 0.019$ \\
$B_4^K$& $1.081\pm 0.006\pm 0.048$ \\
$B_5^K$& $0.853\pm 0.006 \pm 0.049$ \\
$f_k$&$156.2\pm0.2\pm 0.6\pm 0.3$ MeV\\
$M_K$&$497.614\pm0.024$ MeV\\
$m_s + m_d$&$ 135\pm18$ MeV
\\\hline
\end{tabular}
\caption{Parameters relevant for $K-\bar K$ mixing. \label{tab:KKmix} }
\end{table}

\begin{table}[!th]
 {\begin{minipage}{6in}
\centering
\begin{tabular}{c p{.3cm} c}
\begin{tabular}{cl}
\multicolumn{2}{l}{Parameters in $B_d-\bar B_d$  mixing \hfill\cite{Lubicz:2008am,Carrasco:2013zta,Dowdall:2013tga}}\\\hline
$B_1^{d}$& $0.85 \pm 0.03\pm 0.02$  \\
$B_2^d$& $0.73 \pm 0.03 \pm 0.01$\\
$B_3^d$& $0.88\pm 0.12\pm 0.06$ \\
$B_4^d$& $0.95\pm 0.04\pm 0.03$ \\
$B_5^d$& $1.47\pm 0.08\pm 0.09$ \\
$f_{B_d}$&$186\pm 4$ MeV\\
$M_{B_d}$&$5.27942\pm0.00012$ GeV\\
$m_b + m_d$&$ 4.29\pm0.09\pm 0.08 \pm 0.02$ GeV
\\\hline
\end{tabular}
&&
\begin{tabular}{cl}
\multicolumn{2}{l}{Parameters in $B_s-\bar B_s$ mixing \hfill\cite{Lubicz:2008am,Carrasco:2013zta,Dowdall:2013tga}}\\\hline
$B_1^s$& $0.86\pm0.03\pm 0.01$ \\
$B_2^s$& $0.73\pm 0.03\pm0.01$ \\
$B_3^s$& $0.89\pm0.10\pm 0.07$ \\
$B_4^s$& $0.93\pm 0.04\pm 0.01$ \\
$B_5^s$& $1.57\pm 0.07\pm0.08$ \\
$f_{B_s}$&$224\pm 5$ MeV\\
$M_{B_s}$&$5.36668\pm0.00024$ GeV\\
$ m_b+m_s$&$ 4.38 \pm0.09\pm 0.08 \pm 0.02$ GeV   
\\\hline
\end{tabular}
\end{tabular}
\end{minipage}}
\caption{Parameters relevant for $B_{d,s}-\bar B_{d,s}$  mixing. \label{tab:BBmix} }
\end{table}}

\clearpage

{\section{Branching Ratios} \label{app:branchingratios}
\begin{table}[h]
\begin{tabular}{cc}
\includegraphics[width =.45\textwidth]{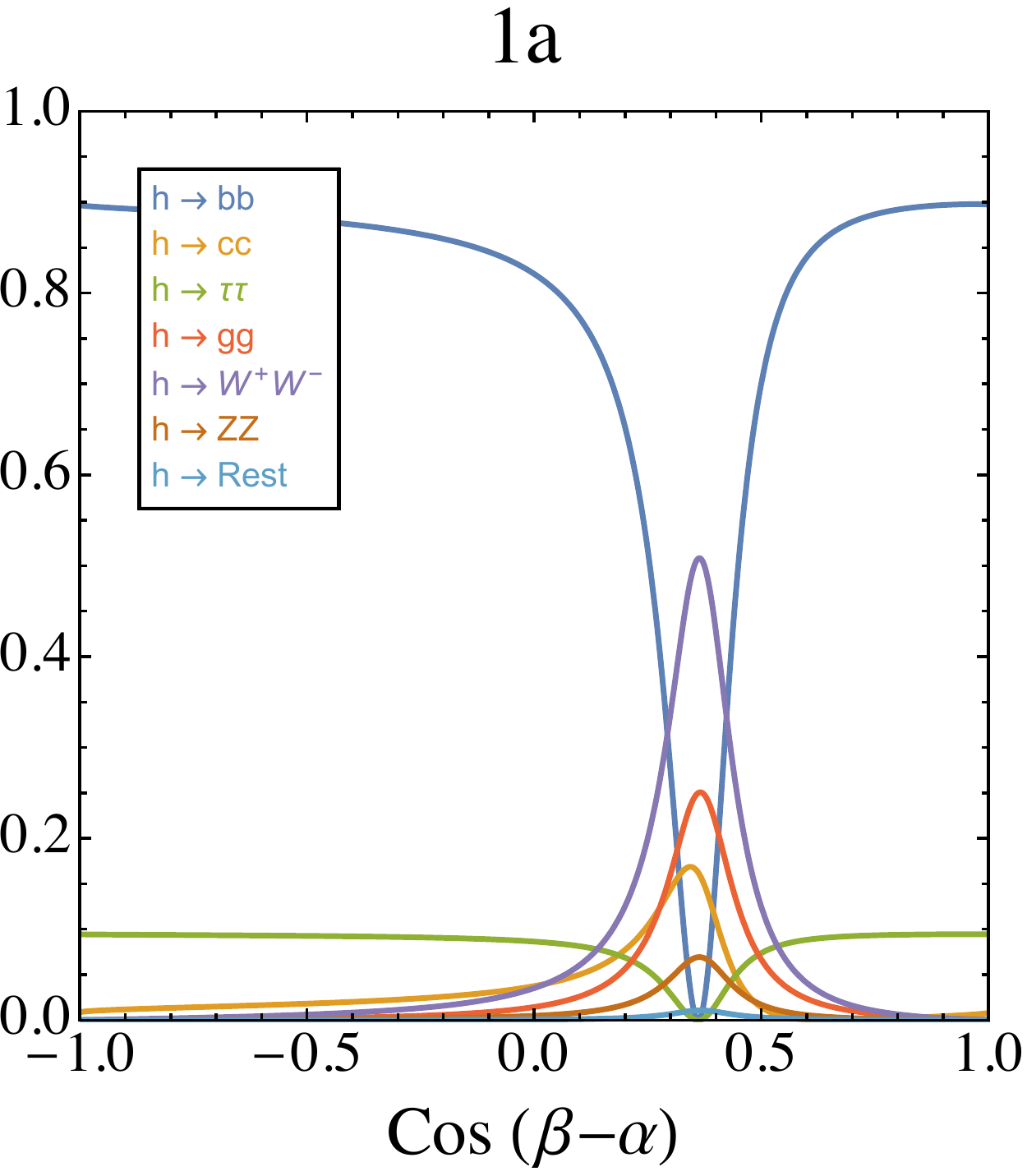}&\includegraphics[width =.45\textwidth]{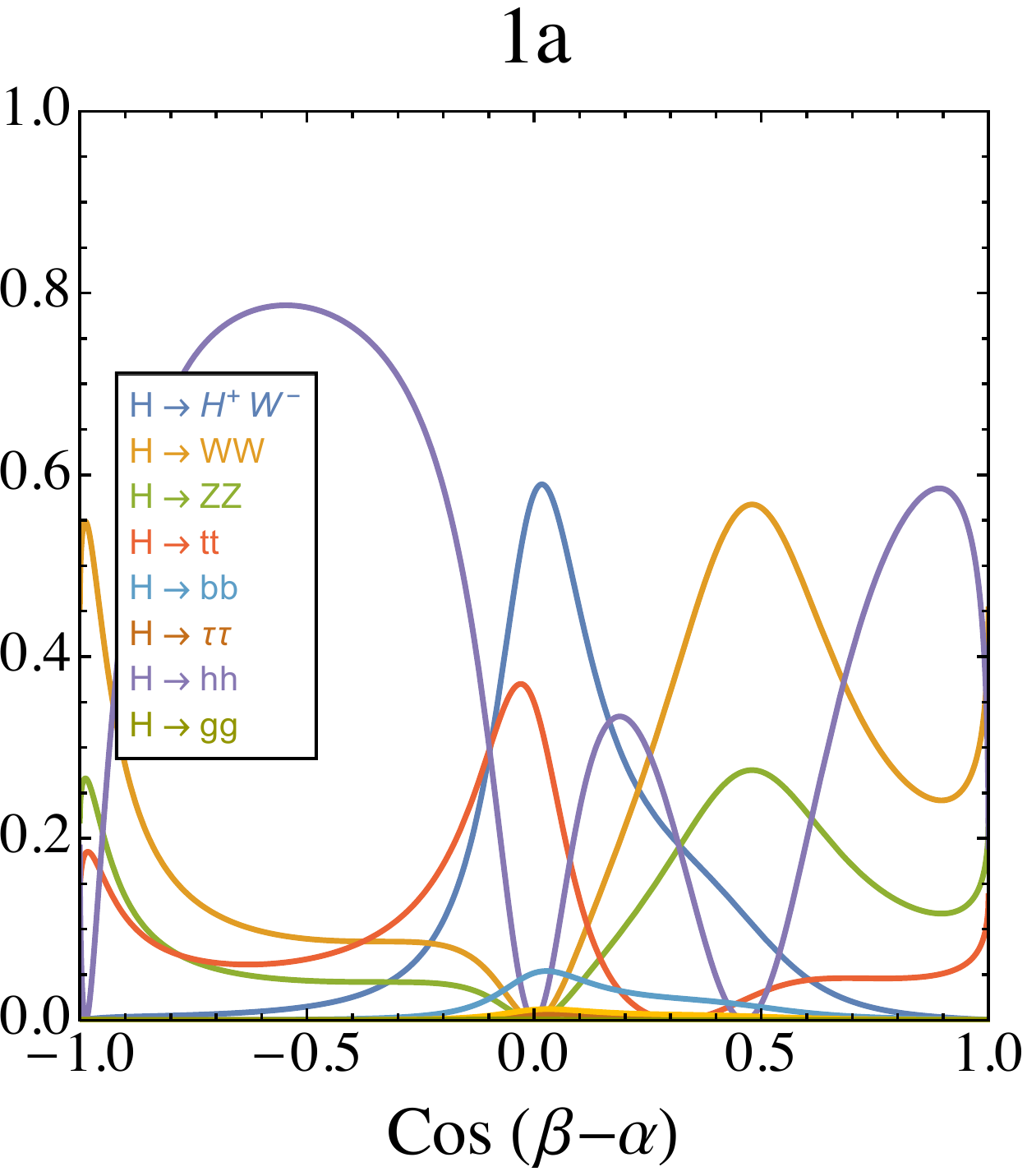}\\

\includegraphics[width =.45\textwidth]{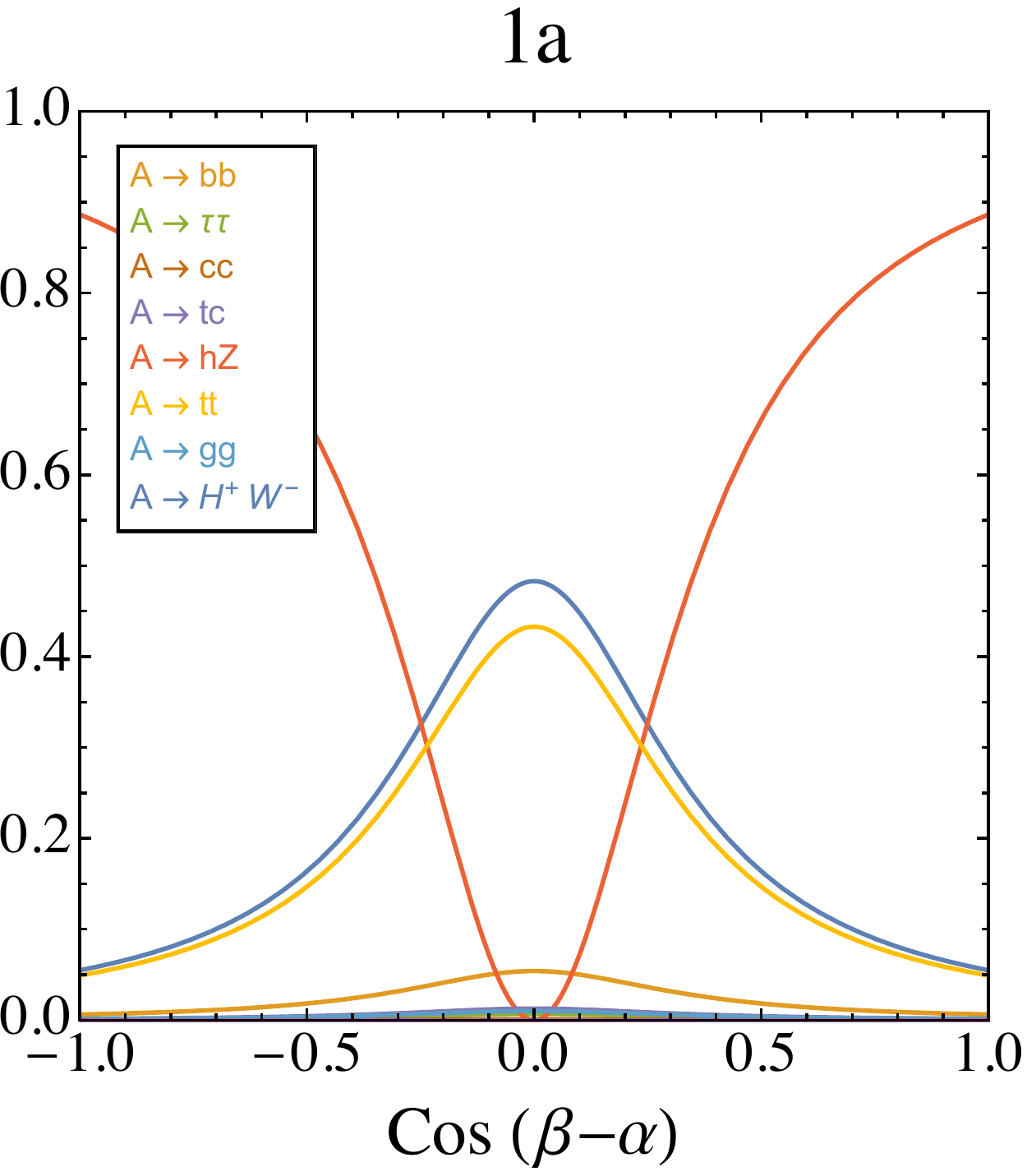}&\includegraphics[width =.45\textwidth]{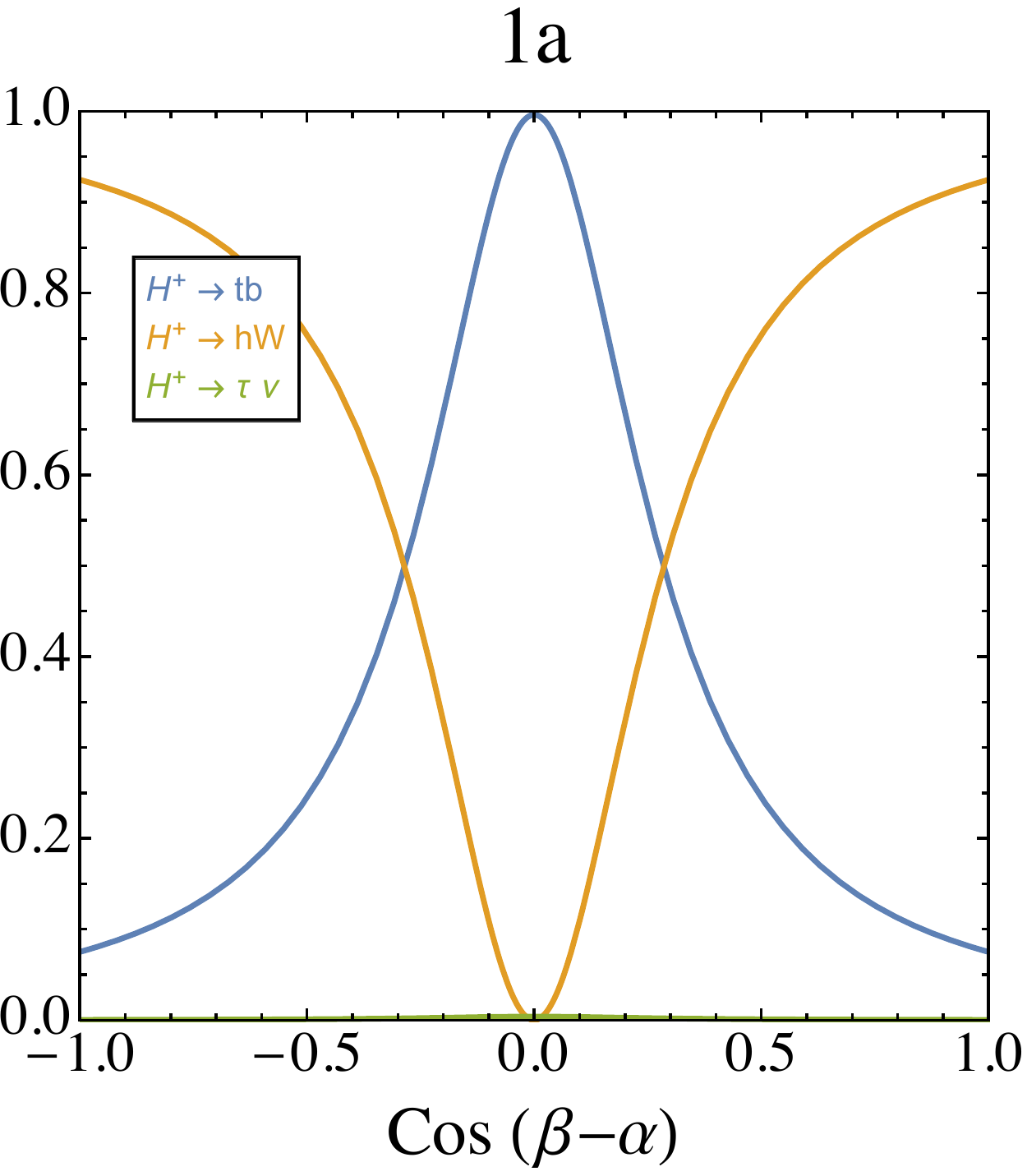}
\end{tabular}

\caption{Branching ratios as a function of $\cos(\beta-\alpha)$ for the light neutral scalar (upper left panel), heavy neutral scalar (upper right panel), pseudoscalar (lower left panel) and charged scalar (lower right panel) for the scalar masses and $\tan\beta$ of the benchmark scenario \textbf{1a} defined in Table \ref{tab:B1}. }
\end{table}

\begin{table}[]
\begin{tabular}{cc}
\includegraphics[width =.45\textwidth]{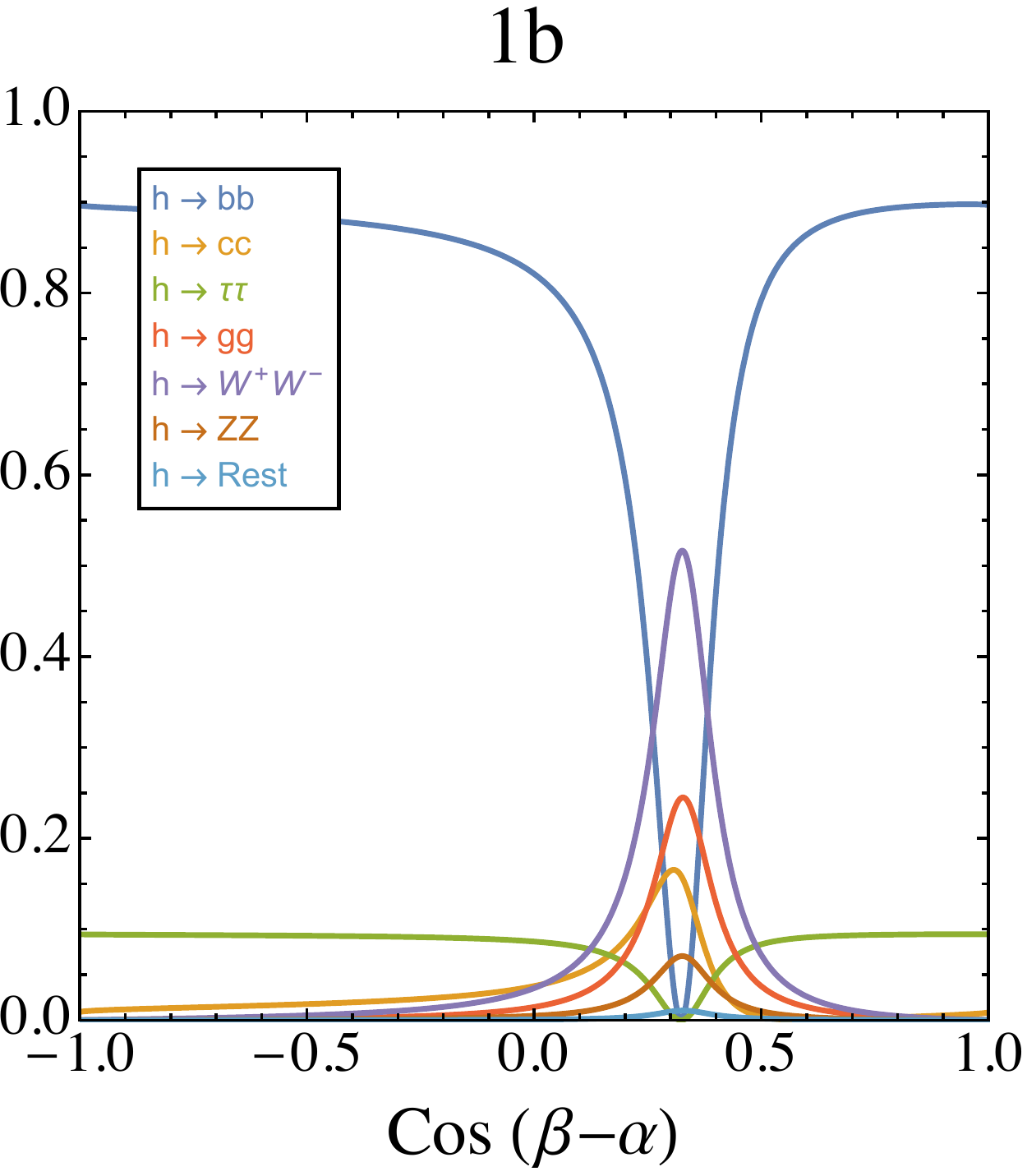}&\includegraphics[width =.45\textwidth]{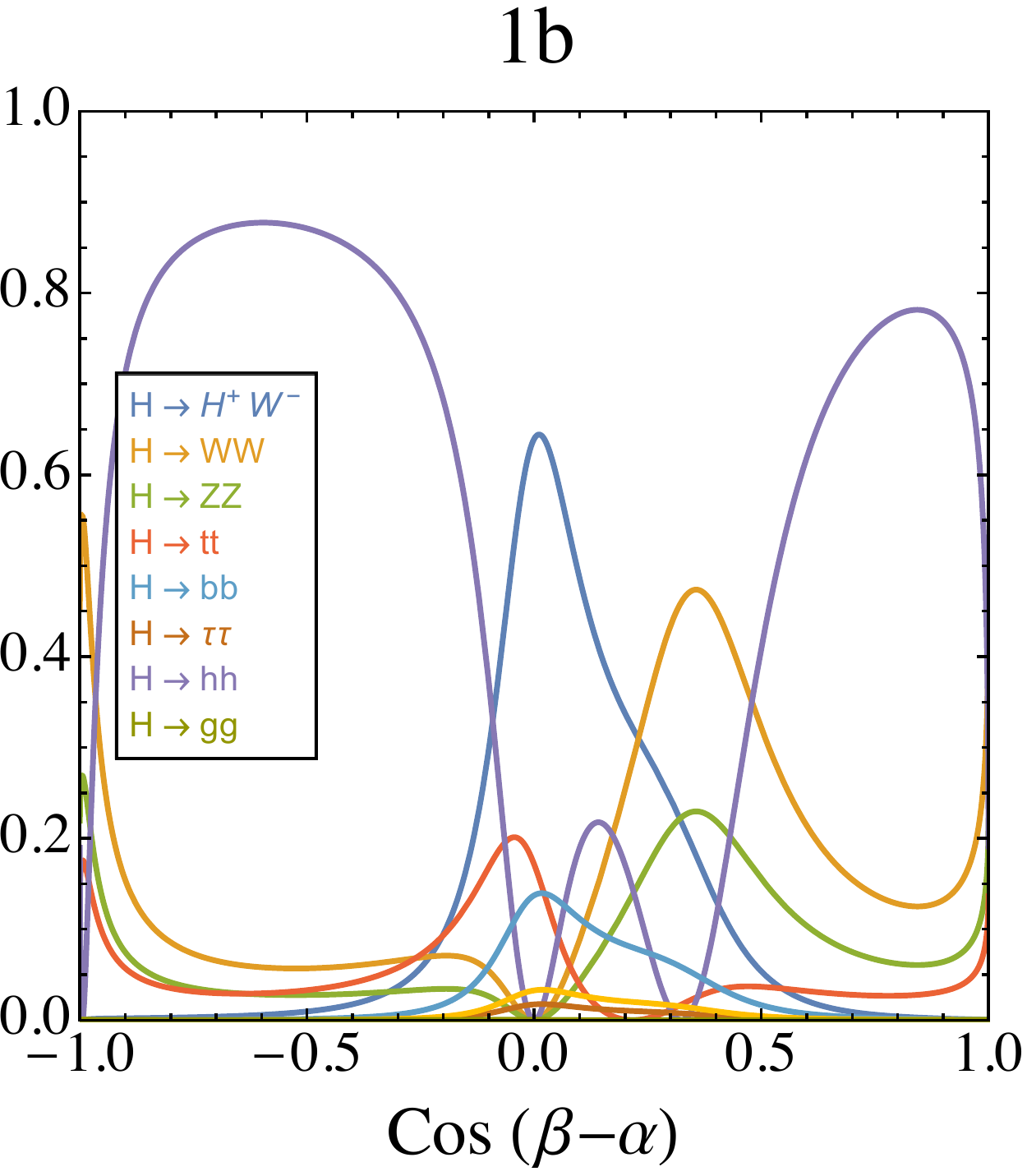}\\

\includegraphics[width =.45\textwidth]{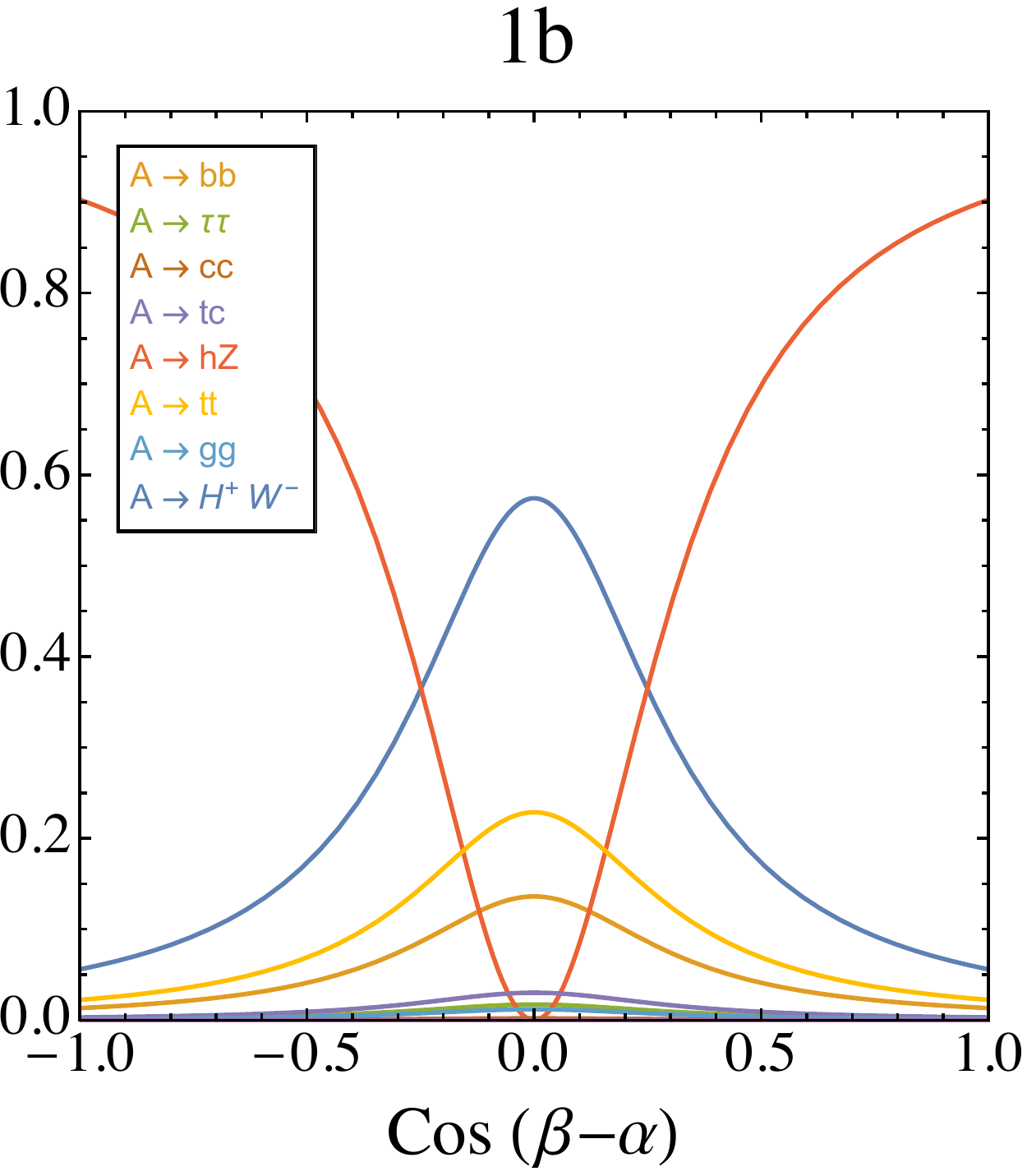}&\includegraphics[width =.45\textwidth]{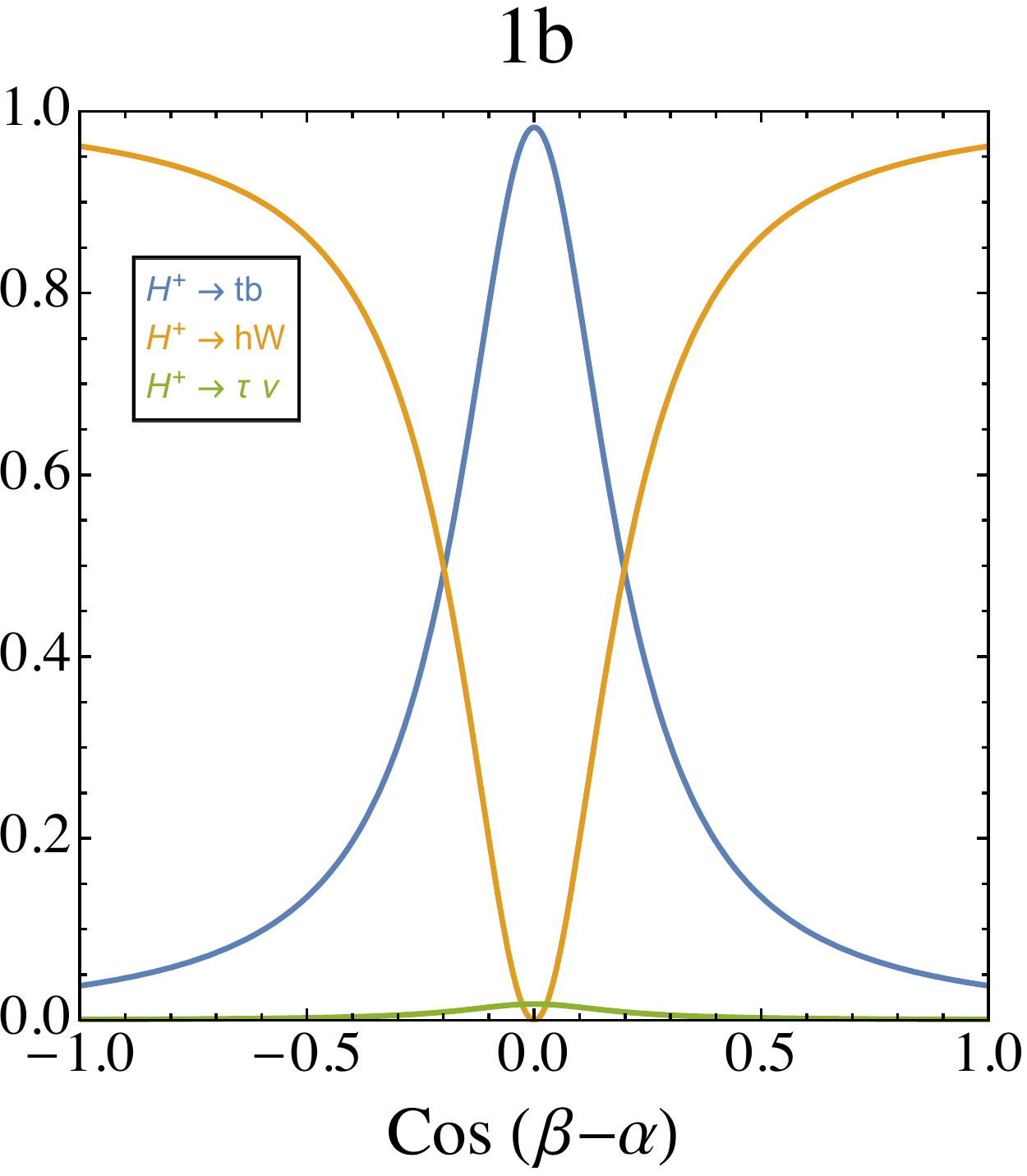}
\end{tabular}

\caption{Branching ratios as a function of $\cos(\beta-\alpha)$ for the light neutral scalar (upper left panel), heavy neutral scalar (upper right panel), pseudoscalar (lower left panel) and charged scalar (lower right panel) for the scalar masses and $\tan\beta$ of the benchmark scenario \textbf{1b} defined in Table \ref{tab:B1}. }
\end{table}

\begin{table}[h]
\begin{tabular}{cc}
\includegraphics[width =.45\textwidth]{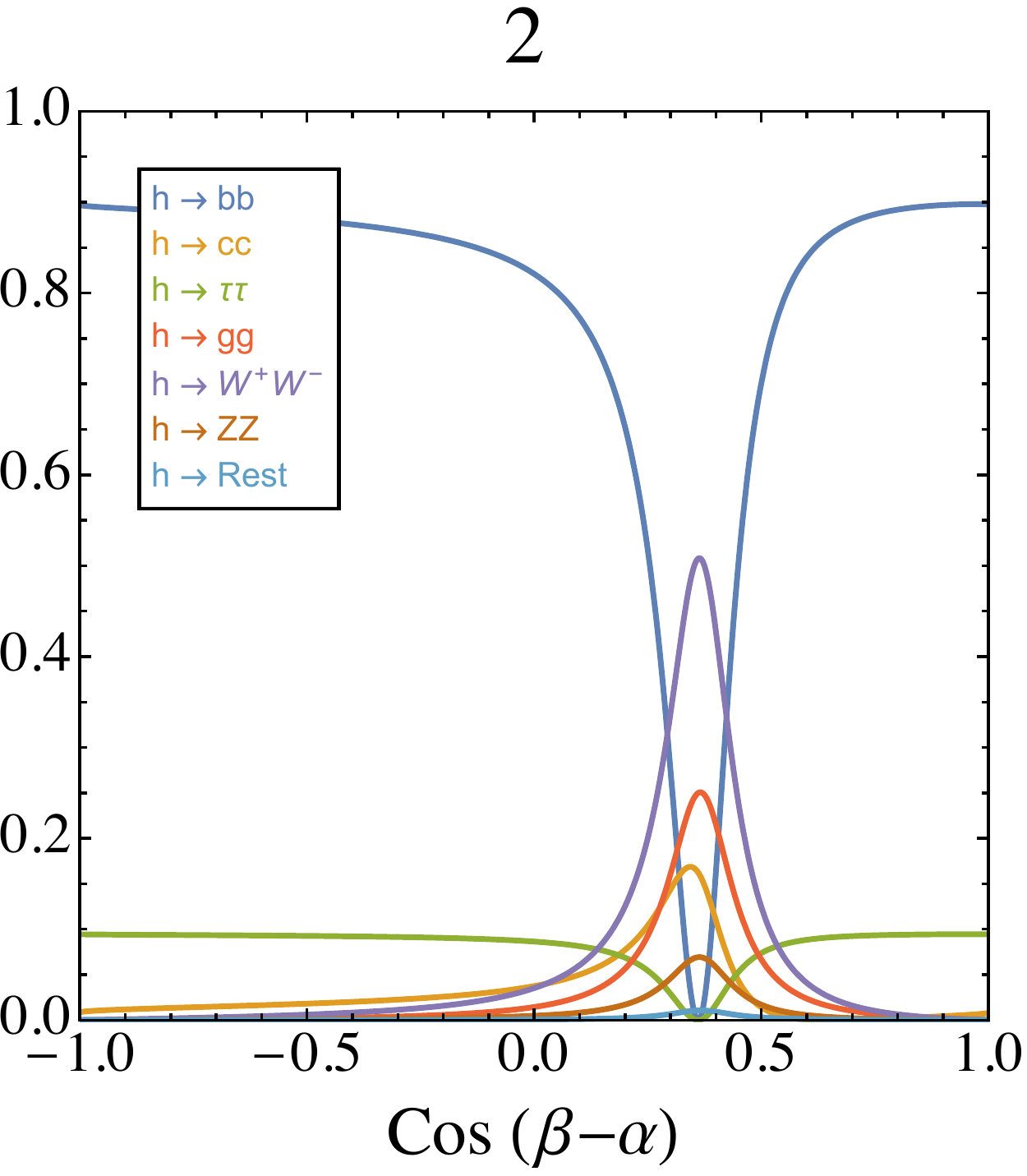}&\includegraphics[width =.45\textwidth]{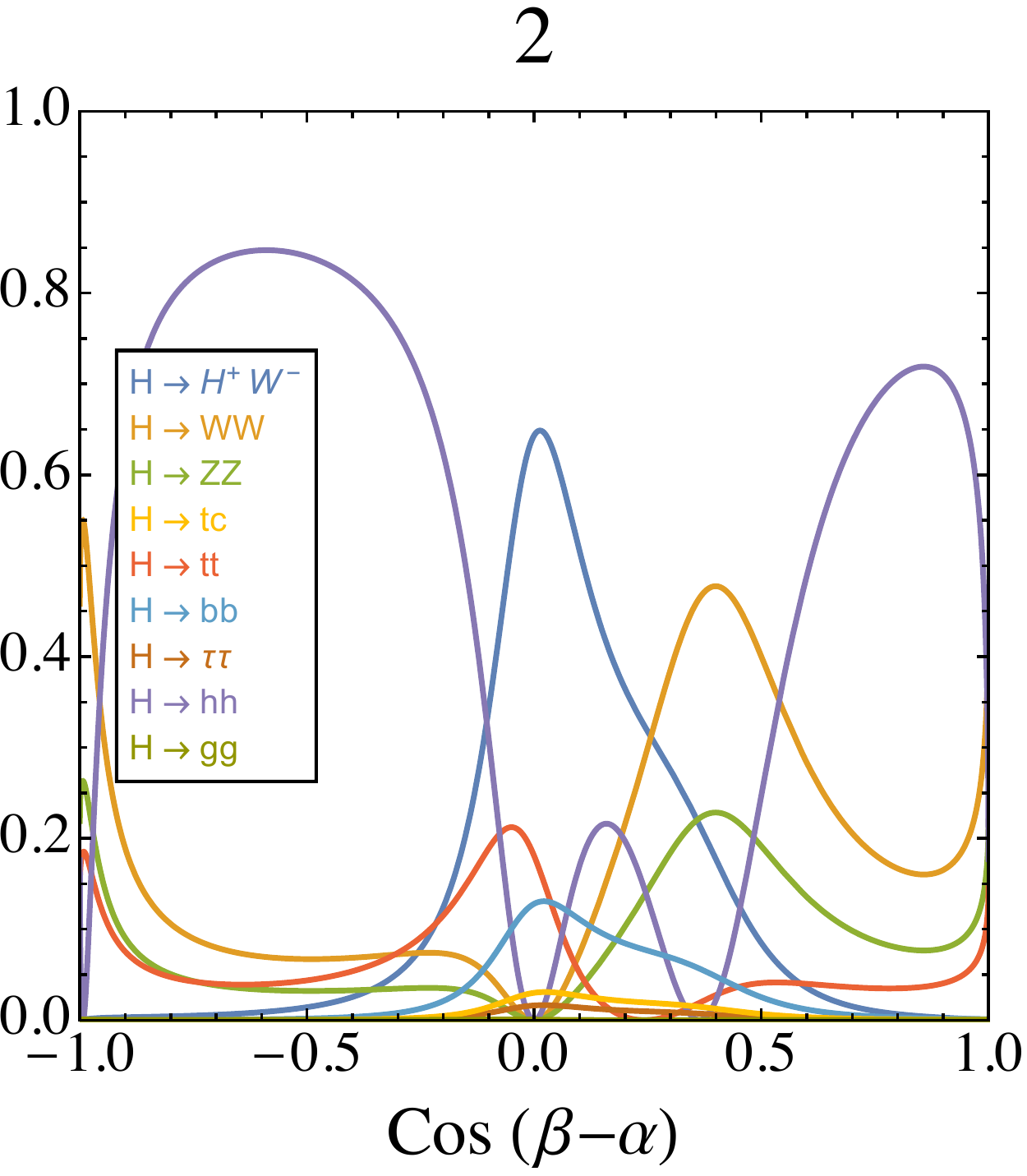}\\

\includegraphics[width =.45\textwidth]{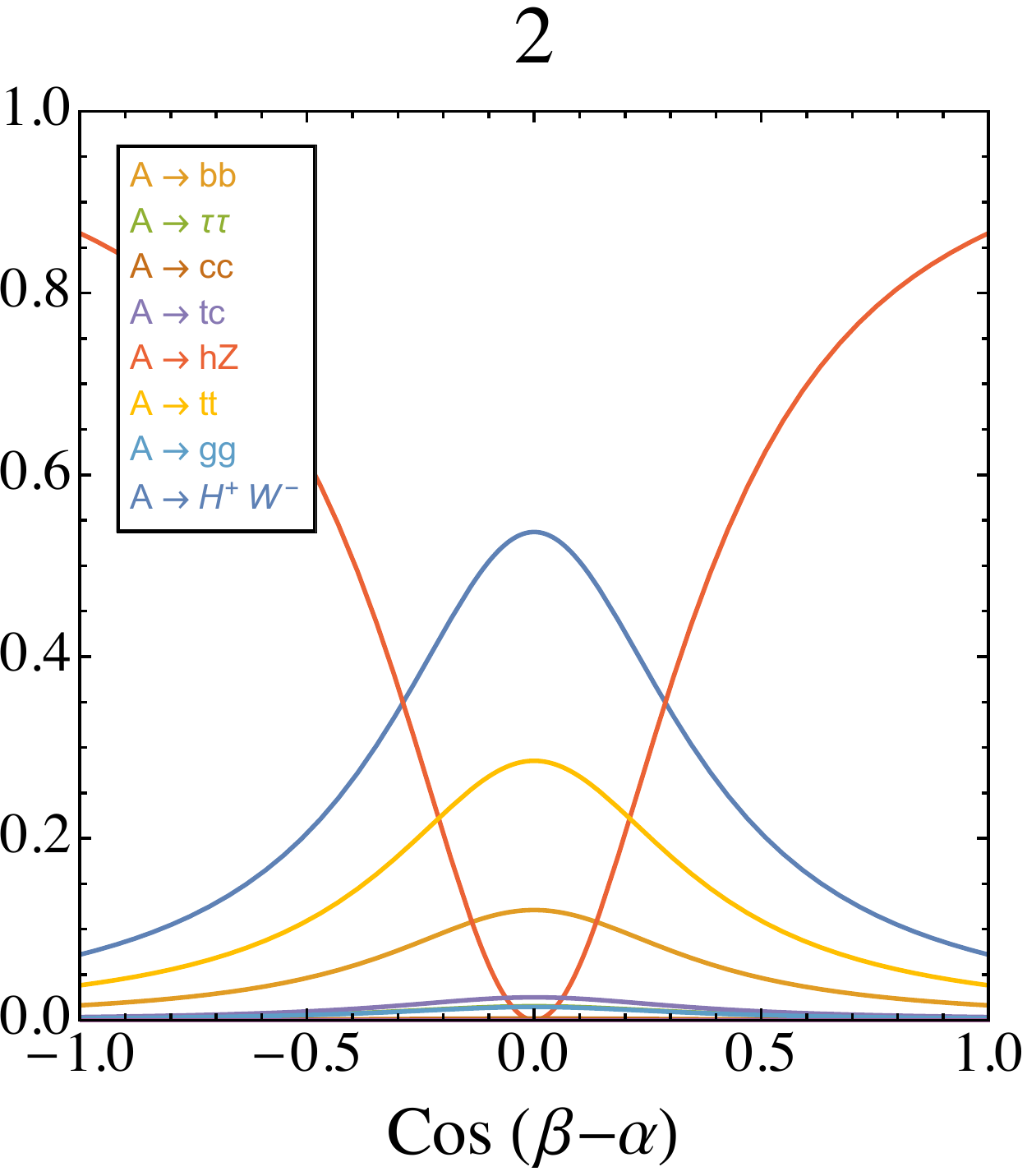}&\includegraphics[width =.45\textwidth]{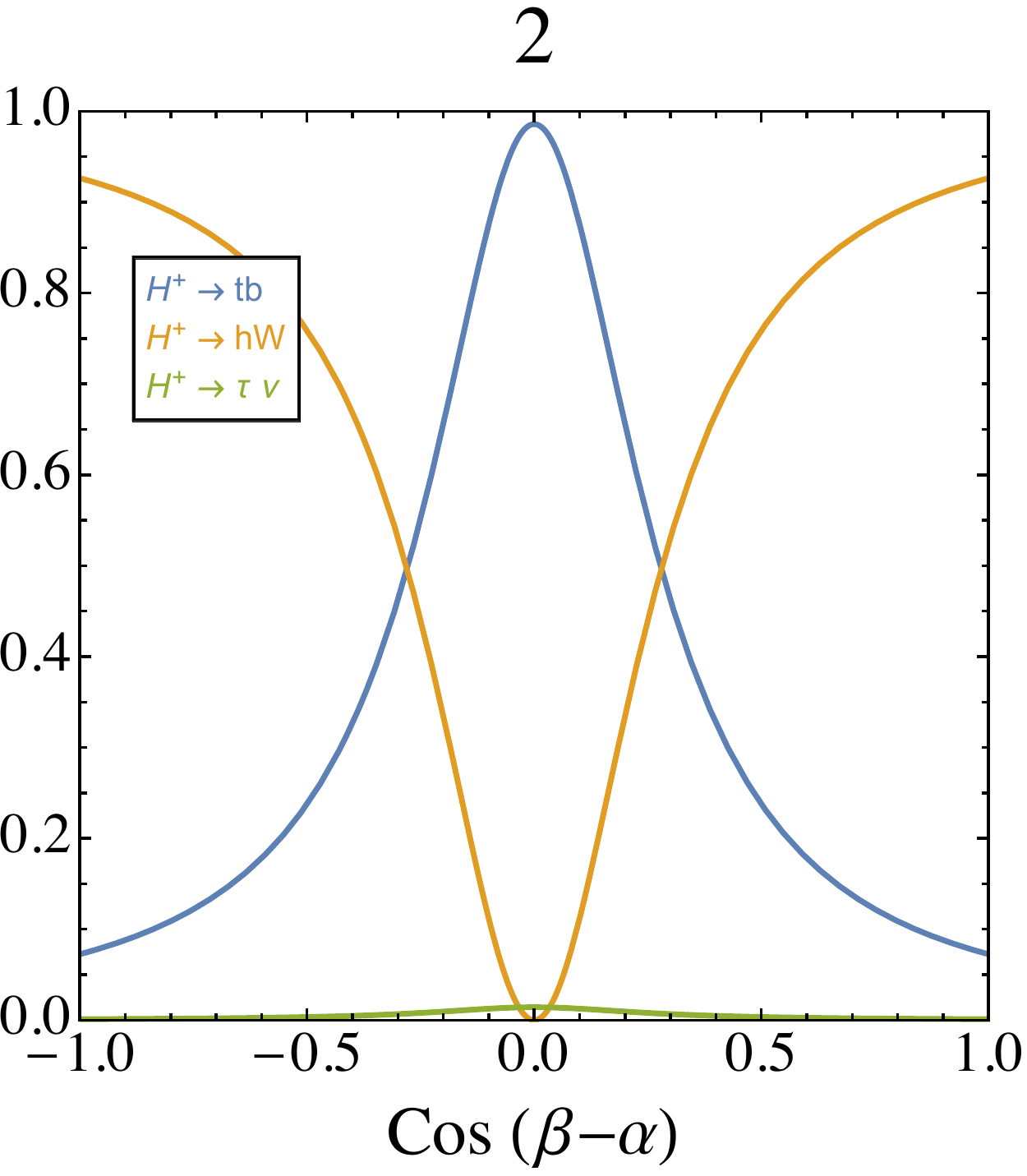}
\end{tabular}

\caption{Branching ratios as a function of $\cos(\beta-\alpha)$ for the light neutral scalar (upper left panel), heavy neutral scalar (upper right panel), pseudoscalar (lower left panel) and charged scalar (lower right panel) for the scalar masses and $\tan\beta$ of the benchmark scenario \textbf{2} defined in Table \ref{tab:B2}. }
\end{table}

}
\end{appendix}

\clearpage

\bibliography{hff.bib}
\bibliographystyle{JHEP}

\end{document}